\documentclass[twocolumn]{aa}
\usepackage{amsmath}
\usepackage{graphicx}
\usepackage{epsfig}
\usepackage{pifont}
\usepackage{color}
\usepackage{txfonts}
\graphicspath{{./}}
\definecolor{turq}{cmyk}{1,0,0,0.5}

\begin{document}

\title{A Nanoflare Distribution Generated by Repeated Relaxations Triggered by Kink Instability}

\author{M. R. Bareford\inst{1}\and P. K. Browning\inst{1}\and R. A. M. Van der Linden\inst{2}}
\offprints{P. K. Browning} \institute{Jodrell Bank Centre for Astrophysics,
Alan Turing Building, School of Physics and Astronomy, The University of Manchester, Oxford Road, Manchester M13 9PL, U.K. \email{michael.bareford@postgrad.manchester.ac.uk} \and
SIDC, Royal Observatory of Belgium, Ringlaan 3, B-1180 Brussels, Belgium}
\date{Received ; accepted }

\abstract
{It is thought likely that vast numbers of nanoflares are responsible for the corona having a temperature of millions of degrees. Current observational technologies lack the resolving power to confirm the nanoflare hypothesis. An alternative approach is to construct a magnetohydrodynamic coronal loop model that has the ability to predict nanoflare energy distributions.}
{This paper presents the initial results generated by a coronal loop model that flares whenever it becomes unstable to an ideal MHD kink mode. A feature of the model is that it predicts heating events with a range of sizes, depending on where the instability threshold for linear kink modes is encountered. The aims are to calculate the distribution of event energies and to investigate whether kink instability can be predicted from a single parameter.}
{The loop is represented as a straight line-tied cylinder. The twisting caused by random photospheric motions is captured by two parameters, representing the ratio of current density to field strength for specific regions of the loop. Instability onset is mapped as a closed boundary in the 2D parameter space. Dissipation of the loop's magnetic energy begins during the nonlinear stage of the instability, which develops as a consequence of current sheet reconnection. After flaring, the loop evolves to the state of lowest energy where, in accordance with relaxation theory, the ratio of current to field is constant throughout the loop and helicity is conserved.}
{There exists substantial variation in the radial magnetic twist profiles for the loop states along the instability threshold. These results suggest that instability cannot be predicted by any simple twist-derived property reaching a critical value. The model is applied such that the loop undergoes repeated episodes of instability followed by energy-releasing relaxation. Hence, an energy distribution of the nanoflares produced is collated. This paper also presents the calculated relaxation states and energy releases for all instability threshold points.}
{The final energy distribution features two nanoflare populations that follow different power laws. The power law index for the higher energy population is more than sufficient for coronal heating.}

{}

\titlerunning{The Energy Distribution of Nanoflares}
\maketitle

\section{Introduction}
The idea that nanoflares are sufficiently numerous to maintain coronal temperatures was first proposed by Parker (1988). This theory implies that the coronal background emission is the result of nanoflares occurring continually throughout the solar atmosphere. Hence, it is probably unfeasible to observe these flares individually. This perhaps explains why observational studies have failed to agree on the importance of nanoflares with regard to coronal heating (Krucker \& Benz 1998; Parnell \& Jupp 2000; Aschwanden \& Parnell 2002; Parnell 2004). The impracticality of nanoflare detection has motivated the development of models that attempt to show how coronal loops might dissipate their magnetic energy in the form of discrete heating events and thereby replenish coronal heating losses. The primary dissipation mechanism is thought to be magnetic reconnection, since strong evidence of this process has been found in observations of large-scale flares (Fletcher 2009; Qiu 2009). Coronal heating requires fast dissipation of magnetic energy, a requirement that is compatible with reconnection timescales.

A loop's excess magnetic energy is introduced via the random convective motions that occur at or below the loop's photospheric boundaries (i.e., footpoints). There are two possibilities for how the kinetic energy of these motions is dissipated within the loop (Klimchuk 2006). Direct current (DC) heating occurs when the Alfv\'en time is small compared to the timescale of photospheric turbulence. Thus, the loop moves quasi-statically through a series of force-free equilibria until the accumulating magnetic stresses are dissipated as heat. If the Alfv\'en time is slow compared to photospheric motions, alternating current (AC) heating takes place: footpoint motions generate waves (acoustic and MHD), some of which may deposit energy within the coronal loop. The waves that can penetrate the corona are thought to be restricted by type and frequency (Narain \& Ulmschieder 1996; Hollweg 1984). Whether or not these waves carry sufficient energy to heat the corona is not certain (Porter et al. 1994). There is considerably less doubt, however, over the sufficiency of DC heating, which is the basis for the model presented here. (AC heating may still be partly reponsible for coronal heating: the reconnection mechanism mentioned above may generate waves within the corona.)

The coronal part of the loop is defined by its magnetic field (plasma beta, $\beta$\,$\approx$\,0.01) and, since the loop's magnetic flux and plasma are frozen together, its magnetic field is continually twisted by random swirling motions at the photosphere. If the driving is slow compared to Alfv\'en timescales, the loop's force-free state can be represented by $\nabla\times\vec{B}\,=\,\alpha (\vec{r})\vec{B}$, where $\alpha\,=\,(\mu_{0}\vec{j}\cdot\vec{B})$/$(|\vec{B}|^{2})$ is the ratio of current density to magnetic field and \textit{$\vec{r}$} is a position vector (Woltjer 1958). 

Many 3D MHD models have shown how such coronal loops exhibit current sheet formation during the nonlinear phase of an ideal kink instability (Baty \& Heyvaerts 1996; Velli et al. 1997; Arber et al. 1999; Baty 2000). Essentially, helical current sheets become the site of Ohmic dissipation, resulting in a heating event. Further simulations have revealed the appropriate correlation between magnetic energy and Ohmic heating (Browning \& Van der Linden 2003; Browning et al. 2008; Hood et al. 2009). The energy released by an instability depends on the complex dynamics of the magnetic reconnection events that occur inside the loop. Obviously, this is difficult to model - although this has been achieved by 3D MHD simulations. The computational expense of these simulations rules them out as a means of exploring fully the relationship between the $\alpha$-profile and the amount of energy released. Fortunately, there is another way to calculate the energy release.

Relaxation theory states that when a magnetic field reaches instability (or is otherwise disrupted) it will evolve towards a minimum energy state such that the total magnetic axial flux and the \textit{global} magnetic helicity are conserved (Taylor 1974, 1986). The relaxed state is the well known constant-$\alpha$ or linear force-free field:
\begin{eqnarray}
  \label{forcefree_field}
  \nabla\times\vec{B} & = & \alpha\vec{B}.
\end{eqnarray}
The original intention of this theory was to explain laboratory plasma phenomena; but latterly, it has been frequently applied to the solar corona (Heyvaerts \& Priest 1984; Browning et al. 1986; Vekstein et al. 1993; Zhang \& Low 2003; Priest et al. 2005). The helicity measures the self-linkage of the magnetic field, see Berger (1999). A modified expression for this quantity needs to be used since the field lines cross the photospheric boundaries, which creates non-zero normal flux and therefore removes gauge invariance.
\begin{eqnarray}
  \label{relative_helicity}
  K & = & \int_V (\vec{A}+\vec{A'})\cdot (\vec{B}-\vec{B\,'})\,\,dV,
\end{eqnarray}
where \textit{$\vec{A}$} is the magnetic potential, \textit{$\vec{B\,'}$} is the potential field with the same boundary conditions and \textit{$\vec{A'}$} is the corresponding vector potential. This relative helicity (Berger \& Field 1984; Finn \& Antonsen 1985) is the difference between the helicity of the actual field and that of a potential field (represented by the dash terms) with the same normal flux distribution. In ideal MHD, the helicity of every flux region is conserved. Taylor proposed that in the presence of reconnection all such local invariants are destroyed, whilst the global helicity is conserved. Helicity is still affected by global resistive diffusion, however, it can still be treated as a conserved quantity, if the change in helicity is substantially smaller than the change in magnetic energy. The results of one aforementioned MHD simulation (Browning et al. 2008) have shown that \textit{$\delta$K/K}\,$\sim$\,10$^{-4}$ and \textit{$\delta$W/W}\,$\sim$\,10$^{-2}$ (it should be noted that these figures are limited by the coarseness of the numerical grid used in the simulation). During relaxation, helicity is redistributed more evenly within the loop; as a consequence $\alpha$ becomes invariant with radius. These properties enable one to locate a loop's relaxed state and hence calculate the energy released during relaxation, i.e., the upper limit of the flare energy.

In applying relaxation theory to coronal heating, it became clear that the effectiveness of the heating depends on how much free energy is stored before a relaxation event occurs (Heyvaerts \& Priest 1984). Browning \& Van der Linden (2003) proposed that relaxation is triggered by the onset of ideal MHD instability. Thus, the coronal field evolves quasi-statically in response to slow photospheric driving, building up free magnetic energy - until the field becomes unstable (i.e., the threshold for linear instability is reached). At this point, a dynamic heating event ensues. It should be noted that ideal instabilities are relevant (as opposed to resistive instabilities) because the time-scales are sufficiently fast. Dissipation and energy release can occur during the nonlinear phase of the instability, as has been demonstrated extensively by numerical simulations of the nonlinear kink instability (Galsgaard \& Nordlund 1997; Velli et al. 1997; Lionello et al. 1998; Baty 2000; Gerrard et al. 2001). The helical deformation of the kink instability generates current sheets in the nonlinear regime, in which fast magnetic reconnection rapidly dissipates magnetic energy. Recently, it has been demonstrated using 3D MHD simulations that this process causes the field to relax towards a state which is closely-approximated as a constant-$\alpha$ state, and the energy release is in good agreement with relaxation theory (Browning et al. 2008; Hood et al. 2009). These papers have investigated field profiles, based on the model of Browning \& Van der Linden (2003), in the unstable region of parameter space and have shown that fast reconnection develops in a current sheet, with dissipation of magnetic energy and relaxation to a new, lower-energy state. Efficient "mixing" of the $\alpha$ profile is facilitated in the later phases of the evolution, during which the current sheet fragments (Hood et al. 2009).

Supported by this evidence, the model developed here is based on the idea that coronal magnetic fields respond to photospheric driving by evolving through force-free equilibria, with a heating event triggered whenever the ideal instability threshold is reached. Following Browning \& Van der Linden (2003), we use a simple cylindrical field model, and represent the nonlinear force-free field using a two-parameter family of current profiles in which $\alpha$ is a piecewise constant. Browning and Van der Linden considered only individual heating events. The primary aim of this paper is to predict a \textit{distribution} of heating events, generated by an ongoing stress-relax cycle. The photospheric driving stresses the coronal magnetic field until it becomes unstable and relaxes; then the driving resumes, and the process repeats, leading to a series of heating events as expected in the nanoflare coronal heating scenario. This is modelled through a monte-carlo approach, with the photospheric footpoint motions treated as random. Recent observational evidence (Abramenko et al. 2006) lends support to the idea that  random, turbulent photospheric motions can provide an energy source for coronal heating in Active Regions.

In calculating the relaxation and distribution of heating events, we need to know the linear instability threshold for line-tied coronal loops over a family of field profiles. As a consequence, we obtain some interesting new results concerning stability properties of cylindrical loops, for a much more extensive range of current profiles than previously studied (including profiles for which the sense of the twist of the field varies across the loop). A secondary aim is thus to explore the properties of the linear kink instability on this family of fields, and in particular, to determine the extent to which stability can be defined by any single quantity such as "critical twist".

The paper is structured in the following manner. Section 2 describes the composition of the loop model, along with the equations used to express the loop's magnetic field. The calculation of the loop's instability threshold is also explained, as well as the procedure for allowing the loop to repeatedly undergo instability and the equations used to determine the energy release associated with each relaxation. Section 3 outlines the analysis of possible critical parameter values for the onset of loop instability. The results of the simulations of energy release are presented in Sect. 4 as nanoflare energy distributions. Finally, in the last section, the results are discussed and our conclusions are given.

\section{Model}
\begin{figure}[h!]
  \vspace{-10pt}
  \center
  \includegraphics[scale=0.32]{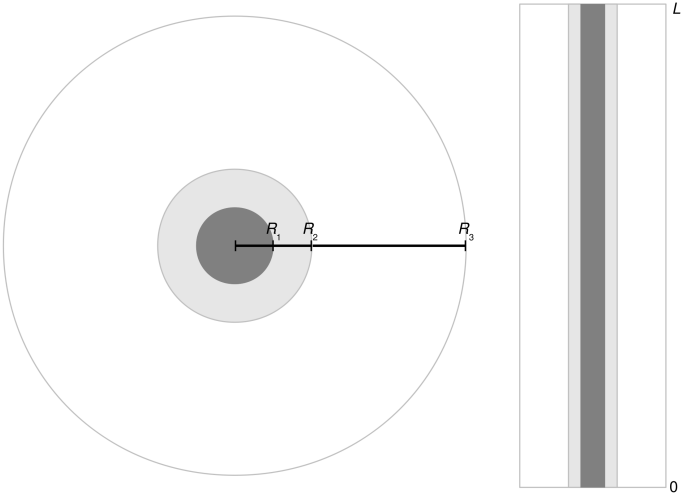}
  \caption{\small{Schematic of a straightened coronal loop in the \textit{r}-$\theta$ plane (left) and in the \textit{r}-\textit{z} plane (right). The loop, comprises a core (dark grey) and an outer layer (light grey); it is embedded in a potential envelope (white). The core radius is half the loop radius and 1/6 the envelope radius (\textit{R}$_1$:\textit{R}$_2$:\textit{R}$_3$ = 0.5:1:3). The loop's aspect ratio (\textit{L}/\textit{R}$_2$) is 20.}}
  \vspace{-10pt}
  \label{schematic}
\end{figure}

The model discussed here was used by Browning \& Van der Linden (2003) and then extended by Browning et al. (2008) to include a potential envelope (Fig. \ref{schematic}). 
First, a loop is considered to evolve through equilibria as it is driven by photospheric footpoint motions. An idealised model of a straight cylindrical loop is used with the photosphere represented by two planes at \textit{z}\,=\,0,\,\textit{L}; however, the essential physics should apply to more complex geometries. The stressed field is line-tied with (in general) a non-uniform $\alpha$(\textit{r}) (where $\alpha\,=\,\mu_{0}\vec{j}_{\parallel}$/$\vec{B}$). This is represented by a two-parameter family of piecewise-constant-$\alpha$ profiles. Secondly, it is proposed that a relaxation event is triggered when the loop's field becomes linearly unstable. The energy released, due to fast magnetic reconnection during the nonlinear development of the instability, can then be calculated using relaxation theory.

The MHD kink instability needs to be ideal in order to be consistent with the observed rapidity of flare occurrences. However, ideal conditions mean there is no resitivity to dissipate magnetic energy; the 3D MHD simulations discussed in the Introduction provide a solution to this impasse. Once a linear instability has achieved a positive growth rate, it will soon become nonlinear: at this point, current sheets will form wherein fast reconnection of the magnetic field can take place. These expectations are justified by the results of the cited numerical simulations. The linear perturbation can be represented as $f\,=\,\tilde{f}(r,z)e^{i\theta}e^{\gamma t}$, where the azimuthal mode number is set to 1; this mode has been found to be the least stable (Van der Linden \& Hood 1999). The effect of such perturbations on the coronal loop are represented by the standard set of linearised ideal MHD equations.

\subsection{Equilibrium fields}
The loop's radial $\alpha$-profile is approximated by a piecewise-constant function featuring two parameters. This design, first proposed by Melrose et al. (1994), is readily extensible: extra layers of constant $\alpha$ can be inserted to obtain more realistic profiles. The ratio of current to magnetic field is $\alpha_1$ in the core, $\alpha_2$ in the outer layer and zero in the potential envelope. Note that the magnetic field is continuous everywhere (though the current has discontinuities). Recent work indicates that these $\alpha$ discontinuities have little discernable effect when compared to similar but continuous $\alpha$-profiles (Hood et al. 2009).

Without an envelope, the loop's outer surface (located at \textit{R}$_2$) acts as a conducting wall. This is unrealistic in the context of the solar corona; the loop would be more stable than it might be otherwise. Browning et al. (2008) plotted the relationship between the growth rate of the instability and the distance to the outer surface of the potential envelope (i.e., a more distant conducting wall). They found that for six unstable loop states the growth rate was invariant once the outer surface of the envelope (\textit{R}$_3$) exceeded $\frac{3}{2}$\,\textit{R}$_2$. The \textit{R}$_3$ boundary was placed at twice this value.

The fields are expressed in terms of the well-known Bessel function model, generalised to the concentric layer geometry (Melrose et al. 1994; Browning \& Van der Linden 2003; Browning et al. 2008). These expressions will change slightly whenever $\alpha_1$ or $\alpha_2$ become negative; these alterations are captured by $\sigma$ symbols: $\sigma_1\,=\,\frac{\alpha_1}{|\alpha_1|}$, $\sigma_2\,=\,\frac{\alpha_2}{|\alpha_2|}$ and $\sigma_{1,2}\,=\,\sigma_1\sigma_2$. (The sign of an $\alpha$ term merely denotes the orientation of the azimuthal field). Thus, the field equations for the three regions (core, outer layer and potential envelope) are as follows:
\begin{eqnarray}
  \label{field_equation_1}
  B_{1z} & = & B_1 J_0(|\alpha_1|r),\\
  \label{field_equation_2}
  B_{1\theta} & = & \sigma_1 B_1 J_1(|\alpha_1|r), \mbox{\hspace{3.6cm}} 0 \leq r \leq R_1,\\ 
  \nonumber &  &\\
  \label{field_equation_3}
  B_{2z} & = & B_2 J_0(|\alpha_2|r) + C_2 Y_0(|\alpha_2|r),\\
  \label{field_equation_4}
  B_{2\theta} & = & \sigma_2(B_2 J_1(|\alpha_2|r) + C_2 Y_1(|\alpha_2|r)), \mbox{\hspace{1.3cm}} R_1 \leq r \leq R_2,\\
  \nonumber &  &\\
  \label{field_equation_5}
  B_{3z} & = & B_3,\\
  \label{field_equation_6}
  B_{3\theta} & = & \sigma_2\frac{C_3}{r}R_2, \mbox{\hspace{4.25cm}} R_2 \leq r \leq R_3.  
\end{eqnarray}
The fields must be continuous at the inner radial boundaries, \textit{R}$_1$ and \textit{R}$_2$. Therefore, the constants \textit{B}$_2$, \textit{B}$_3$, \textit{C}$_2$ and \textit{C}$_3$ can be expressed like so:
\begin{eqnarray}
  B_2 & = & B_1\frac{\sigma_{1,2}J_1(|\alpha_1|R_1)Y_0(|\alpha_2|R_1)-J_0(|\alpha_1|R_1)Y_1(|\alpha_2|R_1)}{\Delta},\\
  \nonumber &  &\\
  C_2 & = & B_1\frac{J_0(|\alpha_1|R_1)J_1(|\alpha_2|R_1) - \sigma_{1,2}J_1(|\alpha_1|R_1)J_0(|\alpha_2|R_1)}{\Delta},\\
  \nonumber &  &\\  
  B_{3} & = & B_2 F_0(|\alpha_2|R_2),\\
  \nonumber &  &\\
  C_{3} & = & B_2 F_1(|\alpha_2|R_2),
\end{eqnarray}
where
\begin{eqnarray}
  \Delta & = & \frac{2}{\pi|\alpha_2|R_1},\\  
  F_{0,1}(x) & = & J_{0,1}(x) + \frac{C_2}{B_2}Y_{0,1}(x).
\end{eqnarray}
At all times, the magnetic flux through the loop and envelope is conserved:
\begin{eqnarray}
  \label{total_axial_flux}
  \nonumber \psi^{*} & = & \int_0^{R_3}2\pi r^* B_z^* \, dr^* = \frac{2\pi B_2}{|\alpha_2|}R_2 F_1(|\alpha_2|R_2)\\
  \nonumber  &   & \,\,\,\,\,\,\,\,\,\,\,\,\,\,\,\,\,\,\,\,\,\,\,\,\,\,\,\,\,\,\,\,\,\,\,\,\,\,\,\,\,\,\,\,\,\,\, +\,2\pi R_1 B_1 J_1(|\alpha_1|R_1)\Bigg(\frac{1}{|\alpha_1|}-\frac{\sigma_{1,2}}{|\alpha_2|}\Bigg)\\
  &   & \,\,\,\,\,\,\,\,\,\,\,\,\,\,\,\,\,\,\,\,\,\,\,\,\,\,\,\,\,\,\,\,\,\,\,\,\,\,\,\,\,\,\,\,\,\,\, +\,\pi B_2 F_0(|\alpha_2|R_2)\Big(R_3^{\,2} - R_2^{\,2}\Big),
\end{eqnarray}
where the asterisks denote dimensionless quantities. Hence, in the model, $\psi^{*}$ is normalised to 1 and \textit{B}$_1$ can be determined (noting that, in equation \ref{total_axial_flux}, $B_2$ is a function of $B_1$). We also normalise with respect to the coronal loop radius, \textit{R}$_2$, see Fig. \ref{schematic}.

As the random motions of the photosphere proceed, the loop evolves through a series of force-free equilibrium states until it becomes linearly unstable. We now discuss the calculation of the instability onset.

\subsection{Linear kink instability threshold}
\begin{figure}[h!]  
  \vspace{-10pt}
  \center
  \includegraphics[scale=0.35]{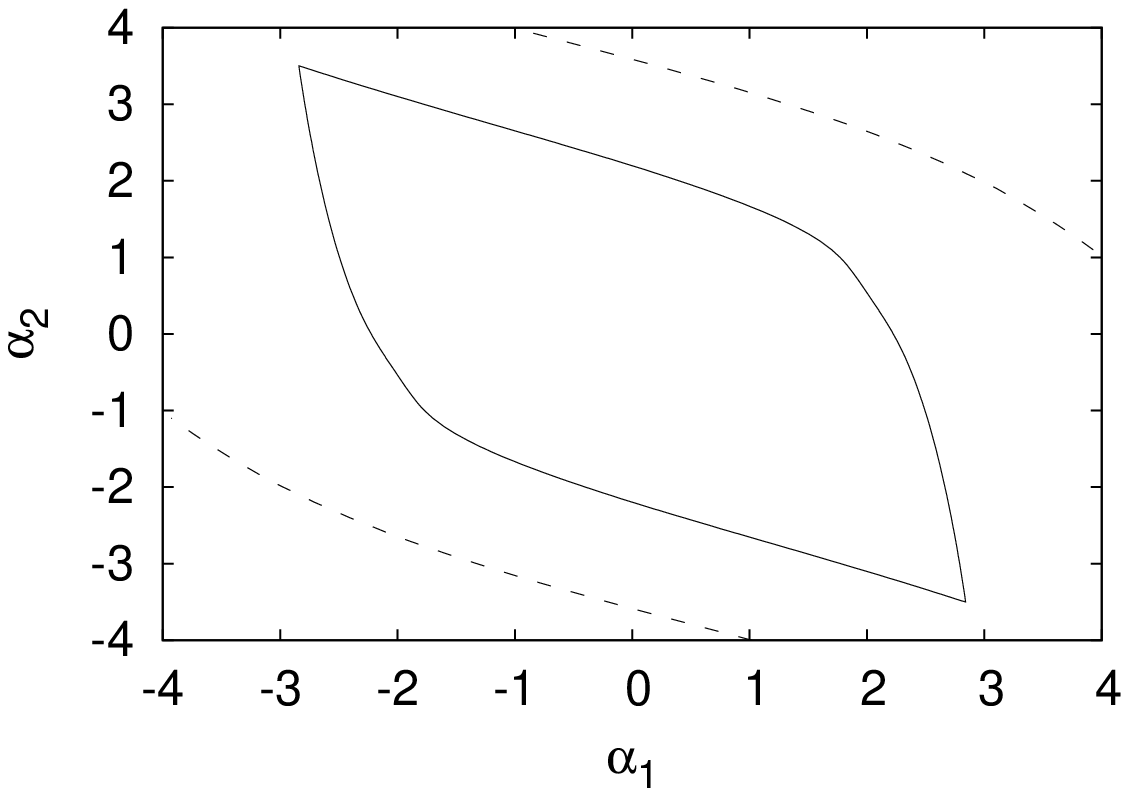}
  \hspace{-5pt}  
  \includegraphics[scale=0.35]{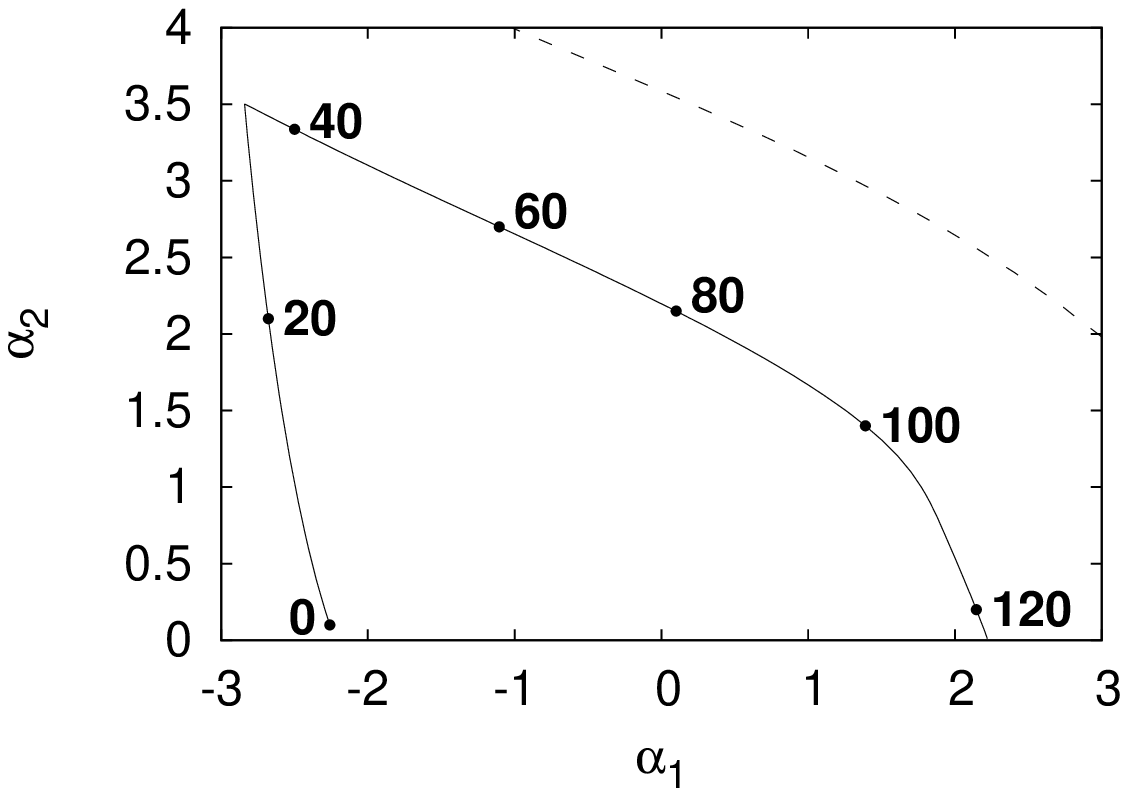}
  \vspace{-5pt}
  \caption{\small{The left panel shows the closed instability threshold (solid) with the \textit{B}$_z$ reversal lines (dashed). The top half of the threshold (where $\alpha_2\,>\,0$), annotated with threshold point numbers, is shown in the right panel.}}
  \vspace{-10pt}
  \label{it}
\end{figure}

A coronal loop's instability is constrained by the line-tying of the photospheric footpoints (Hood 1992). Hence, all perturbations are required to vanish at the loop ends (\textit{z}\,=\,0,\,\textit{L}). When the growth rate of a perturbation transitions from a negative value to a positive one, the loop has reached the threshold of an ideal linear instability. The instability threshold is a curve in 2-dimensional $\alpha$-space ($\alpha_1$, $\alpha_2$). The properties of the loop (e.g., $\alpha_1$ and $\alpha_2$) at these threshold points can be found by substituting the perturbation function into the linearised MHD equations, leading to an eigenvalue equation for the growth rates (Chap. 7, Priest 1987). The growth rates and eigenfunctions of the most unstable modes are found numerically, for line-tied fields, with the CILTS code, described in Browning \& Van der Linden (2003) and Browning et al. (2008). CILTS can be configured such that one of the loop's $\alpha$ parameters is fixed whilst the other is incremented. The code terminates as soon as the real part of the eigenfunction falls below zero, i.e., the loop is no longer unstable to kink perturbations. The left panel of Fig. \ref{it} shows the closed instability threshold curve mapped by the CILTS code (see also Fig. 5 of Browning et al. 2008). The threshold curve has symmetry: it is invariant when rotated by $\pi$ radians. Thus, it is sufficient to show how various properties (e.g., magnetic twist and energy release) vary along the top half of the threshold curve. For ease of plotting we can convert this half of the threshold curve to a one dimensional form: the filled circles and bold numbers shown in the right panel of Fig. \ref{it} represent the tic marks and labels for the 1D threshold point axis, see Figs. \ref{it_a2p_tw_0_r1_r2}\,-\,\ref{it_a2p_k_w_arx_wr}.

It is important to remember that the threshold only applies to the specific loop geometry outlined above. A new threshold would need to be calculated should the loop's proportions, composition or envelope change as a consequence of some activity.
Another caveat is that there are some points in $\alpha$-space that yield singularities when calculating quantities such as helicity or magnetic energy. These arise when the axial field (\textit{B}$_z$) has a significant region of reversal; since $\psi^{*}$ is normalised to 1 and conserved, the helicity and energy terms will diverge when unnormalised $\psi^{*} \to 0$ (see Browning \& Van der Linden 2003). Fortunately, the instability threshold does not enter the region where \textit{B}$_z$ begins to pass through zero (see dashed lines in Fig. \ref{it}), so these singularities are not encountered. 

\subsection{Random walk}
\label{sec:random_walk}
When a loop is twisted by turbulent photospheric motions, it performs a random walk through the $\alpha$-space enclosed by the instability threshold (Fig. \ref{it_rl_random_walks}). This traversal of $\alpha$-space is random in direction but constant in length (we set the dimensionless step-length to be $\delta\alpha$\,=\,0.1 for most of the results presented here). Clearly, the nature of this random walk depends on the statistical properties of the driving photospheric motions; in future, different forms of this random driving will be investigated, but for now, we take the simplest assumptions. 

The time unit $\tau$ is the step time, the time taken for $\alpha$ to change by 0.1/\textit{R}$_2$ (in dimensional units). We may estimate, roughly, a timescale for this process as follows. Based on axial values, a change $\delta\alpha$ corresponds to a change in magnetic twist $\delta\phi$\,$\approx$\,(\textit{L}/2)($\delta\alpha$/\textit{R}$_2$); taking \textit{L}/\textit{R}$_2$\,=\,20 gives $\delta\phi$\,$\approx$\,1. If this is caused by photospheric twisting motions of magnitude \textit{v}$_{\theta}$ for a time interval $\tau$, we find $\tau$\,$\approx$\,($\delta\phi$)\textit{R}$_f$/\textit{v}$_{\theta}$, where \textit{R}$_f$ is the footpoint radius. With typical values of \textit{R}$_f$\,=\,200\,km and \textit{v}$_{\theta}$\,=\,1\,km\,s$^{-1}$, we obtain $\tau$\,$\approx$\,200 s; note that this is consistent with quasi-static evolution, justifying \textit{a posteriori} our choice of random-walk step size. The step time ($\tau$) may be identified with the correlation time of photospheric motions, which is likely to be rather longer than the value given above; for example, a granule lifetime of 1000 s may be appropriate (Zirker \& Cleveland 1993). The effect of increasing the step length ($\delta\alpha$) is considered in Sect. \ref{sec:random_walk_step_size}

Eventually, the field will reach the instability threshold: it will become linearly unstable. At this point, the field releases energy and transitions to a lower-energy state defined by Taylor relaxation: helicity is conserved and the $\alpha$-profile relaxes to a single value. 

\subsection{Energy release calculation}
\label{sec:energy_release_calculation}
We have extended the model (Browning \& Van der Linden 2003; Browning et al. 2008) to allow a loop to \textit{repeatedly undergo relaxation} as it evolves within $\alpha$-space. Initially, a loop starts from a randomly-selected stable state. The field profile then undergoes a random walk until it crosses the instability threshold; whereupon, the loop relaxes and the profile transitions to the relaxation line ($\alpha_1$\,=\,$\alpha_2$). The constant $\alpha$-value ($\alpha_{rx}$) will, of course, vary depending on where the threshold was crossed; $\alpha_{rx}$ is found by helicity conservation (Browning \& Van der Linden 2003). In mathematical terms, we find the roots of the following equation:
\begin{eqnarray}
  \label{helicity_root}
  K(\alpha_{rx}) - K(\alpha_{\textit{it1}},\alpha_{\textit{it2}}) & = & 0,
\end{eqnarray}
where $\alpha_{it1}$ and $\alpha_{it2}$ are the coordinates of the instability threshold crossing (conservation of axial flux is assured through the normalisation $\psi^{*}$\,=\,1). The helicity can be expressed as follows:
\begin{eqnarray}
  \label{helicity}
  K & = & 2L \int_{0}^{R_3}\frac{I\psi (r)}{r} \, dr,  
\end{eqnarray}
where \textit{I} is the current and \textit{L} is the loop length (Finn \& Antonsen 1985). The equation for the magnetic energy contained within the loop and envelope is straightfoward:
\begin{eqnarray}
  \label{energy}
  W & = & \frac{L\pi}{\mu_0} \int_{0}^{R_3}rB^{\,2} \, dr,
\end{eqnarray}
where \textit{L} is normalised to 20 (since \textit{R}$_2$\,=\,1) and $\mu_0$ is set to 1. See Appendix A for the full expressions. The energy difference between the unstable and relaxed states can be calculated thus:
\begin{eqnarray}
  \label{energy_release}
  \delta W & = & W(\alpha_{\textit{it1}},\alpha_{\textit{it2}}) - W(\alpha_{rx}).
\end{eqnarray}
This is the relaxation energy: the energy released as heat during the event.

\begin{figure}[h!]
  \vspace{-10pt}
  \center
  \includegraphics[scale=0.64]{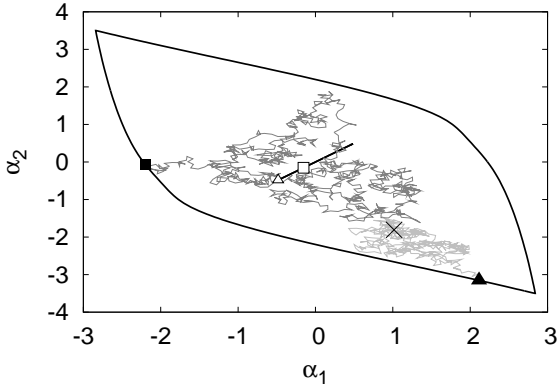}
  \vspace{-5pt}
  \caption{\small{The instability threshold encloses the relaxation line, which is a subsection of the $\alpha_1$\,=\,$\alpha_2$ line centred on the origin. The annotations illustrate the initial stages of a simulation that begins at the position marked by the cross. The first random walk is shown in light grey; it ends at the threshold position marked $\blacktriangle$ and the associated relaxation point is indicated by $\triangle$, which is the starting point for a second random walk (dark grey). This walk attains instability onset at the position marked $\blacksquare$ and relaxes to the point labelled $\square$ - the starting point for a third walk.}}
  \vspace{-5pt}
  \label{it_rl_random_walks}
\end{figure}

After relaxation, the loop resumes its random walk until it reaches the threshold and the process repeats. Fig. \ref{it_rl_random_walks} gives an illustration of this process.
We thus generate a sequence of energy release events, which are collated to produce a nanoflare energy distribution, see Sect. \ref{sec:flare_energy_distribution}. 

When the loop relaxes, the $\alpha$-profile throughout the loop and envelope becomes constant, i.e., $\alpha_1$\,=\,$\alpha_2$\,=\,$\alpha_3$\,$\neq$\,0. The envelope is no longer potential; it has acquired a residual current. In principle, as the main portion of the loop is twisted again by ongoing motions, a new equilibrium would develop with varying current ($\alpha_1$,\,$\alpha_2$) in the loop and a non-zero current ($\alpha_3$) in the envelope. For some threshold sections the consequent residual current is so small that the threshold shape would remain unchanged. However, the validity of the simulation process can only be assured by including an extra stage, wherein the envelope dissipates its helicity so that it becomes potential again. The loop can now resume its random walk with respect to the same threshold.

Although the primary purpose of this model is to calculate the distribution of energy releases, in achieving this we also obtain some interesting new results on linear stability, which are summarised in the next section. This is because, in order to explore the full parameter space of equilibrium current profiles, we have calculated the linear stability properties of a much wider family of fields than previously investigated: in particular, fields with reversed twists.

\section{Instability threshold and critical twist}
\label{sec:instability_threshold_and_critical_twist}
The evolution of the field profile through $\alpha$-space is determined by photospheric perturbations which map to changes in the magnetic field and hence to changes in $\alpha_1$ and $\alpha_2$. A loop's magnetic twist is directly related to rotational photospheric motions:
\begin{eqnarray}
  \label{magnetic_twist}  
  \varphi & = & \frac{LB_{\theta}(r)}{rB_z(r)}, \mbox{\hspace{1cm}} 0 \leq r \leq R_3.
\end{eqnarray}
Thus, the photospheric motions directly determine a loop's $\varphi$-profile, which in turn determines $\alpha$(\textit{r}) (and hence, in our model, $\alpha_1$ and $\alpha_2$). The magnetic twist of coronal loops is an observable feature (Kwon \& Chae 2008). Portier-Fozzani et al. (2001) have even observed a loop's twist decreasing over time - evidence perhaps of a loop evolving towards a state of minimum energy.

\subsection{Criteria for instability}
\label{sec:criteria_for_instability}
Many workers have looked at the idea that the magnetic twist of a loop can be used as a proxy for the onset of kink instability with some \textit{critical twist} parameter determining instability. Hood \& Priest (1979) performed a MHD stability analysis on a variety of straightened line-tied coronal loops. They showed that the critical twist, $\varphi_{crit}$, varies according to the aspect ratio (\textit{L}/\textit{R}$_2$), plasma beta ($\beta$) and the transverse magnetic structure (i.e., the nature of the variable twist profile). In general, short fat loops (low aspect ratio) have low critical twists, whereas long thin loops (high aspect ratio) have high critical twists. Subsequent work (Hood \& Priest 1981) revealed that a loop of \textit{uniform twist}, aspect ratio 10 and zero $\beta$ possessed a critical twist of 2.49$\pi$. This figure is often quoted as a general result.

The question arises as to whether there is any single parameter (such as peak or average twist) which determines instability onset for \textit{all} twist profiles. Indeed, more generally, it would be desirable to have a single quantity determining instability onset even for more complex (non-cylinderical) fields (Malanushenko et al. 2009). Incidental to our task, we have calculated the stability properties of an extensive family of equilibria, including fields with reversed twist (as well as simple monotonic-twist profiles as used by other authors). This provides a very useful test for any proposed criteria for instability onset.

Loops with variable-twist profiles have been previously studied; however, all such profiles generally have a similar form (Velli et al. 1990; Miki\'c et al. 1990; Baty 2001): the axial twist is the maximum, then the twist declines to a negligible value at the loop boundary. Velli et al. (1990) calculated that instability occurred when $\varphi_0$\,=\,$\varphi$(\textit{r}=0)\,=\,2.5$\pi$; this agrees with Hood and Priest's result for a uniform twist profile. Miki\'c et al. (1990) calculated a critical axial twist of 4.8$\pi$, the loop's average twist however, was $\sim$2.5$\pi$.

The idea of using magnetic twist as a marker for instability relies on the existence of some twist-derived parameter having a constant value for all the points on the instability threshold. Baty (2001) used a MHD stability code to show that for a small set of equilibria the average twist at instability is the same for several different magnetic configurations. However, there are several differences between Baty's work and the model presented here. Firstly, the twist profiles defined for each equilibrium are all positive and none contain multiple peaks (Fig.1, Baty 2001). Furthermore, critical twist convergence arises when the normalised distance, \textit{d}, is greater than 5 (Fig.5, Baty 2001).
\begin{eqnarray}
  \label{normalised_distance}
  d & = & \frac{\varphi_0 R_2}{L}.
\end{eqnarray}
In our equilibria, $d\,=\,\varphi_0/20$, the threshold values of the absolute axial magnetic twist vary from zero to 9$\pi$, which means \textit{d} varies from 0 to 1.42, thus the average twists will not be in the regime where the critical axial twist converges to 2.5$\pi$. In fact, the large $d$ regime cannot be attained.

The idea of a single critical average twist seems unlikely when one examines the threshold presented here. Clearly, at some threshold points $\alpha_{1}$ and $\alpha_{2}$ are of opposite sign; thus, one or two places on the threshold will have an average twist equal to zero, and yet they are unstable. Perhaps critical twist is achieved within a subsection of the loop; this idea is explored further in Sects. \ref{sec:radial_twist_profiles_and_linear_eigenfunctions} and \ref{sec:critical_twist_parameters}.

\subsection{Radial twist profiles and linear eigenfunctions}
\label{sec:radial_twist_profiles_and_linear_eigenfunctions}
\begin{figure}[h!]
  \vspace{-10pt}
  \center
  \includegraphics[scale=0.35]{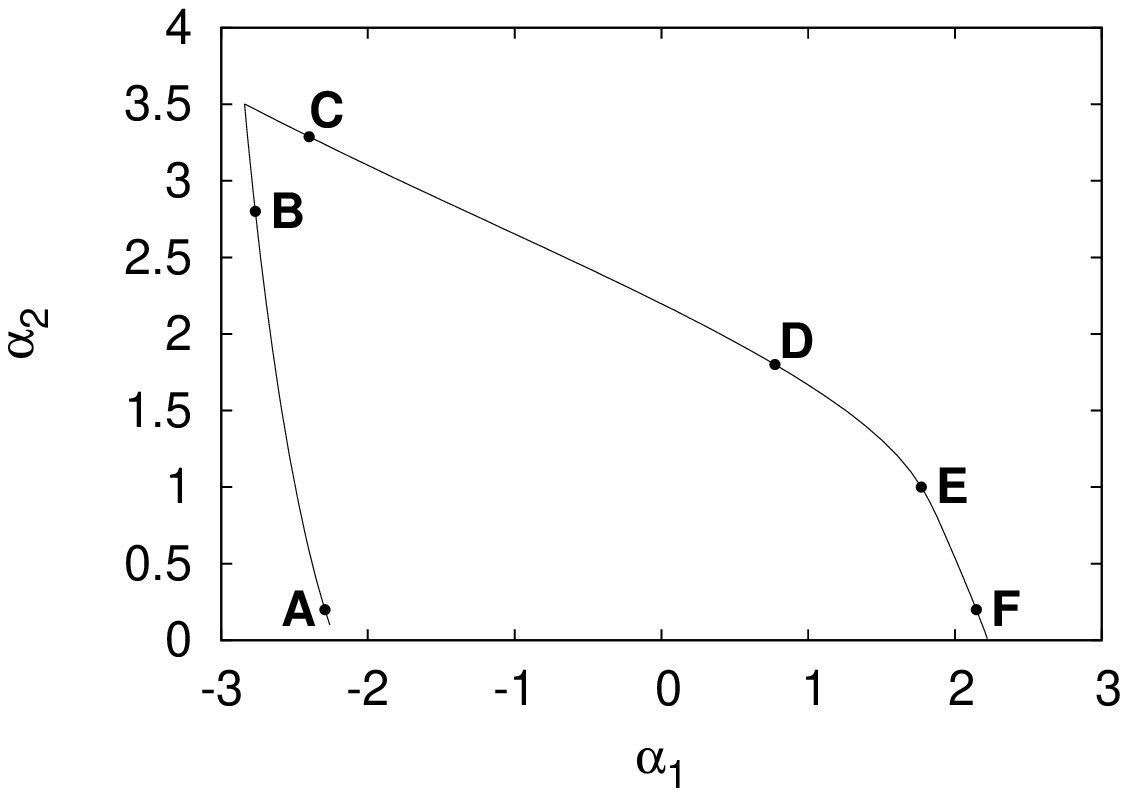}
  \includegraphics[scale=0.35]{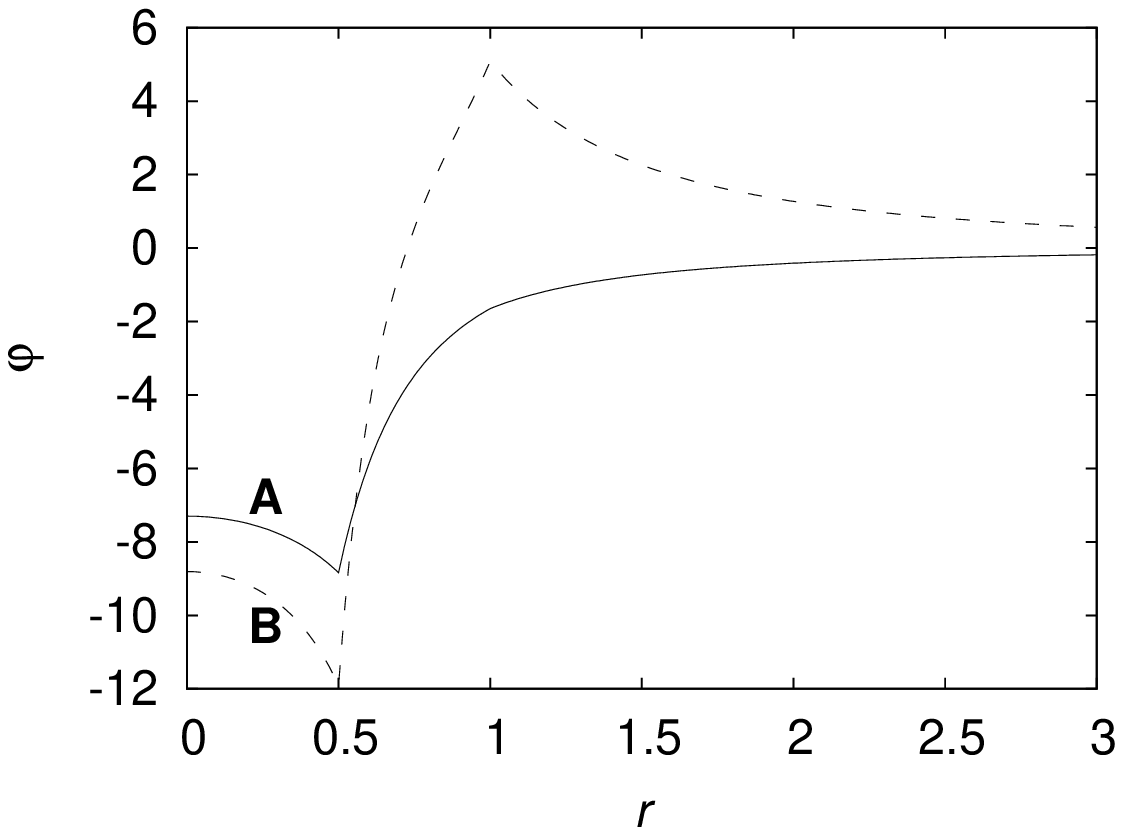}
  \includegraphics[scale=0.35]{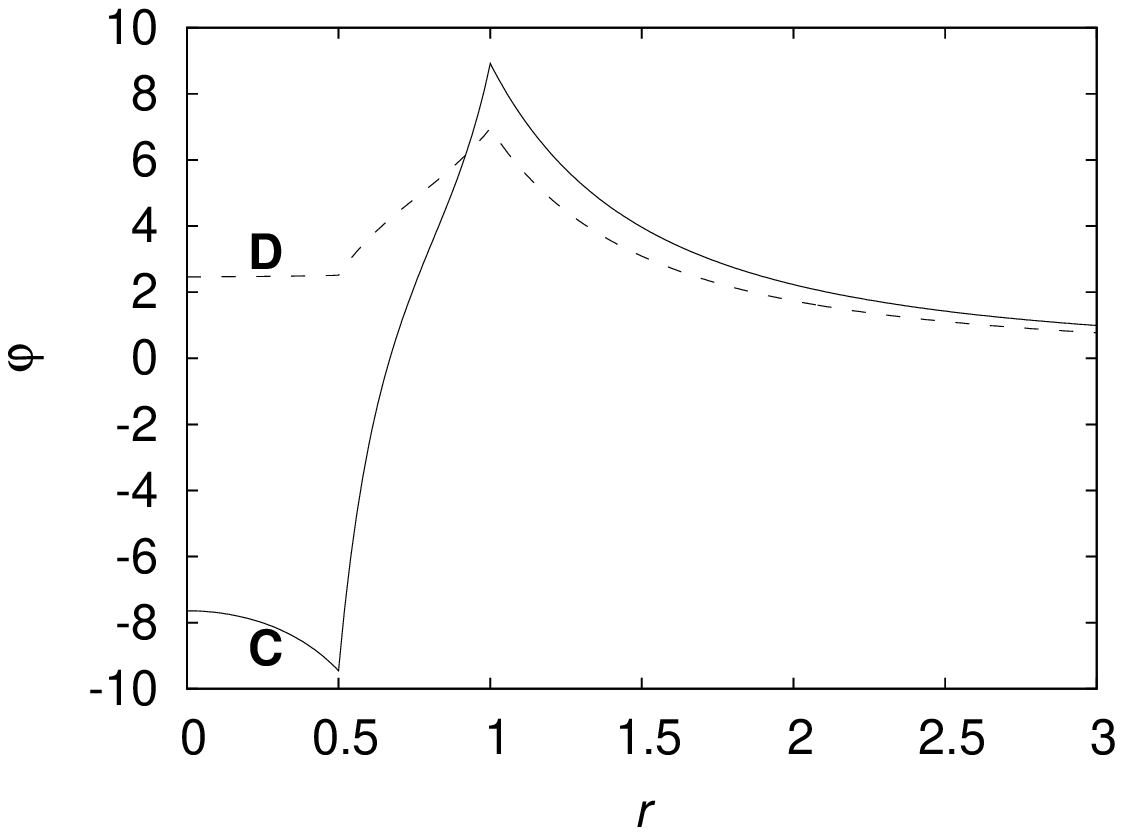}
  \includegraphics[scale=0.35]{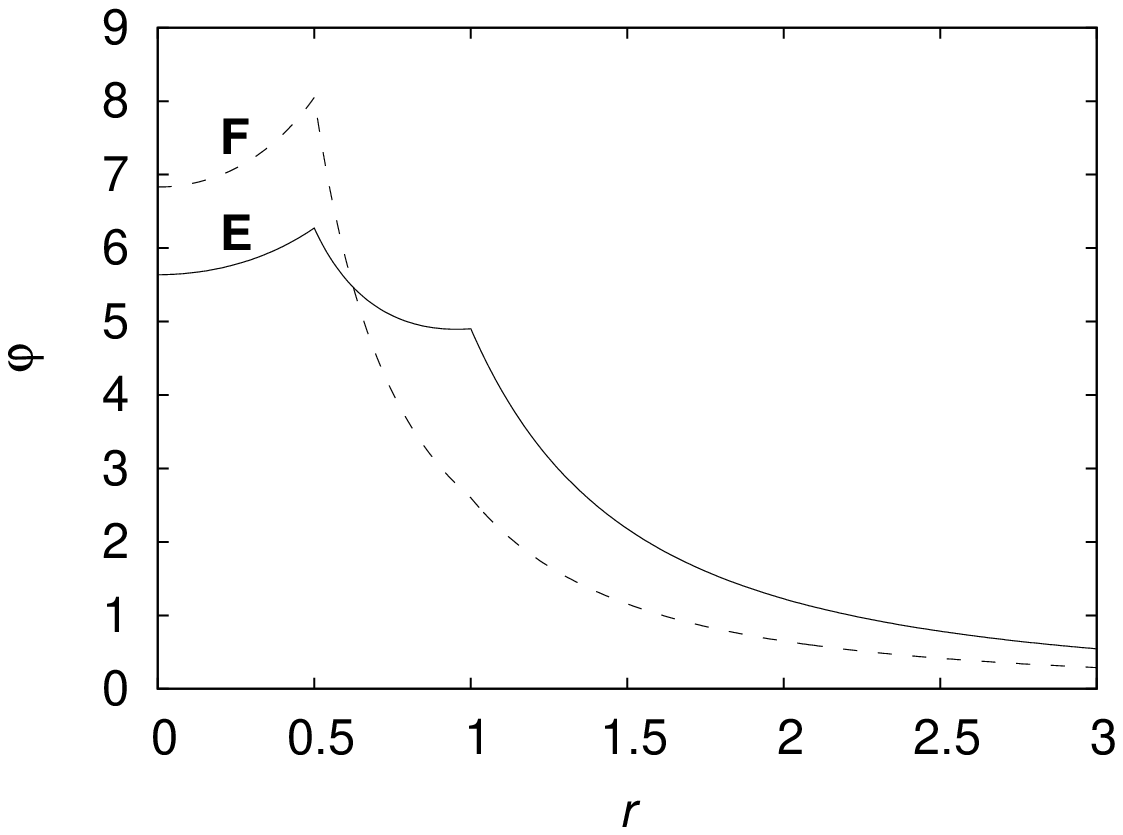}
  \caption{\small{The loop's radial twist profile at specific points along the threshold (labelled A-F).}}
  \label{rdtwpf}
  \vspace{-10pt}
\end{figure}

In order to understand the nature of the instability, we investigate the twist profiles and eigenfunctions of the unstable mode, for different parts of the instability threshold. The magnetic twist profiles for a selection of points just outside the threshold curve exhibit considerable variation, as Fig. \ref{rdtwpf} illustrates.
For the unstable equilibrium labelled as point B ($\alpha_1$\,$<$\,0, $\alpha_2$\,$>$\,0), the corresponding twist profile (also labelled B) shows that the core field has a strong negative twist, whilst in most of the outer layer and in all of the envelope the field has positive twist. As one moves from A to B, the twist in the core becomes more negative whilst the twist in the outer layer moves in the opposite direction. It appears that the increase in $\alpha_2$ stabilises the negative core twist by providing additional reversed (i.e., positive) twist in the outer layer. The sharp corner at the top left of the threshold marks the point where instabilities driven within the core intersect those that originate from within the outer layer. Profiles C to D therefore, suggest instabilities driven in the outer layer, since $|\alpha_2| > |\alpha_1|$. The peak twist in the outer layer reduces as the core twist moves from negative to positive. Further along the threshold, where $\alpha_1 > \alpha_2$, the instabilities are likely to be driven in the core. Note that profiles D, E and F are always positive in sign; D has a twist peak near the loop edge while E and F are roughly monotonically decreasing. The corresponding magnetic field profiles for the six points A-F are given in Appendix B.

It seems that instabilities are driven mainly on or near the peak of largest absolute twist. A twisted field region may be stabilised by an enclosing region of opposite twist. Furthermore, a twisted outer layer may need less twist to achieve instability if the core has the same twist orientation. To investigate these ideas further, we plot the unstable eigenfunctions obtained from CILTS for the same $\alpha$-space points, A to F.

\begin{figure}[h!]
  \vspace{-10pt}
  \center
  \includegraphics[scale=0.25]{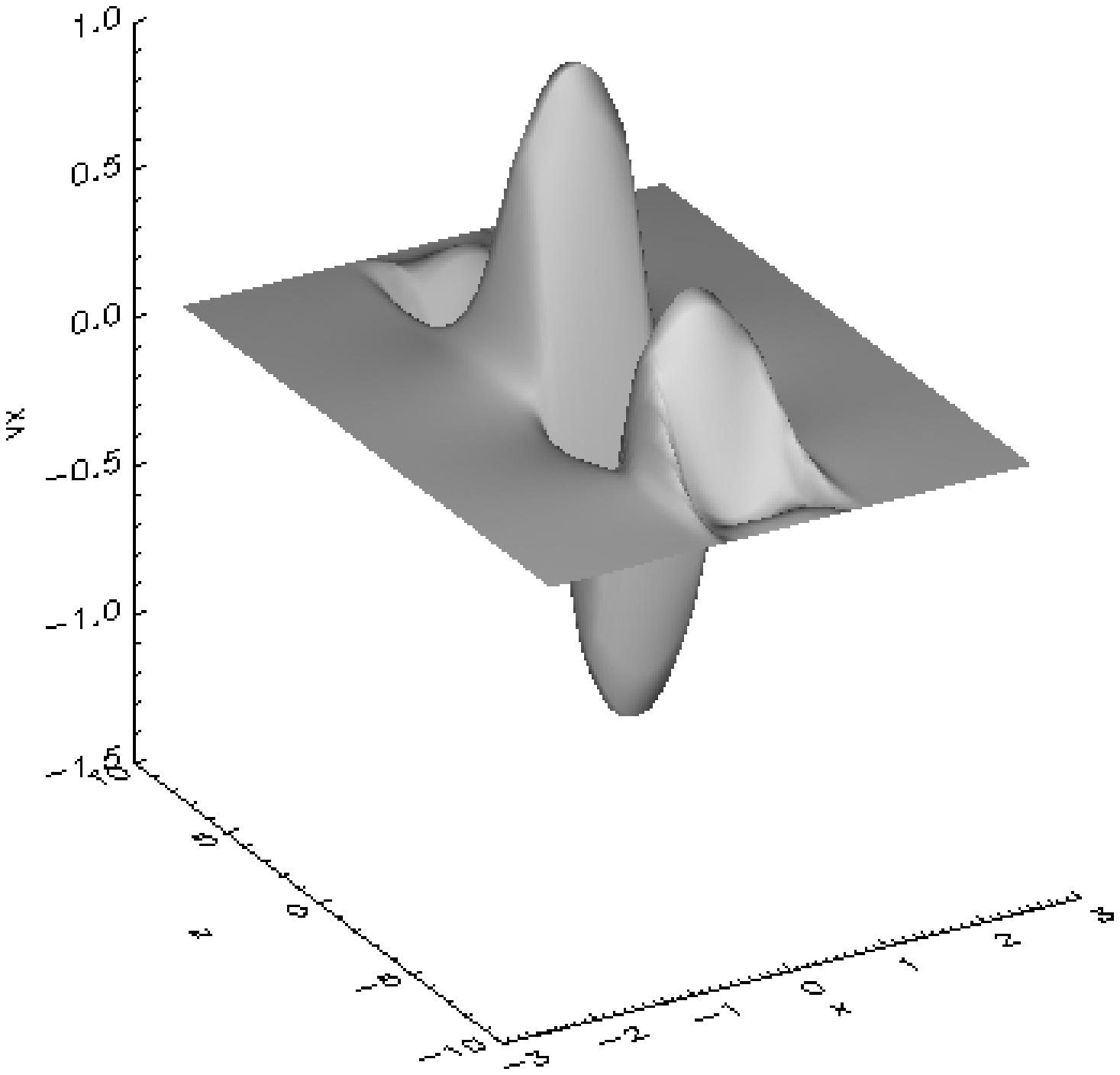}
  \hspace{10pt}
  \includegraphics[scale=0.25]{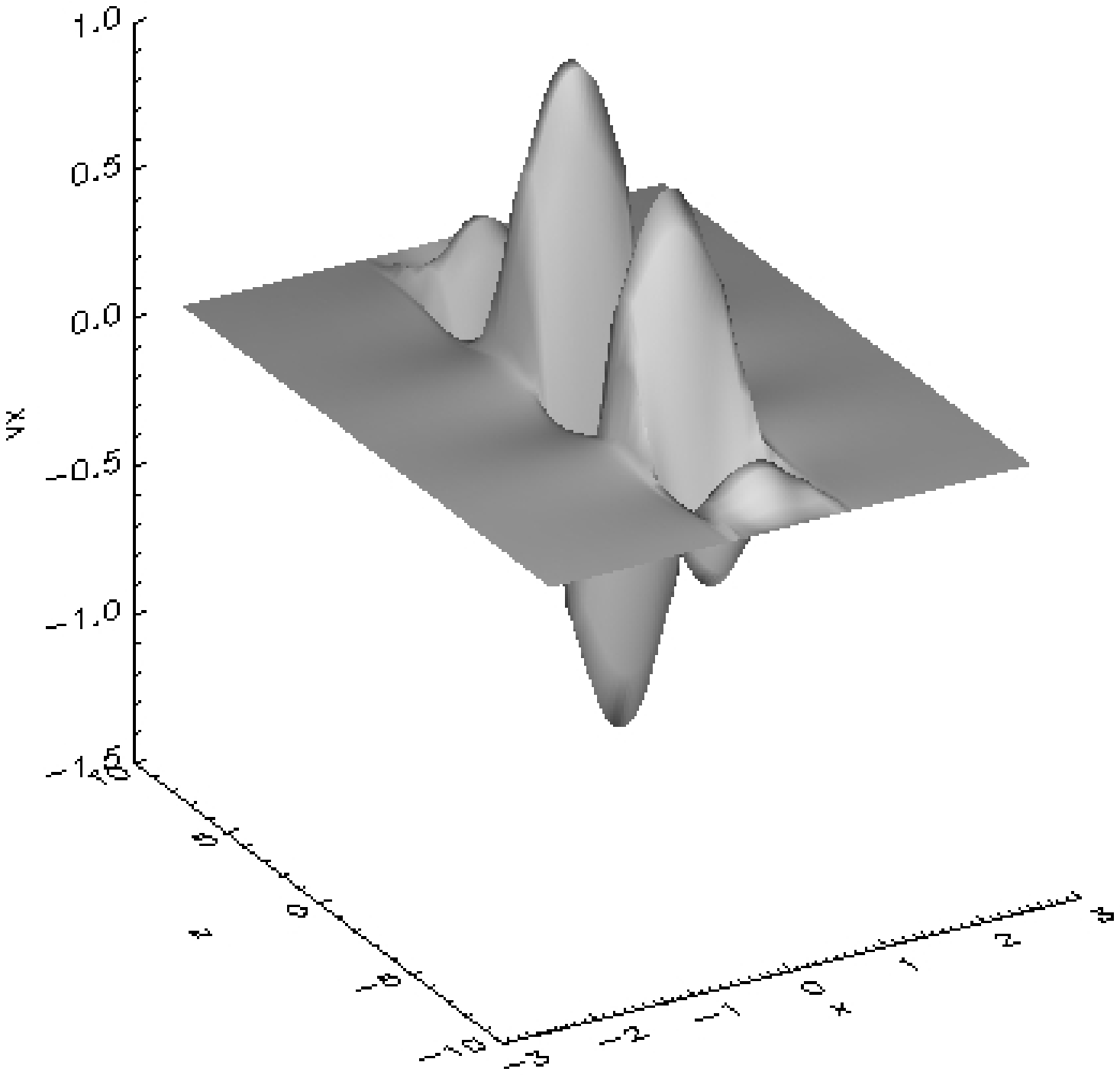}
  \caption{\small{The linear eigenfunction, \textit{V}$_x$(\textit{x},\,\textit{y}=0,\,\textit{z}), for the $\alpha$-space points A (left) and B (right) profiled in Fig. \ref{rdtwpf}. Cartesian coordinates are used, hence, the x-axis is equivalent to the radial axis.}}
  \label{eigenfuncsAandB}
  \vspace{-10pt}  
\end{figure}
The eigenfunctions for profiles A and B (Fig. \ref{eigenfuncsAandB}) show that the amplitude is strongest in the core. Interestingly, the amplitude for profile A has dropped to zero long before the envelope boundary at \textit{R}$_3$, which suggests that in the subsequent relaxation only inner regions will be affected, with little change in the potential envelope. 
\begin{figure}[h!]
  \vspace{-10pt}
  \center
  \includegraphics[scale=0.25]{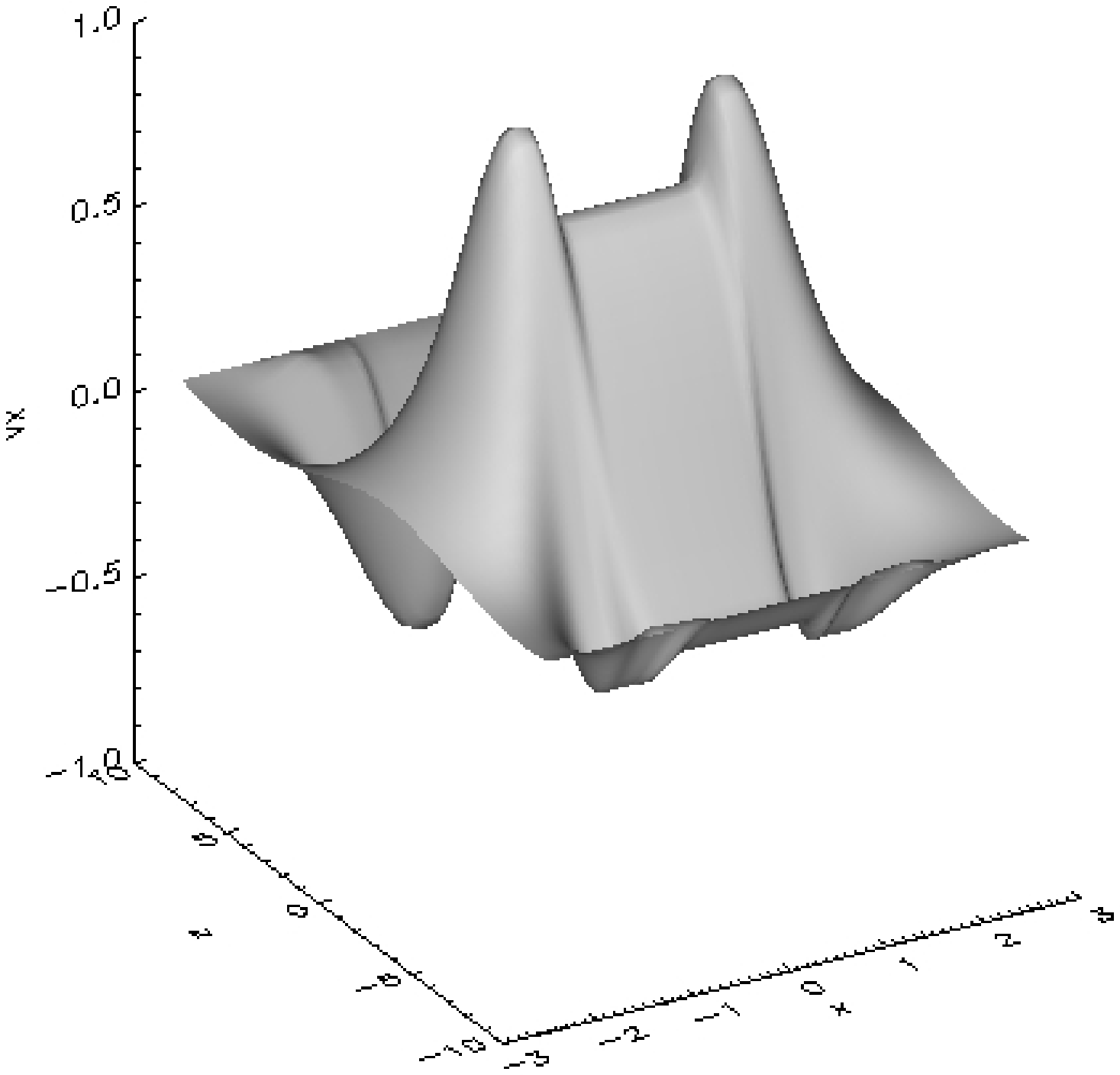}
  \hspace{10pt}
  \includegraphics[scale=0.25]{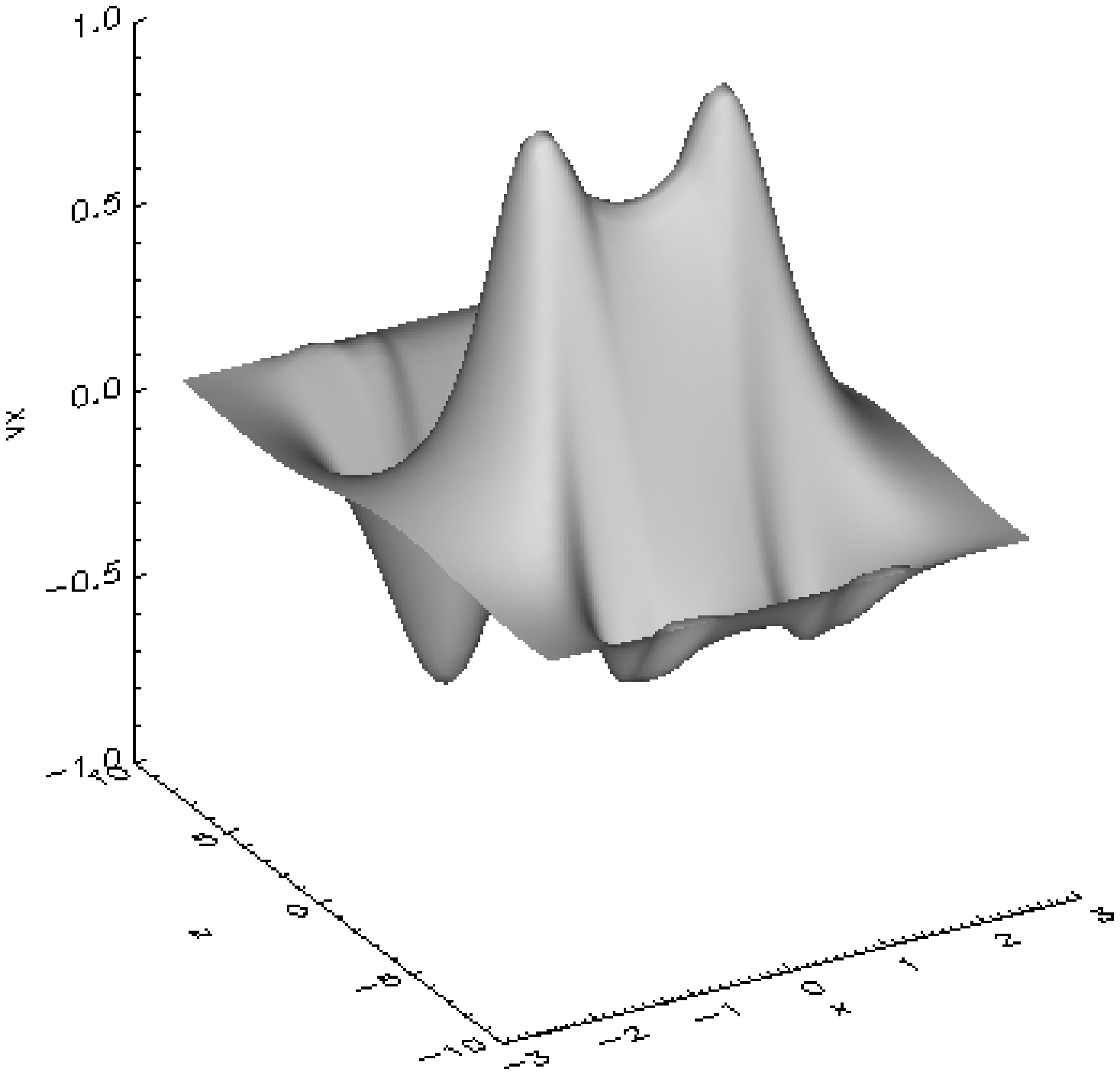}
  \caption{\small{The linear eigenfunction, \textit{V}$_x$(\textit{x},\,\textit{y}=0,\,\textit{z}), for the $\alpha$-space points C (left) and D (right) profiled in Fig. \ref{rdtwpf}.}}
  \label{eigenfuncsCandD}  
  \vspace{-10pt}
\end{figure}
There is a strong similarity between the eigenfunctions for profiles C to D (Fig. \ref{eigenfuncsCandD}) and the amplitude is highest near the \textit{R}$_2$ boundary, indicating an outer layer instability. Finally, the two plots in Fig. \ref{eigenfuncsEandF} (profiles E and F), clearly show a progression towards the eigenfunction calculated for A (albeit with a \textit{V}$_x$ of opposite sign).
\begin{figure}[h!]
  \vspace{-10pt}
  \center
  \includegraphics[scale=0.25]{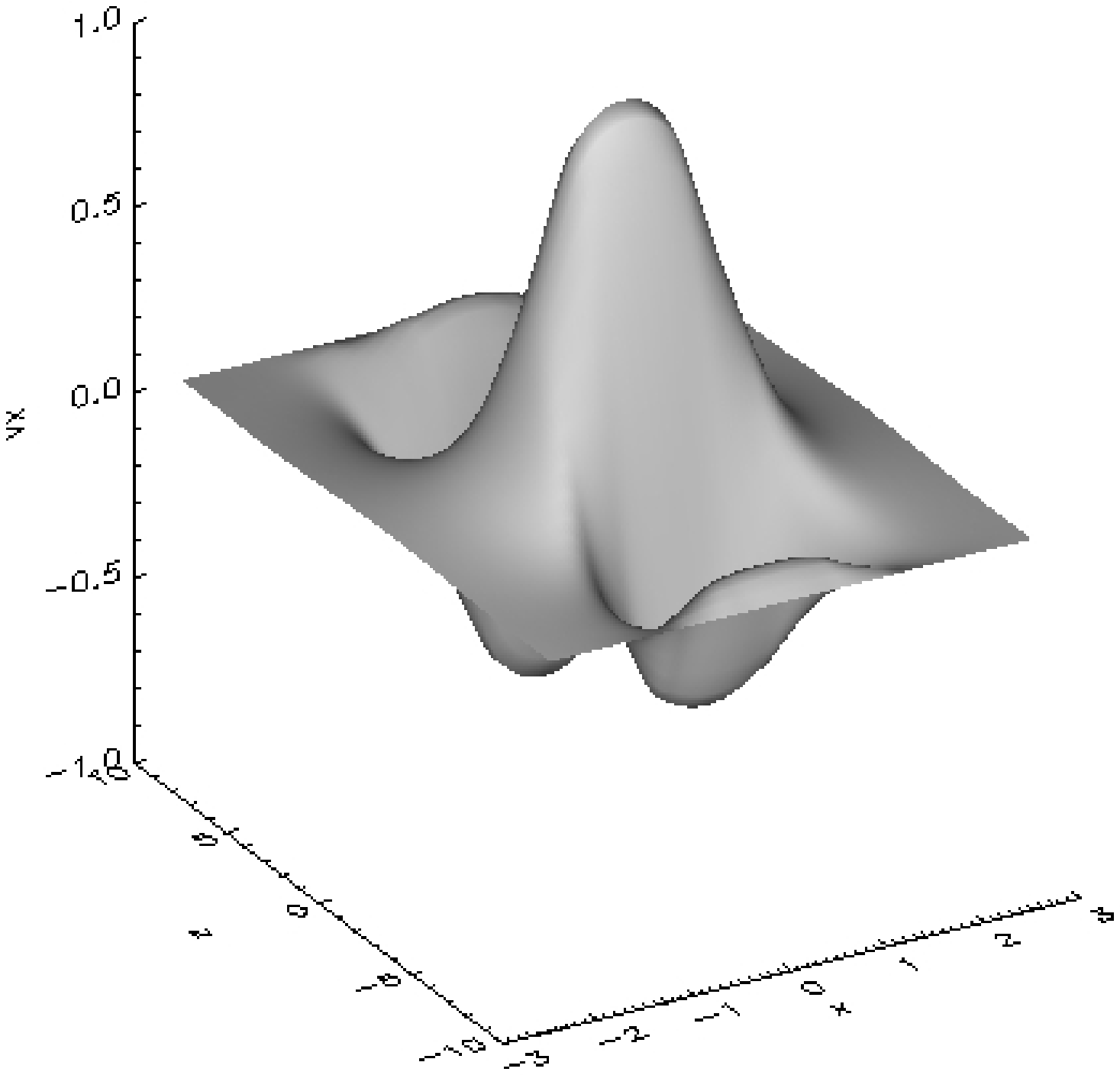}
  \hspace{10pt}
  \includegraphics[scale=0.25]{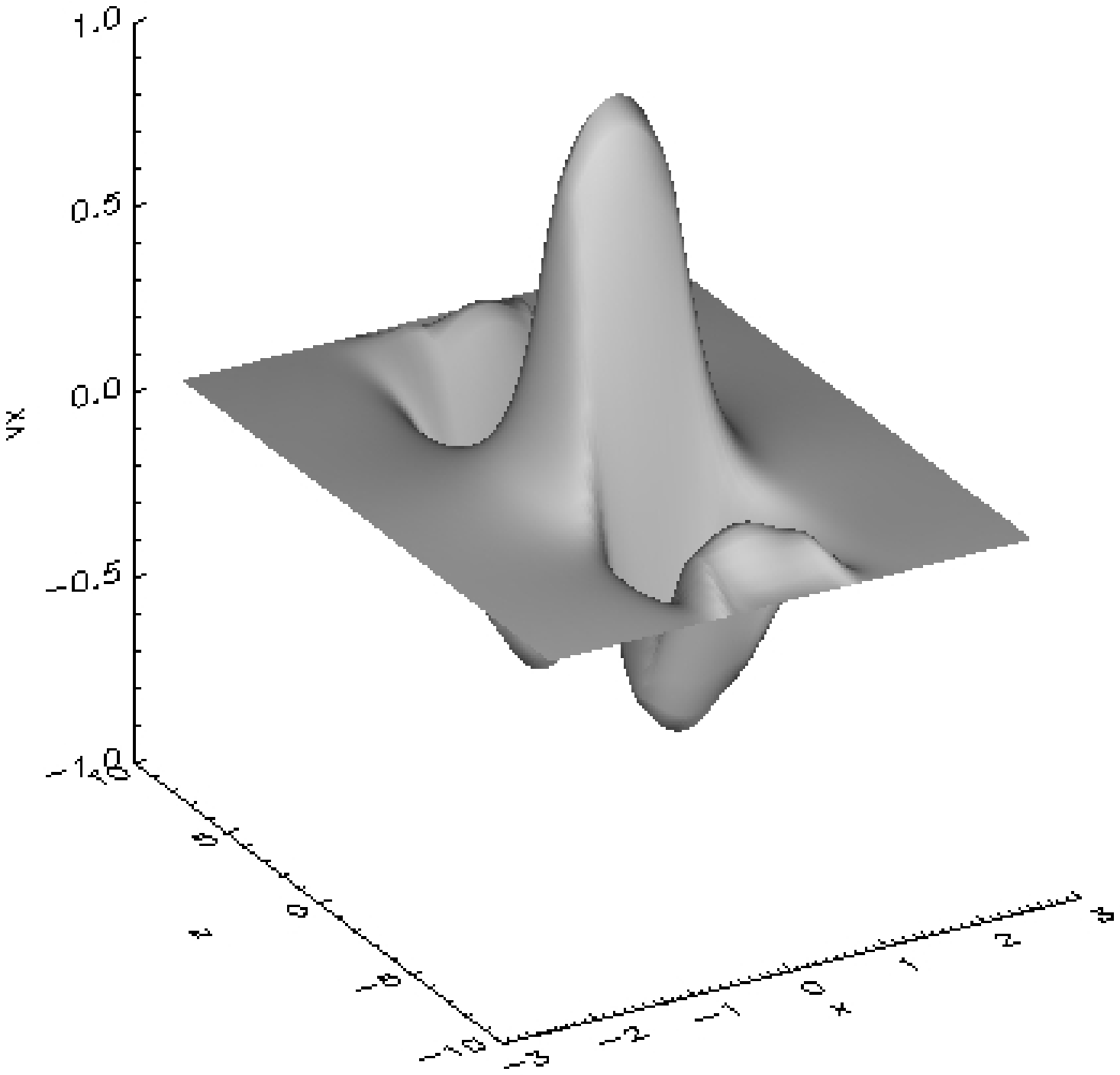}
  \caption{\small{The linear eigenfunction, \textit{V}$_x$(\textit{x},\,\textit{y}=0,\,\textit{z}), for the $\alpha$-space points E (left) and F (right) profiled in Fig. \ref{rdtwpf}.}}
  \label{eigenfuncsEandF} 
  \vspace{-10pt} 
\end{figure}
Notice also that the form of the eigenfunction changes significantly between B and C, indicating that different modes are going unstable. This is to be expected, since the $\alpha$-space positions labelled B and C (Fig. \ref{rdtwpf}, top left) are either side of the intersection point formed by the two curves that describe the instability threshold.

\subsection{Critical twist parameters}
\label{sec:critical_twist_parameters}
\begin{figure}[h!]
  \vspace{-10pt}
  \center
  \includegraphics[scale=0.55]{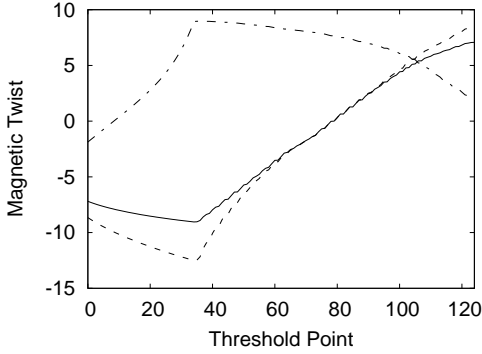}
  \caption{\small{The variation in magnetic twist around the instability threshold (the threshold points are defined in Fig. \ref{it}) for three radial positions. The solid line represents the variation in axial twist, $\varphi_0$; the dashed line is the variation in twist at the boundary between the core and the outer layer, $\varphi$(\textit{R}$_1$); and the long-short dashed line is the variation in twist at the boundary between the outer layer and the potential envelope, $\varphi$(\textit{R}$_2$). The twist values are plotted in units of $\pi$.}}
  \label{it_a2p_tw_0_r1_r2}
  \vspace{-10pt} 
\end{figure}

Next, we look for a twist-related parameter that takes on a critical value whenever the loop reaches the threshold. As expected, the variation in axial twist, $\varphi_0$, is similar to the variation in $\varphi$(\textit{R}$_1$), see Fig. \ref{it_a2p_tw_0_r1_r2}. 
None of the quantities suggest any single (constant) critical value. Perhaps, the average twist is less variable around the threshold? There are several ways to calculate this expression;
\begin{eqnarray}
  \label{average_velli_twist}
  {\langle\tilde{\varphi}\rangle}_0^{R_i} & = & \frac{\int_0^R LB_{\theta}(r)\,dr}{\int_0^R rB_z(r)\,dr}\,\,,\\
  \nonumber &   &\\
  \label{average_baty_twist}
  {\langle\hat{\varphi}\rangle}_0^{R_i} & = & \frac{1}{R}\int_0^R \frac{LB_{\theta}(r)}{rB_z(r)}\,dr\,\,,\\
  \nonumber &   &\\
  \label{average_wght_twist}
  {\langle\varphi\rangle}_0^{R_i} & = & \frac{1}{\pi R^{2}}\int_0^R 2\pi r \hspace{0.1cm} \frac{LB_{\theta}(r)}{rB_z(r)}\,dr\,\,,
\end{eqnarray}
where \textit{R}$_i$ is a radial upper bound (e.g., \textit{R}$_2$ or \textit{R}$_3$). Equation \ref{average_wght_twist} is the average twist weighted by area. The other two equations (\ref{average_velli_twist} and \ref{average_baty_twist}) have been used by Baty (2001) and Velli et al. (1990). Note, Equation \ref{average_velli_twist} can be calculated analytically, see Appendix A. ${\langle\varphi\rangle}_0^{R_2}$ denotes the average twist, weighted by area, over the core and outer layer. Similarly, ${\langle\varphi\rangle}_0^{R_3}$ denotes the same quantity but over the loop and potential envelope. The tilde ($\sim$) and hat ($\wedge$) symbols are used to indicate the other equations.

When $\alpha_1$ and $\alpha_2$ are equal there is no distinction between the loop regions; this occurs on the threshold when $\alpha_1$\,=\,$\alpha_2$\,$\approx$\,1.4 - at this point ${\langle\varphi\rangle}_0^{R_2}$\,$\approx$\,5$\pi$. When $\alpha_2$\,=\,0 and $\alpha_1$\,$\approx$\,2.2 the loop is identical to the core with a bigger potential envelope and ${\langle\varphi\rangle}_0^{R_1}$\,$\approx$\,7.7$\pi$. Thus, as expected, fatter loops like the first case, have lower instability twists than thinner ones. When the loop's core is potential (i.e., when $\alpha_1$\,=\,0 and $\alpha_2$\,$\approx$\,2.2), ${\langle\varphi\rangle}_{R_1}^{R_2}$\,$\approx$\,4.5$\pi$ is the average outer layer twist. In this configuration, the loop is less stable than the case when only the core is non-potential. 

\begin{figure}[h!]
  \vspace{-5pt} 
  \center
  \includegraphics[scale=0.35]{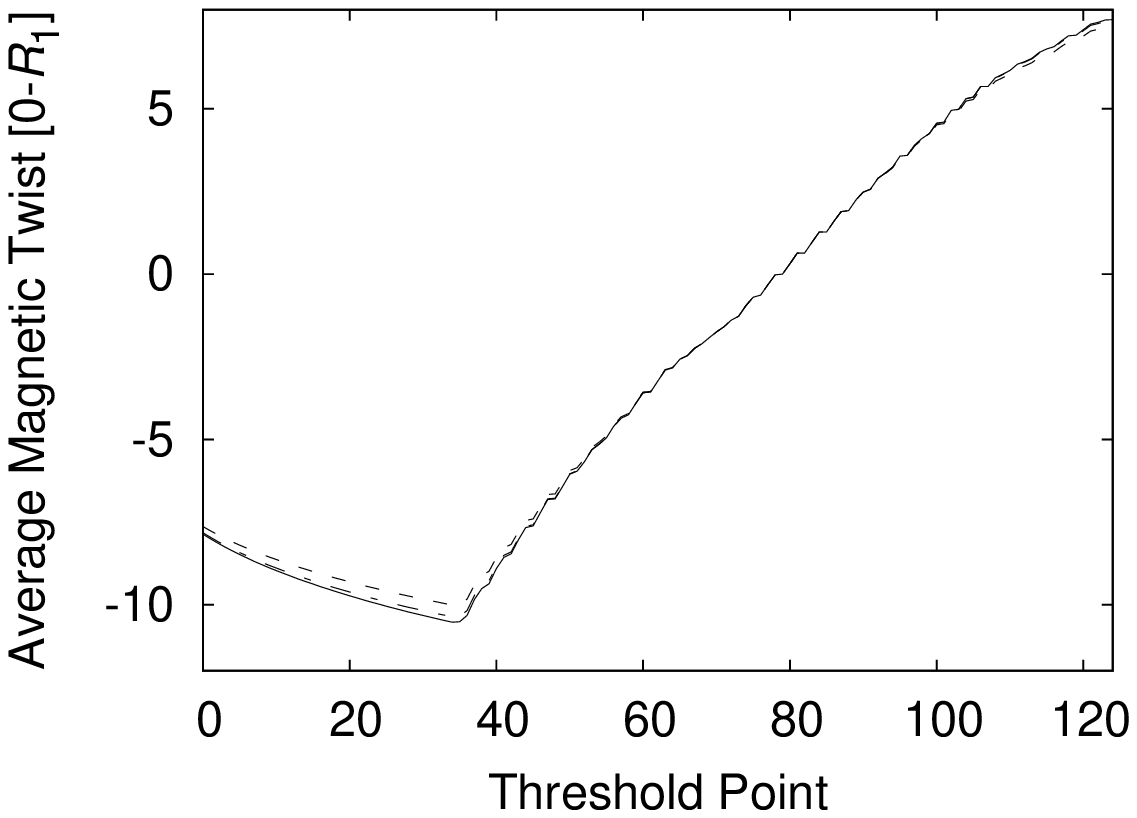}
  \includegraphics[scale=0.35]{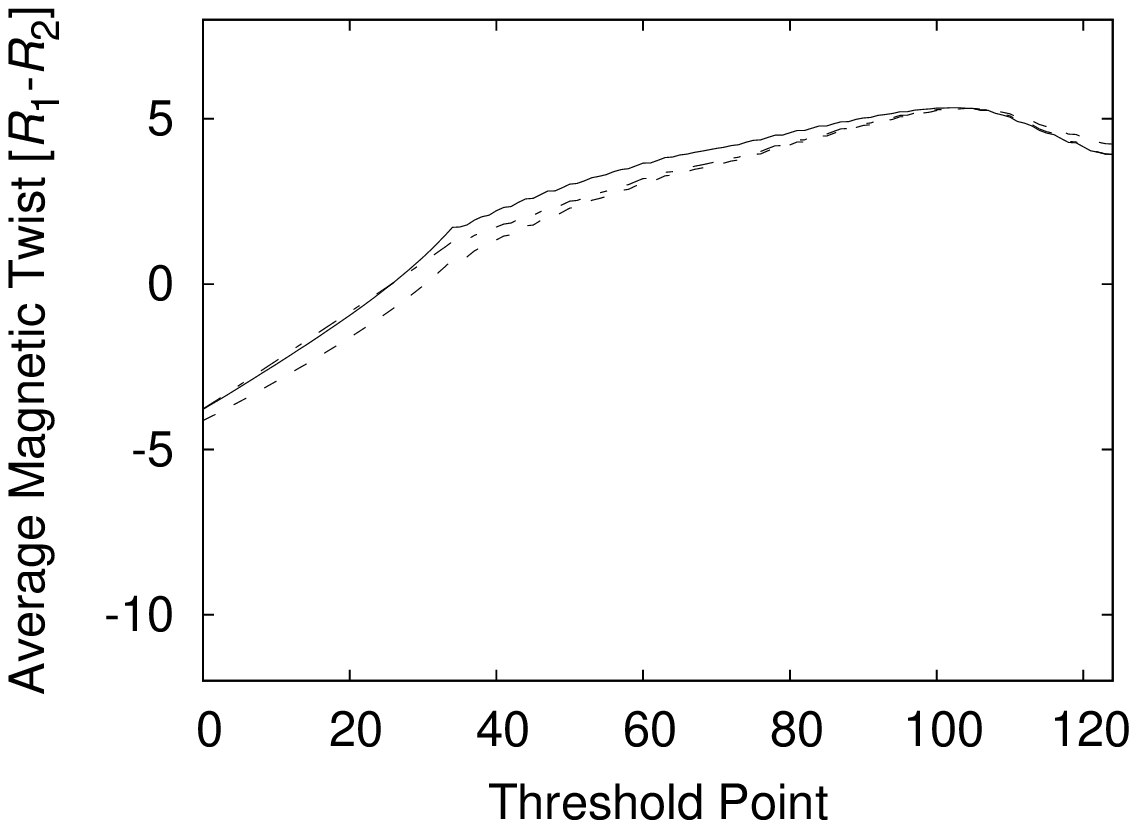}
  \center
  \includegraphics[scale=0.35]{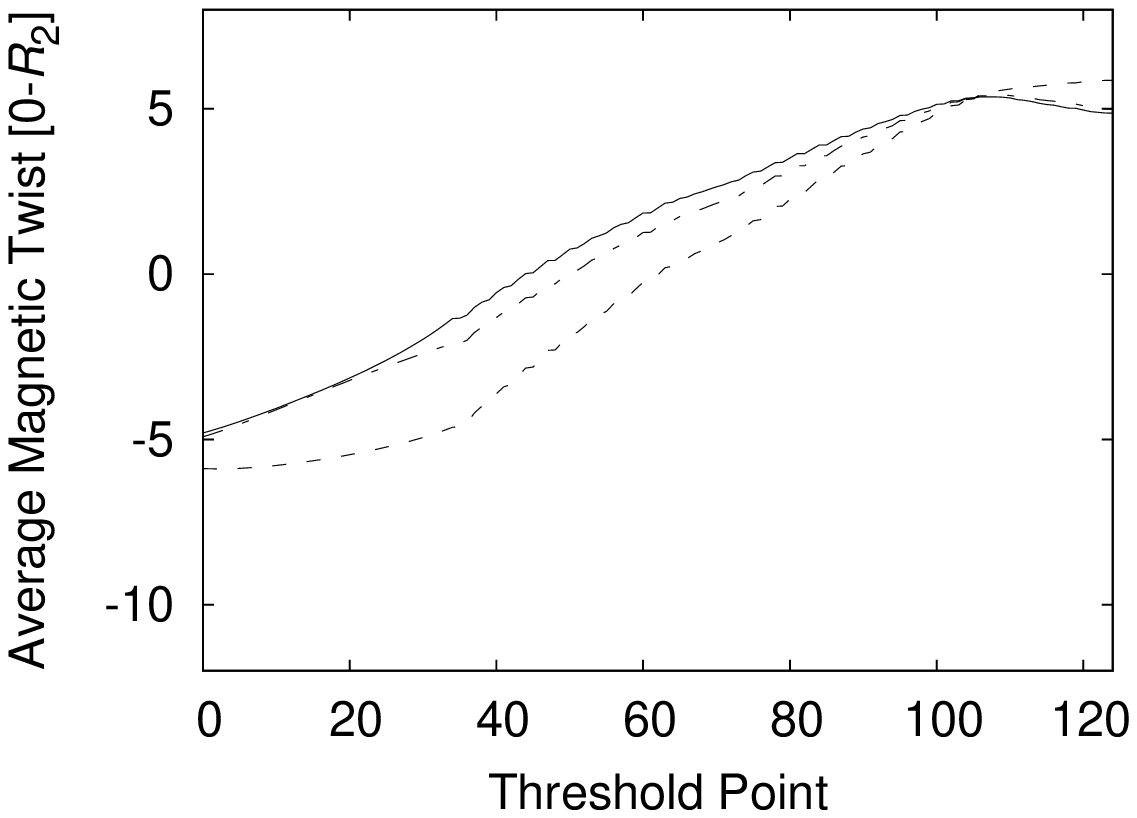}
  \includegraphics[scale=0.35]{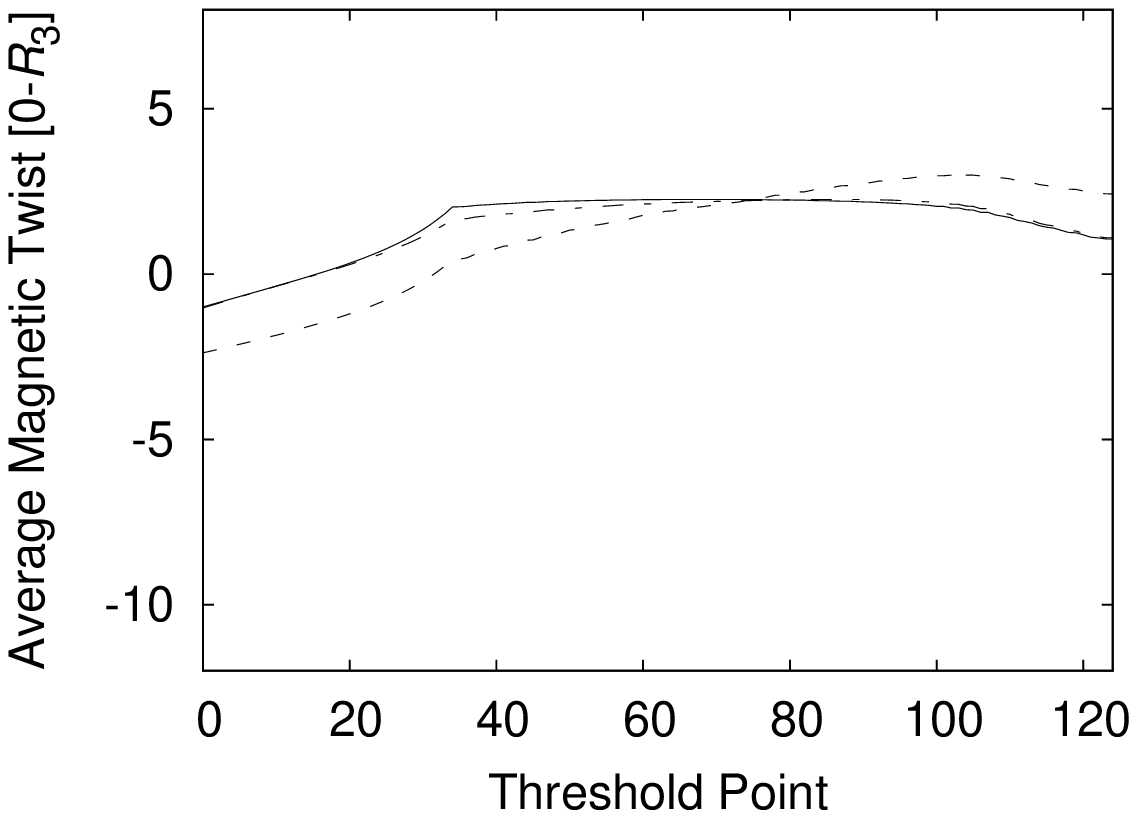}
  \caption{\small{The variation in the average twist over the core, 0\,-\,\textit{R}$_1$ (top left), the outer layer, \textit{R}$_1$\,-\,\textit{R}$_2$ (top right), the loop, 0\,-\,\textit{R}$_2$ (bottom left) and the loop and envelope, 0\,-\,\textit{R}$_3$ (bottom right). The solid lines were calculated according to Eqn. \ref{average_wght_twist}; the dashed according to Eqn. \ref{average_baty_twist} and the long-short dashed according to Eqn. \ref{average_velli_twist}. Again, the twist values are plotted in units of $\pi$.}}
  \label{it_a2p_avtw}
  \vspace{-30pt} 
\end{figure}

None of the twist averages (Fig. \ref{it_a2p_avtw}) is invariant along the entire threshold. Although, ${\langle\varphi\rangle}_0^{R_3}$ and ${\langle\tilde{\varphi}\rangle}_0^{R_3}$ have approximately the same value ($\approx 2.2\pi$) between threshold points 40 and 90. All of these points show a positive peak twist at \textit{R}$_2$. Hence, the envelope's contribution dominates the overall average twist. This is especially true for the ${\langle\varphi\rangle}_0^{R_3}$ case: the higher the radial coordinate the greater the weight of the calculated twist. Within the envelope, the twist declines as 1/r and so, the inclusion of the envelope twist averages out the final result.

Notice also, that all the twist averages go through zero when the core and outer layer have opposite twists. It seems that these quantities do not reveal the detail necessary to understand why a particular loop configuration is on the point of instability, nor where in the loop that instability originates. 

Finally, we consider the proposal of Malanushenko et al. (2009), that a critical value of normalised helicity (equivalent, in our terms, to the normalised loop helicity, $\textit{K}/\psi^2$, over the range 0\,-\,\textit{R}$_2$) indicates instability onset. 
\begin{figure}[h!]
  \vspace{-15pt} 
  \center
  \includegraphics[scale=0.55]{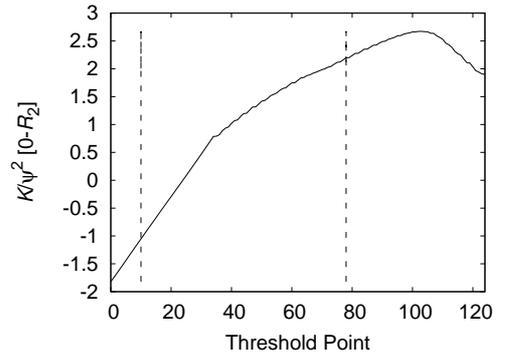}
  \caption{\small{The variation in \textit{K}/$\psi^2$ (over the range 0\,-\,\textit{R}$_2$) along the instability threshold. The threshold states between the vertical dashed lines feature reverse twist; outside the lines the twist is single-signed.}}
  \label{it_a2p_kopsi2_0_r2}
  \vspace{-15pt} 
\end{figure}
In fact, the normalised helicity is certainly not the same for every threshold point; this quantity passes through zero because $\alpha_1$ and $\alpha_2$ take on values of opposite sign along some sections of the threshold. For fields with single-signed twist, the normalised helicity gives an approximate threshold, but even here, the (absolute) critical value ranges from about 1.5\,-\,2.5. Figure \ref{it_a2p_kopsi2_0_r2} shows that the idea of such a critical value breaks down for loops that feature regions of reversed twist. 

\section{Distribution of energies and coronal heating considerations}
\label{sec:distribution_of_energies_and_coronal_heating_considerations}
We now proceed to the main task of our work, which is to calculate the distribution of magnitudes of the sequence of heating
events generated by random photospheric driving.

\subsection{Helicity and energy}
\label{sec:helicity_and_energy}
\begin{figure}[h!]
  \vspace{-10pt}
  \center
  \includegraphics[scale=0.35]{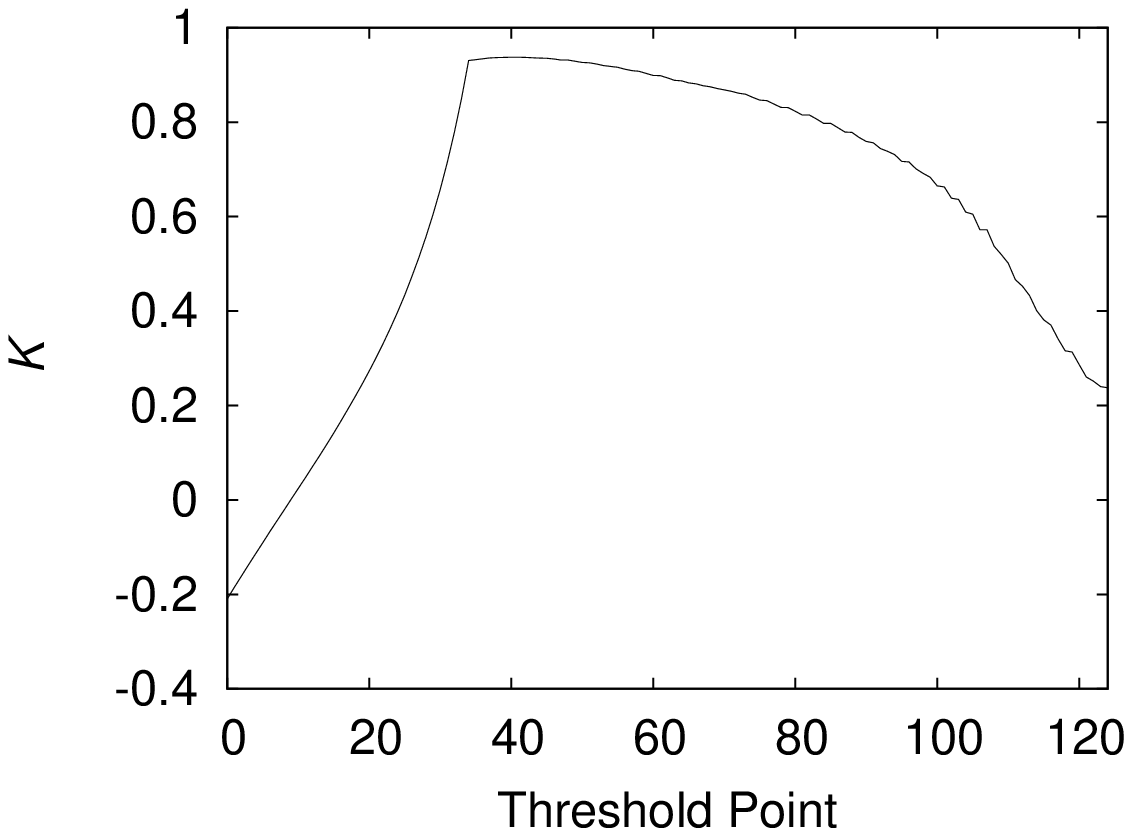}
  \includegraphics[scale=0.35]{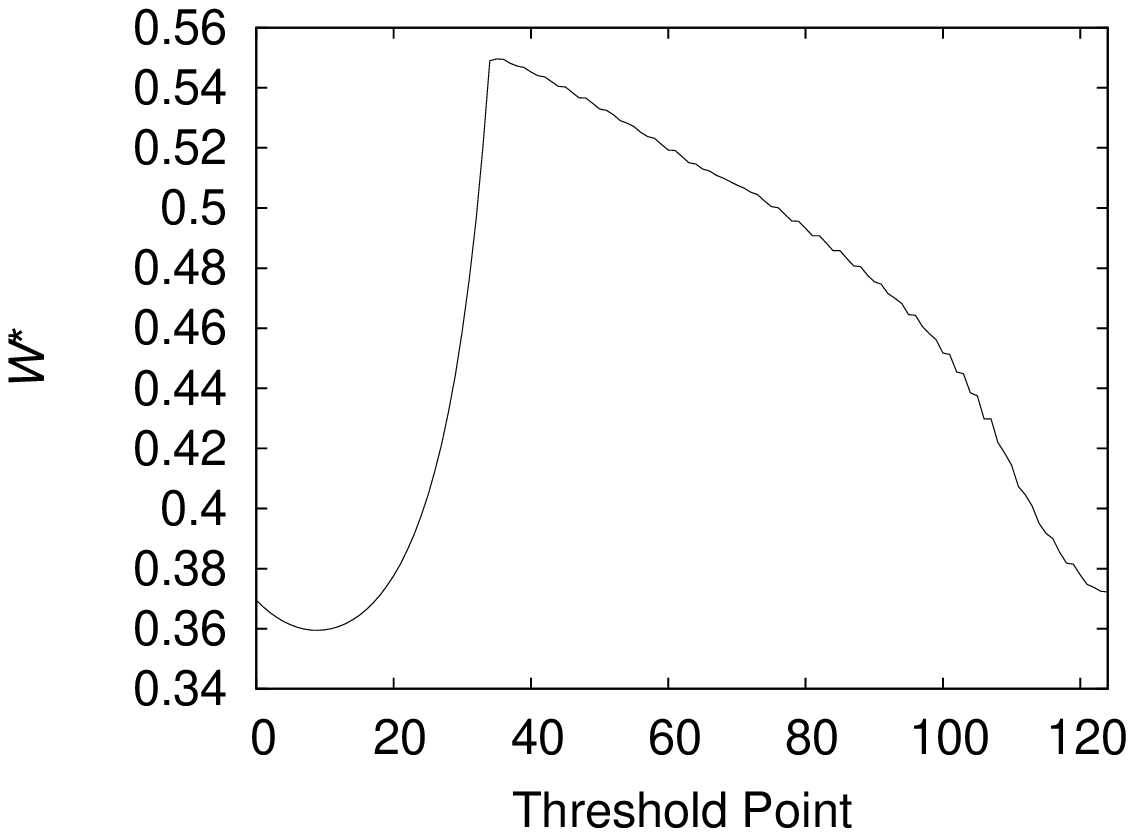}
  \includegraphics[scale=0.35]{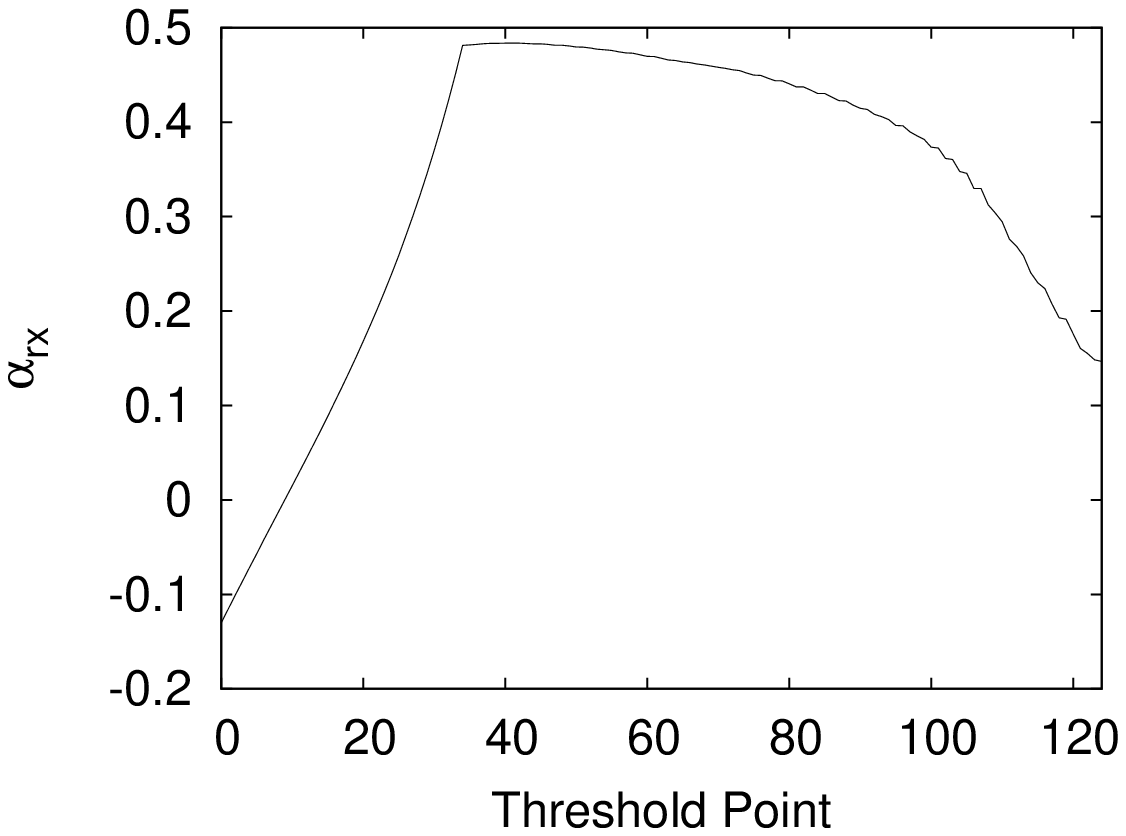}
  \includegraphics[scale=0.35]{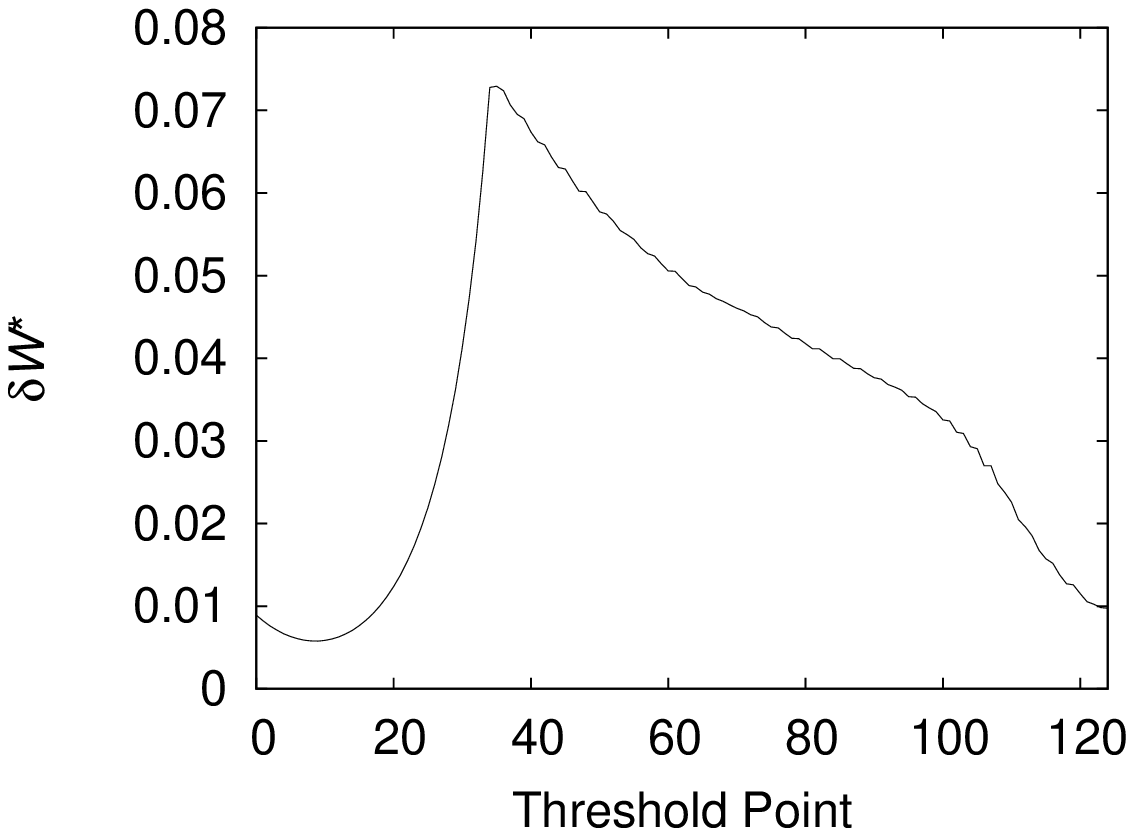}
  \caption{\small{Helicity (top left), magnetic energy (top right), relaxed alpha (bottom left) and energy release (bottom right) along the 1D representation of the instability threshold. The energies, \textit{W}* and $\delta$\textit{W}$^{*}$, are dimensionless quantities.}}
  \label{it_a2p_k_w_arx_wr}
  \vspace{-10pt}
\end{figure}

The top left panel of Fig. \ref{it_a2p_k_w_arx_wr} plots total helicities of the threshold states. A total helicity (or flux) is one calculated over the range 0\,-\,\textit{R}$_3$, i.e., the loop and envelope. 
None of the threshold states have sufficient helicity for the relaxed state to feature helical modes (Taylor 1986) and so all relaxed states are cylinderically symmetric. The bottom left panel confirms that each threshold state corresponds to a relaxed state. The maximum $\alpha_{rx}$ is about 0.48. Hence, there is a good chance that the envelope current after relaxation will not be insignificant. This result confirms the need to investigate how this current could be dissipated (although the maximum value is still significantly less than typical $\alpha_1$ and $\alpha_2$ threshold values).

The energies shown in Fig. \ref{it_a2p_k_w_arx_wr} (top right and bottom right) are given as dimensionless quantities, to calculate the dimensional energy we first give the dimensional flux though the loop and envelope,
\begin{eqnarray}
  \label{dim_axial_flux}	
  \psi = \int_0^{3R_c}2\pi r B_z\,dr & = & 9\pi\,R_c^2\,B_c\,,
\end{eqnarray}
where \textit{R}$_c$ is the coronal loop radius (the dimensionalised \textit{R}$_2$) and \textit{B}$_c$ is the \textit{mean} axial coronal field. Since we have normalised the magnetic field, such that $\psi^*$\,=\,1, the field strength can be dimensionalised thus,
\begin{eqnarray}
  \label{dim_field_strength}	
  B & = & \frac{\psi}{R_c^2} B^* = 9\pi B_c B^*,
\end{eqnarray}
where the asterisk is again used to denote dimensionless values. Hence the dimensional energy release becomes 
\begin{eqnarray}
  \label{dim_energy_release}	
  \delta W & = & \frac{1}{\mu_0}\Bigg(\frac{\psi}{R_c^2}\Bigg)^2 \hspace{0.05cm} R_c^3 \hspace{0.1cm} \delta W^{\hspace{0.02cm}*} = \frac{81\pi^2}{\mu_0} \hspace{0.1cm} R_c^3 B_c^2 \hspace{0.1cm} \delta W^{\hspace{0.02cm}*}\hspace{0.1cm}.
\end{eqnarray}
This expression differs slightly from the one used by Browning \& Van der Linden (2003): their dimensionless energy release is calculated per unit length and \textit{R}$_3$\,=\,\textit{R}$_2$. Assuming typical values (\textit{R}$_c$\,=\,1 Mm and \textit{B}$_c$\,=\,0.01 T), we obtain dimensional energy values of $6\,\times\,10^{22}\,\delta\textit{W}^{*}$ J\,\,$\equiv$\,\,$6\,\times\,10^{29}\,\delta\textit{W}^{*}$ erg. Thus, the top end of the $\delta$\textit{W}$^{*}$ scale ($\approx$\,0.073) is equivalent to $4\,\times\,10^{28}$ erg. This is in the microflare range, but nanoflare energies will be obtained for weaker fields or for smaller loops.

\subsection{Flare energy distribution}
\label{sec:flare_energy_distribution}
Every time a loop's random walk reaches the instability threshold, the loop's relaxed state (i.e., its position on the relaxation line) and energy release are calculated. The relaxed state is the start of a new random walk which will lead to another relaxation (and energy release). If this process is repeated often enough it can be shown that certain energy release sizes are more common than others. The flare energy distributions converge as the number of relaxation events simulated (see Sect. \ref{sec:energy_release_calculation}) is increased. 

\begin{figure}[h!] 
  \vspace{-10pt}
  \center
  \includegraphics[scale=0.35]{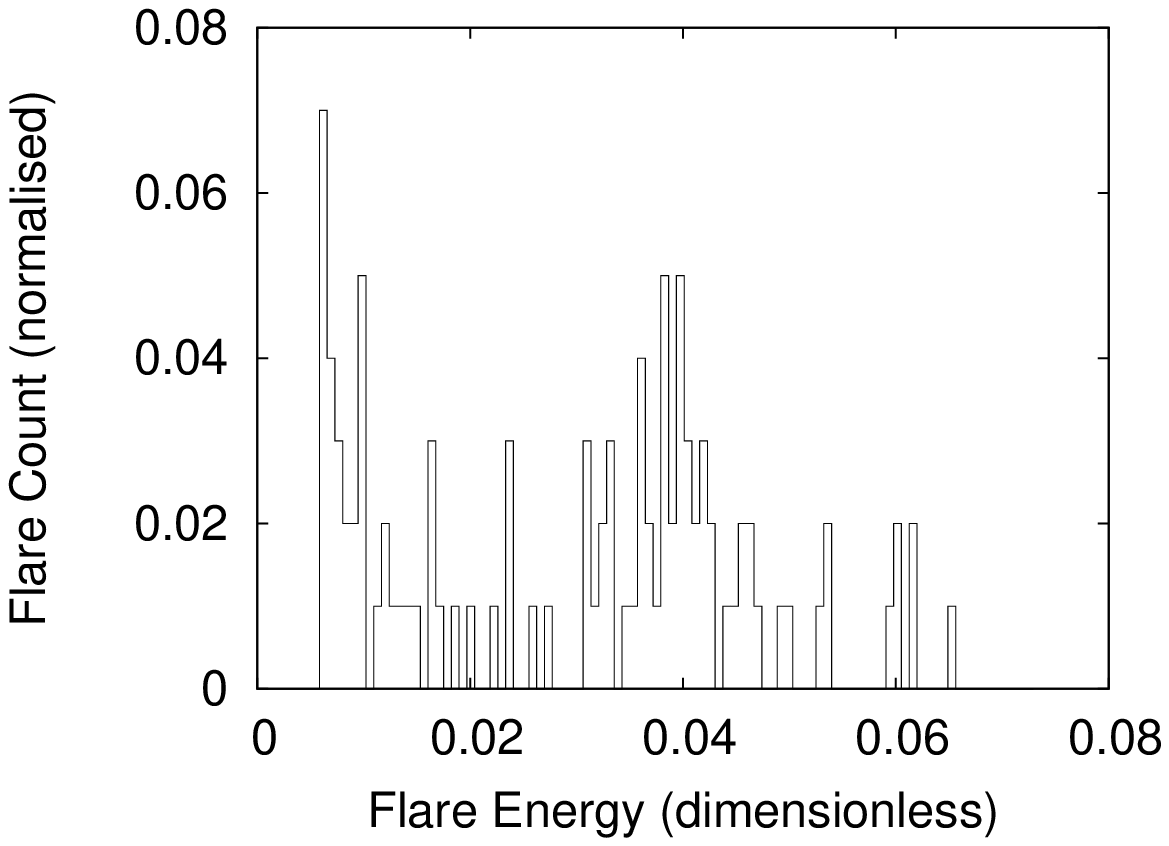}
  \includegraphics[scale=0.35]{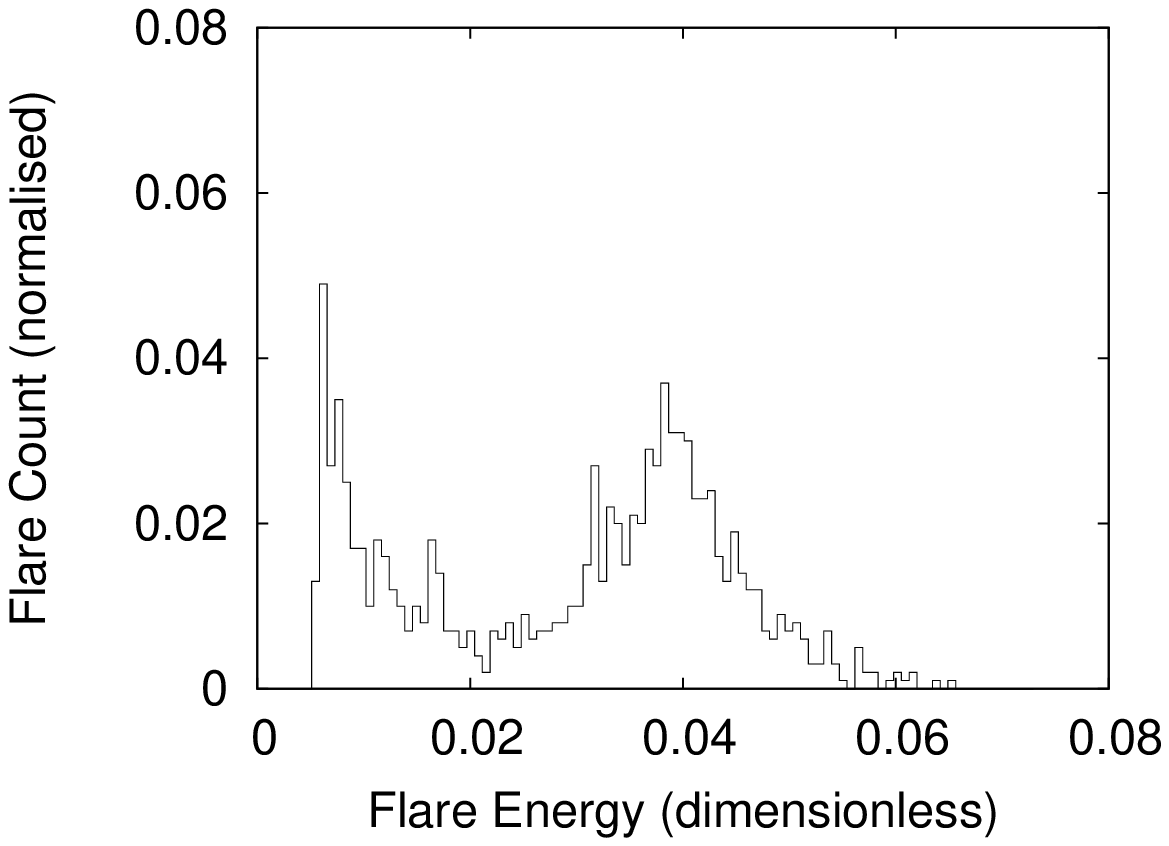}
  \includegraphics[scale=0.35]{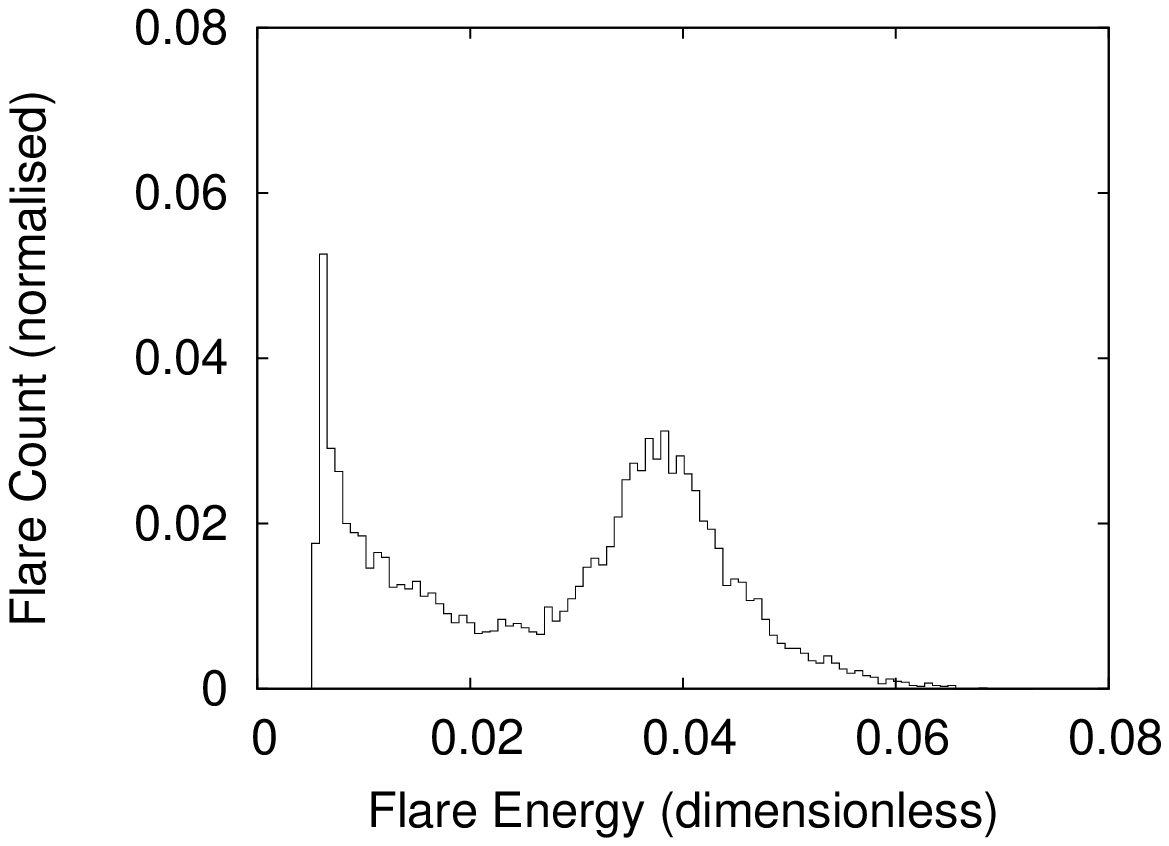}
  \includegraphics[scale=0.35]{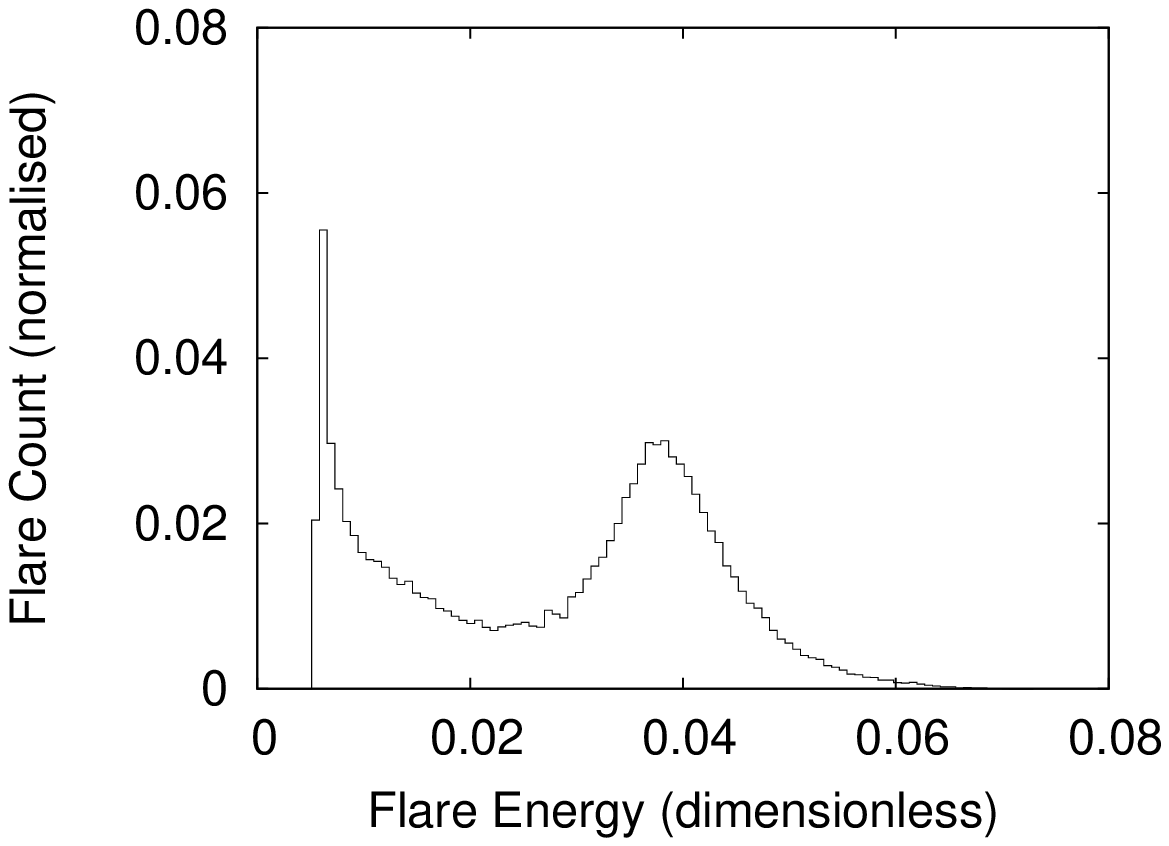}
  \caption{\small{Flare energy distributions for 100 (top left), 1000 (top right), 10$^4$ (bottom left) and 10$^5$ (bottom right) relaxation events.}}
  \label{wrpf}
  \vspace{-10pt} 
\end{figure}

\begin{figure}[h!]
  \vspace{-30pt}
  \center
  \includegraphics[scale=0.62]{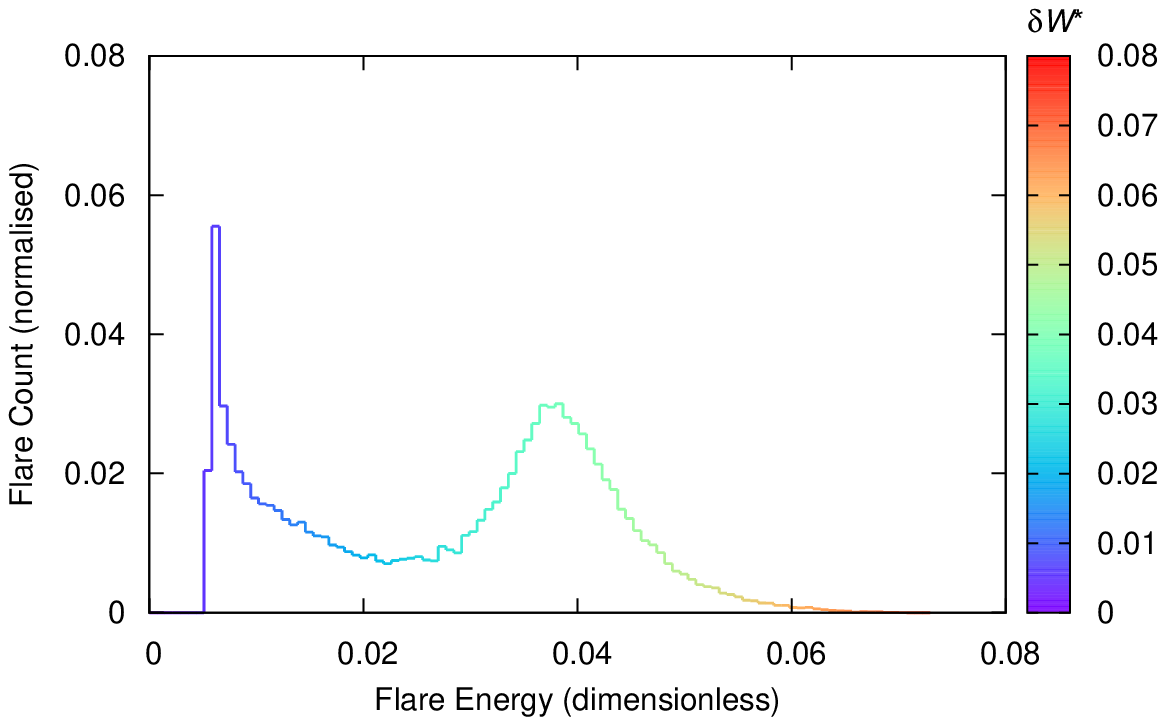}
  \vspace{-25pt} 
  \center
  \includegraphics[scale=0.62]{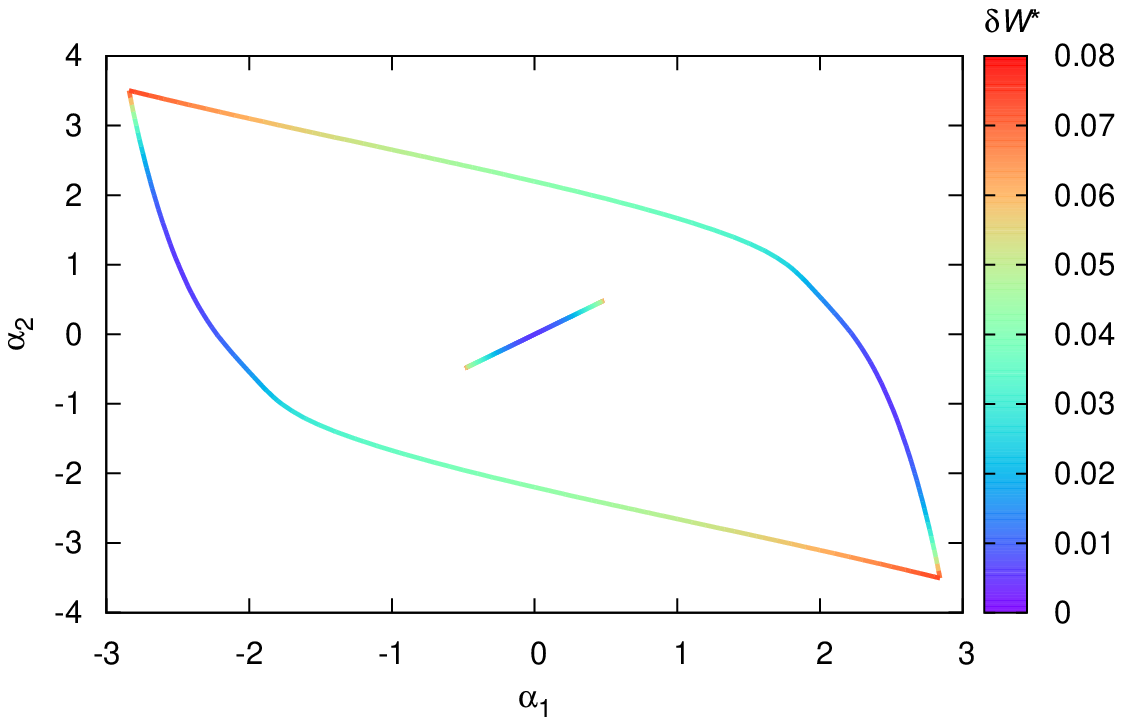}
  \vspace{-10pt}
  \caption{\small{The flare energy distribution for 10$^5$ relaxation events and the instability threshold (with relaxation line). All lines are colour-coded according to energy release.}}
  \label{wrpf_it_cl}
  \vspace{-20pt}
\end{figure}

The gross features of the converged energy distribution can be explained by presenting the energy distribution for the highest number of flare events (Fig. \ref{wrpf}, bottom right) alongside the instability threshold in Fig. \ref{wrpf_it_cl}. Both the distribution and the threshold are colour-coded according to event energy. The colour of the threshold point encountered by a coronal loop indicates the energy of the resulting flare and also the part of the distribution where the heating event falls. Dimensionalised values for the flare energies have been recovered by using typical loop parameters (see above).

The energy release changes as one moves along the threshold, i.e., there is an energy release gradient. This gradient is small for low energies, therefore, only a few bins cover the low energy sections of the threshold, see Fig. \ref{wrpf_it_cl}. The low energy release events are divided amongst a small number of bins, hence, these bins contain many more events. The threshold sections corresponding to the profile minima are slightly longer than those associated with the first peak ($\approx$ 1.5 times), however, the minima sections have a much higher energy release gradient. The energy release events are divided amongst a higher number of bins and so each of these bins contain fewer events. As one moves into the threshold sections that correspond with the second profile peak (denoted by blue-green shades) the energy release gradient decreases. Hence, one would expect the associated bins to have higher event counts. This increase is accentuated however by the proximity of the corresponding relaxation points. Once a loop achieves a blue-green instability, it will relax to a point (also coloured blue-green), which happens to be closest to the green section of the threshold. Subsequent walks will have a higher chance of crossing that section, thus, the corresponding distribution bins will contain even more events. The highest energy releases available on the threshold are farthest away from the relaxation line. In this part of the distribution, the chance of an energy release event is inversely proportional to the size of the energy release.

\begin{figure}[h!] 
  \vspace{-5pt}
  \center
  \includegraphics[scale=0.35]{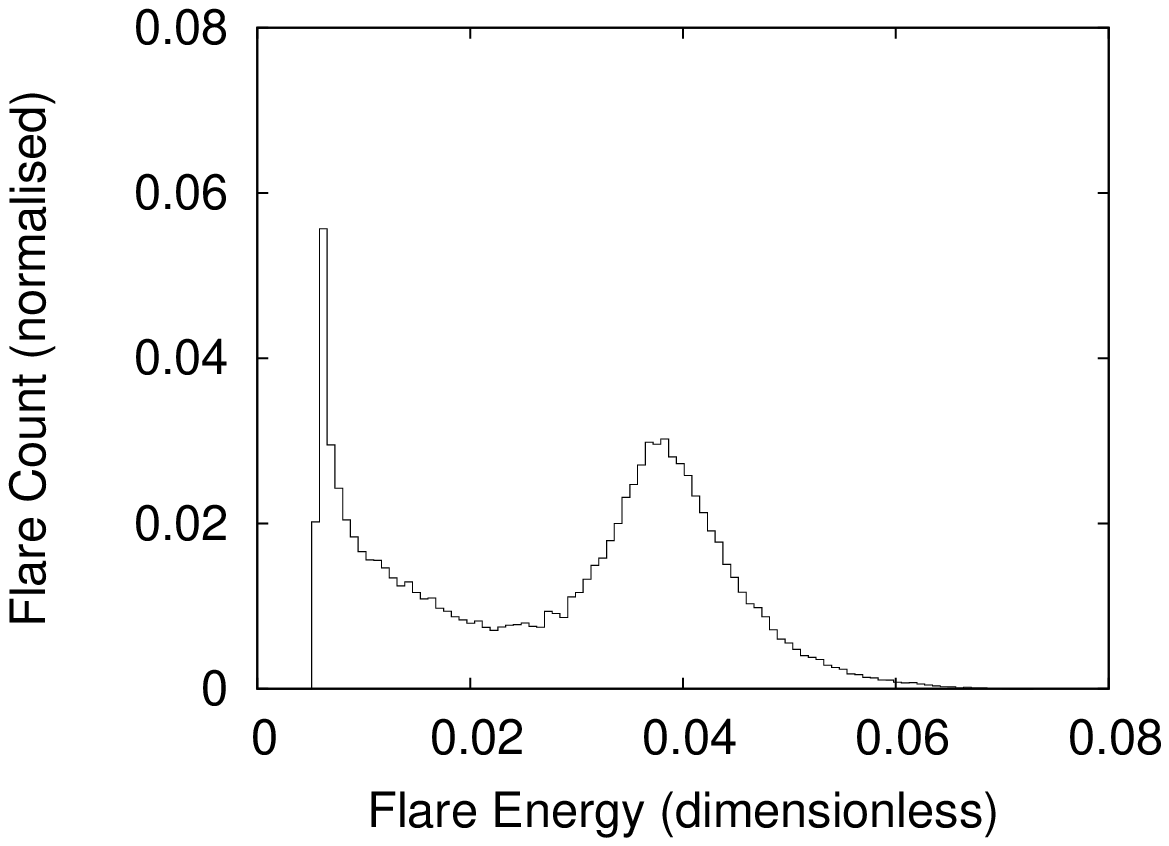}
  \includegraphics[scale=0.35]{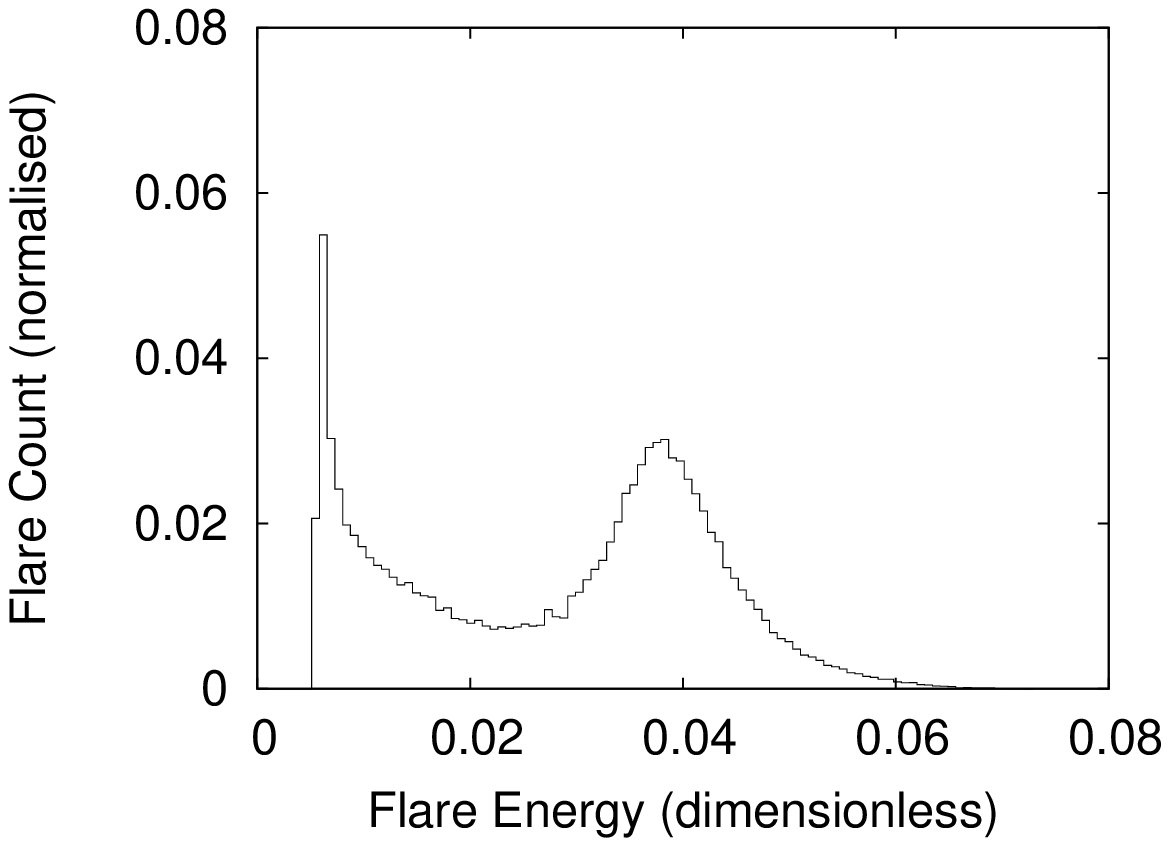}
  \includegraphics[scale=0.35]{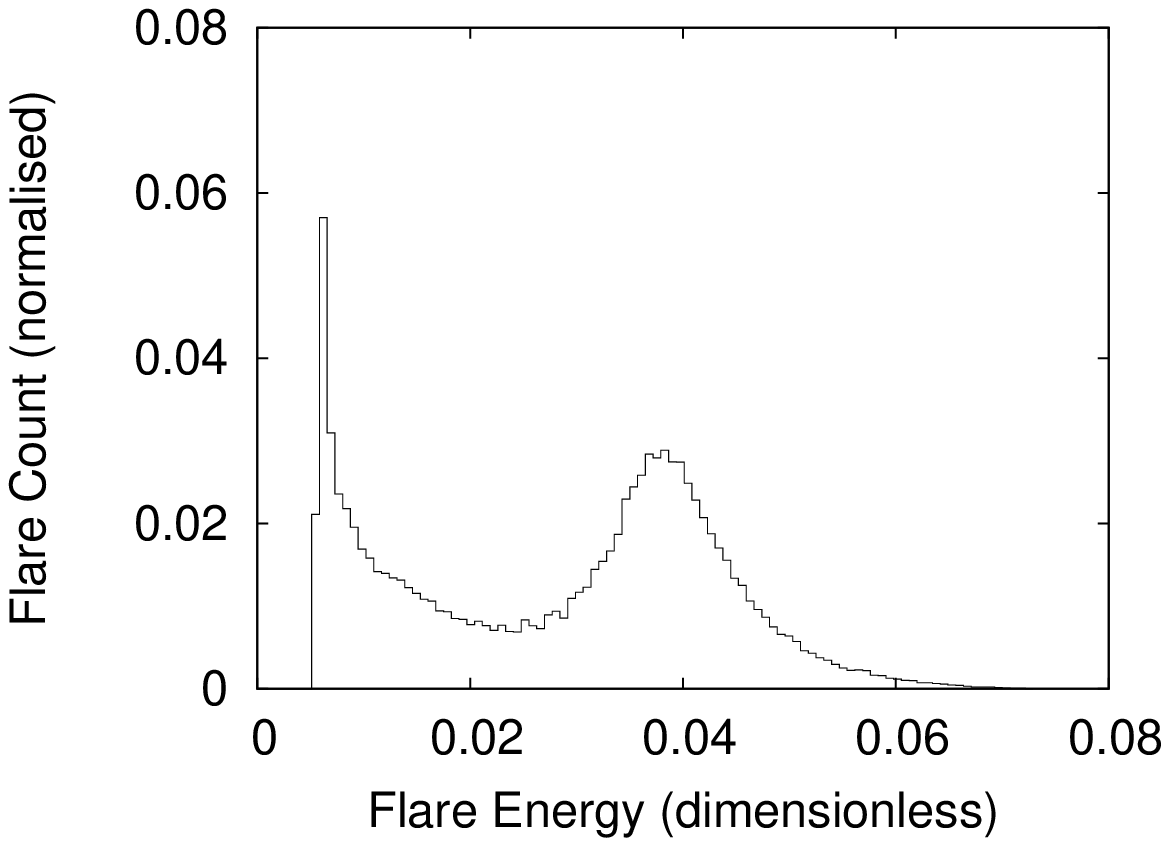}
  \includegraphics[scale=0.35]{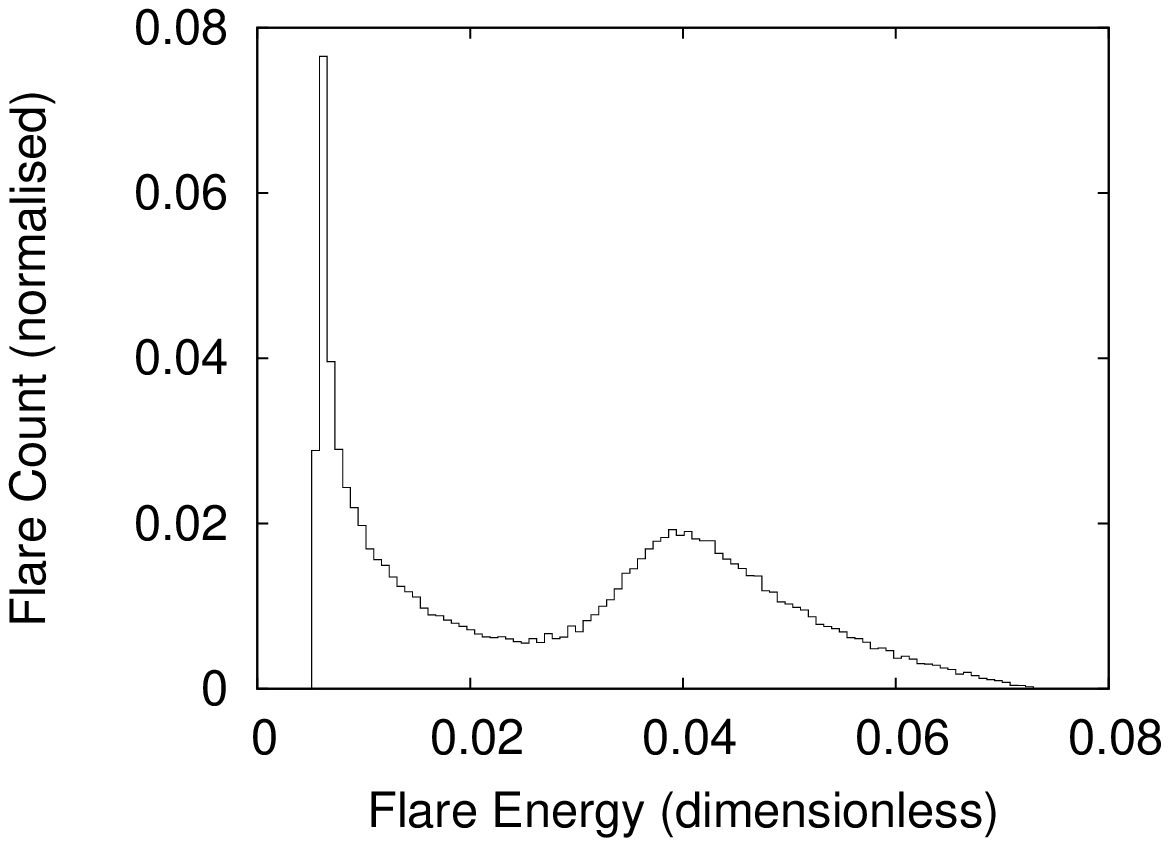}
  \caption{\small{Flare energy distributions over 10$^5$ relaxation events for a variety of loop lifetimes. The lifetimes are 1000 relaxation events (top left), 100 relaxation events (top right), 10 relaxation events (bottom left) and 1 relaxation event (bottom right).}}
  \label{wrpf_lts}
  \vspace{-10pt} 
\end{figure}

These results are not strongly tied to the loop lifetime, which is the number of relaxations a loop undergoes during the simulation. We can simulate loop replacement by randomly selecting a new position within the stability region after a certain number of relaxations. The top left distribution of Fig. \ref{wrpf_lts} is generated from a loop that is replaced after every 1000 relaxation events. As the loop lifetime is reduced so is the height of the second peak. However, this peak reduction is only noticeable for small lifetimes, e.g., $<$ 10 relaxations. A minimum lifetime of 1 relaxation (Fig. \ref{wrpf_lts}, bottom right) still yields a two-peaked distribution, albeit with a reduced second peak. This result corresponds to finding the energy release from an ensemble of identical loops. It is also the most conservative in terms of total energy released.

The concept of a loop ensemble (a collection of 10$^5$ loops flaring simultaneously) is particularly useful, since it allows us to sidestep the complications that come with allowing loops to survive many relaxations. Otherwise, we would need to understand how a loop is affected by its energy release. A loop may shrink or implode after flaring (Janse \& Low 2007). The instability threshold would be invalidated should the loop's aspect ratio be altered as a consequence of post-flare shrinkage. Therefore, the next section will examine in more detail the ensemble distribution.

\subsubsection{Distribution of "nanoflares"}
\label{sec:distribution_of_nanoflares}
The ensemble distribution (Fig. \ref{wrpf_lts}, bottom right) does not yield a simple inverse power-law when converted to a log scale: there are two peaks in the profile. The trailing edge of the first peak equates to a power-law slope of $\approx$\,-1.5. Its internal structure cannot currently be resolved since one would need to increase the threshold point density. The trailing edge of the second peak gives a slope of $\approx$\,-8.3; this is much greater than the critical gradient for nanoflare heating, $\textit{m}\,\leq\,-2$ (Hudson 1991).

These power-law figures are provisional and are likely to change as the model is enhanced. For example, a more realistic $\alpha$-space traversal function - one where $\delta\alpha_2$ is correlated with $\delta\alpha_1$ - will prefer walks parallel to the relaxation line; this will alter the distribution of heating events. Furthermore, we consider here only a single loop of fixed dimensions (or an ensemble of identical loops). In reality, in any given large-scale loop structure, sub-loops of varying radii will be generated, depending on the horizontal scale-length of the driving photospheric motions. The distribution must then be averaged over a distribution of radii, as well as considering variations in length and field strength.

\subsubsection{Heating Flux}
The primary aim of our model is to calculate the distribution of nanoflares. In general terms, the heating rate will be similar to other calculations in the literature based on random photospheric twisting (Abramenko et al. 2006; Zirker \& Cleveland 1993; Berger 1991; Sturrock \& Uchida 1981). Within our approach, all the energy input from the photosphere must be dissipated, in a long-term time-average over many events, since the build up of coronal magnetic field is limited by the instability threshold.

The energy flux, \textit{F}, can be expressed using Eqn. \ref{dim_energy_release};
\begin{eqnarray}
  \label{dim_energy_flux}	
  F & = & \frac{81\pi^2}{\mu_0} \hspace{0.1cm} \frac{1}{N\tau} \hspace{0.1cm} \frac{1}{2\pi R_c^2} \hspace{0.1cm} R_c^3 B_c^2 \hspace{0.1cm} \langle\delta W^{\hspace{0.02cm}*}\rangle = \frac{81\pi}{2\mu_0} \hspace{0.1cm} \frac{R_c B_c^2}{\tau} \hspace{0.1cm} \frac{\langle\delta W^{\hspace{0.02cm}*}\rangle}{N} \hspace{0.1cm},
\end{eqnarray}
where \textit{N} is the average number of steps taken to reach the threshold, $\tau$ is the time taken to complete each step in the random walk and $\langle\delta W^{\hspace{0.02cm}*}\rangle$ is the average dimensionless energy release, given by
\begin{eqnarray}
  \label{av_energy_release}	
  \langle\delta W^{\hspace{0.02cm}*}\rangle & = & \frac{1}{10^5} \sum_{i=1}^{10^5}\delta W^{\hspace{0.02cm}*} \hspace{0.1cm}.
\end{eqnarray}

For the ensemble case, $\langle\delta W^{\hspace{0.02cm}*}\rangle$\,$\approx$\,0.0293 and \textit{N}\,$\approx$\,264.
Applying previously used values (\textit{B}$_c$\,=\,0.01 T, $\tau$\,=\,200 s and \textit{R}$_c$\,=\,1 Mm), yields a total flux of 6\,$\times$\,10$^6$ erg cm$^{-2}$ s$^{-1}$. This result is applicable to Active Regions (a value for the Quiet Sun can be obtained by setting \textit{B}$_c$\,=\,0.001 T; this simply lowers \textit{F} to 6\,$\times$\,10$^4$ erg cm$^{-2}$ s$^{-1}$). These results are slightly lower than Withbroe \& Noyes (1977) measurements of coronal heating losses, which are 10$^7$ erg cm$^{-2}$ s$^{-1}$ (Active Regions) and 3\,$\times$\,10$^5$ erg cm$^{-2}$ s$^{-1}$ (Quiet Sun). This shortfall vanishes however if the random walk is conducted using larger step sizes, see following section.

\subsubsection{Random walk step size}
\label{sec:random_walk_step_size}
\begin{figure}[h!] 
  \vspace{-5pt}
  \center
  \includegraphics[scale=0.35]{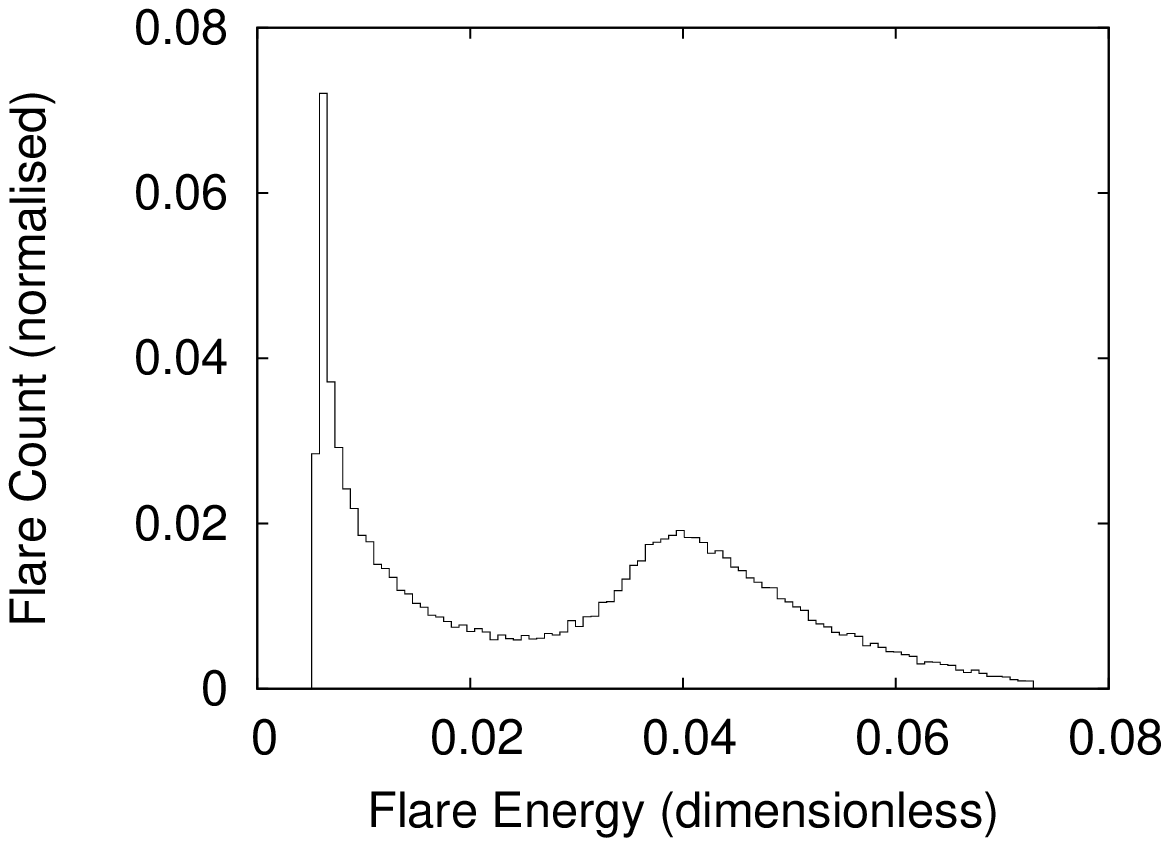}
  \includegraphics[scale=0.35]{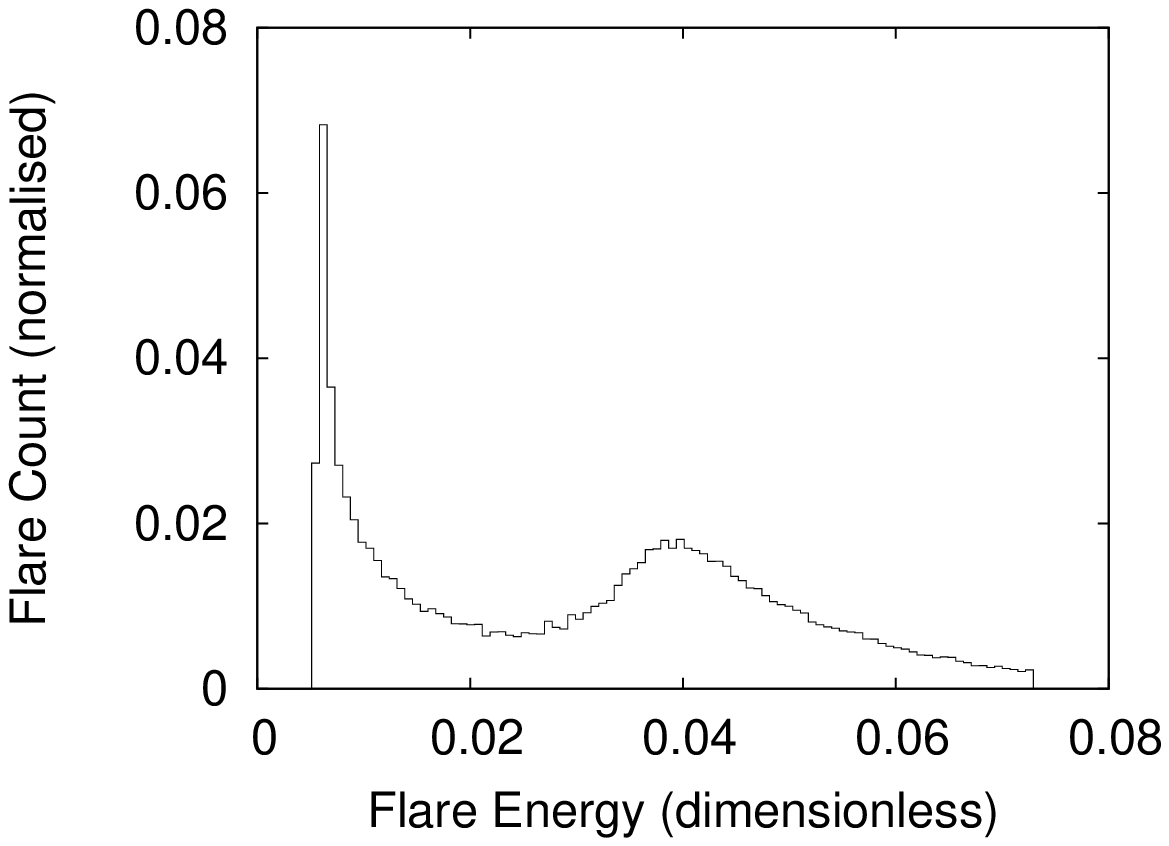}
  \caption{\small{Flare energy distributions over 10$^5$ relaxation events for a step size of $\delta\alpha$\,=\,1 (left) and $\delta\alpha$\,=\,4 (right). Loop lifetime is one relaxation event.}}
  \label{wrpf_step}
  \vspace{-5pt} 
\end{figure}

So far, we have assumed photospheric motions to be somewhat temporally incoherent, with a short random walk step, $\delta\alpha$\,=\,0.1, corresponding to $\tau$\,=\,200 s. Here, we consider the effect of varying this step length; in other words, varying the coherence time of the photospheric motions. Fig. \ref{wrpf_step}, which should be compared with the bottom right panel of Fig. \ref{wrpf_lts}, shows the distributions that result when $\delta\alpha$ is increased by factors of 10 and 40. For the latter case, the step is sufficiently long that, on average, just one step is required to reach the instability threshold: this corresponds to a temporally coherent twisting of the loop (although the spatial profile of the twisting varies randomly between heating events).

It can be seen that the energy distribution is virtually independent of the random walk step size. However, the heating flux \textit{does} depend on this quantity. It is expected that the heating flux should be proportional to $\tau$ (Berger 1991; Zirker \& Cleveland 1993). Eqn. \ref{dim_energy_flux} shows this to be the case, since the average number of steps, \textit{N}, for the random walk to reach the threshold scales as \textit{N}\,$\propto$\,$\tau^{-2}$. In particular, if we consider the limiting case of coherent twisting ($\tau$\,=\,8000 s, \textit{N}\,$\approx$\,1.13 and $\langle\delta W^{\hspace{0.02cm}*}\rangle$\,$\approx$\,0.0305) we find that the flux increases significantly, \textit{F}\,$\approx$\,3\,$\times$\,10$^7$ erg cm$^{-2}$ s$^{-1}$ for Active Regions.

\subsubsection{Random walks in twist space}
The instability threshold can be expressed in terms of $\varphi$(\textit{R}$_1$) and $\varphi$(\textit{R}$_2$). This is more representative of the twist profile (see Sect. \ref{sec:instability_threshold_and_critical_twist}), which is directly generated by photospheric motions. We use $\varphi$(\textit{R}$_1$) and $\varphi$(\textit{R}$_2$) to obtain a 2-parameter representation of the family $\varphi$(\textit{r}), since the twist profiles of Fig. \ref{rdtwpf} usually show peak twists at the radial boundaries \textit{R}$_1$ (0.5) and \textit{R}$_2$ (1.0). The instability threshold can thus be plotted in \textit{twist space} rather than \textit{alpha space}, as in Fig. \ref{it_rl_random_walks}. Given that the process is driven by turbulent photospheric motions, it is more realistic to consider that the twist randomly evolves, rather than the $\alpha$ profile. We thus repeat our calculations with a random walk in $\varphi$-space.
\begin{figure}[h!]
  \vspace{-5pt}
  \center
  \includegraphics[scale=0.35]{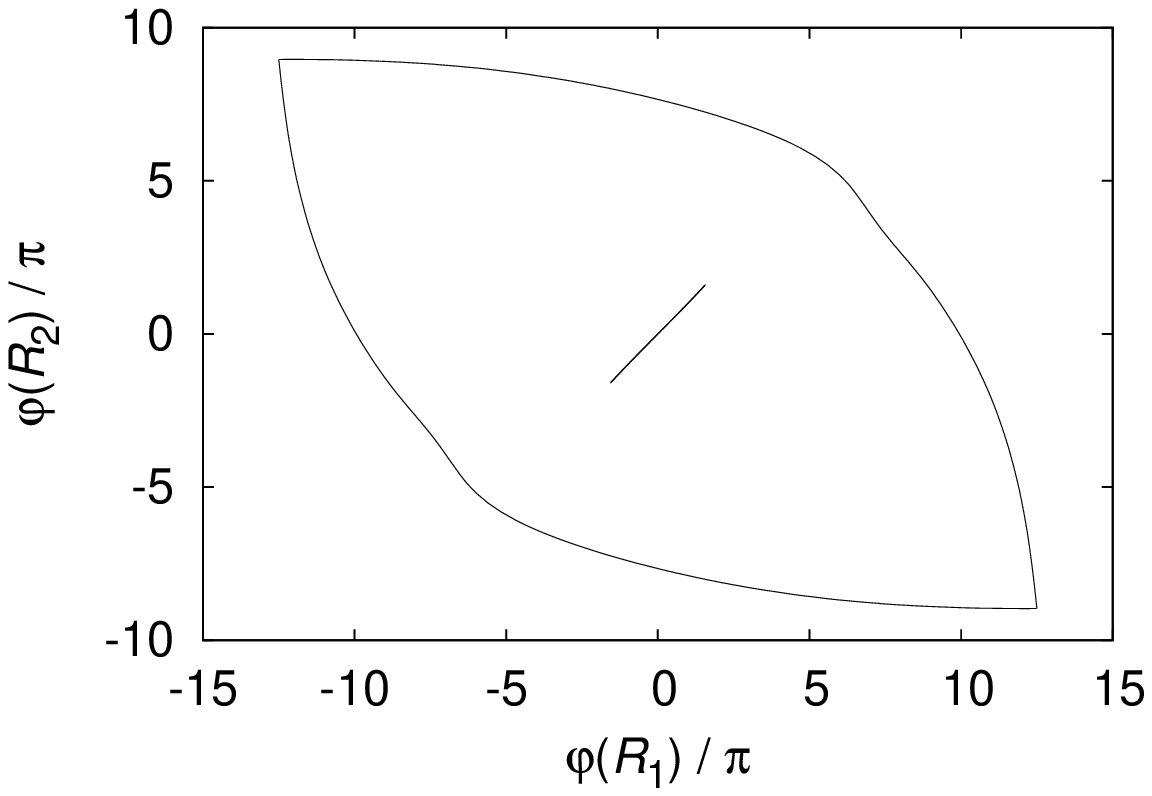}
  \includegraphics[scale=0.35]{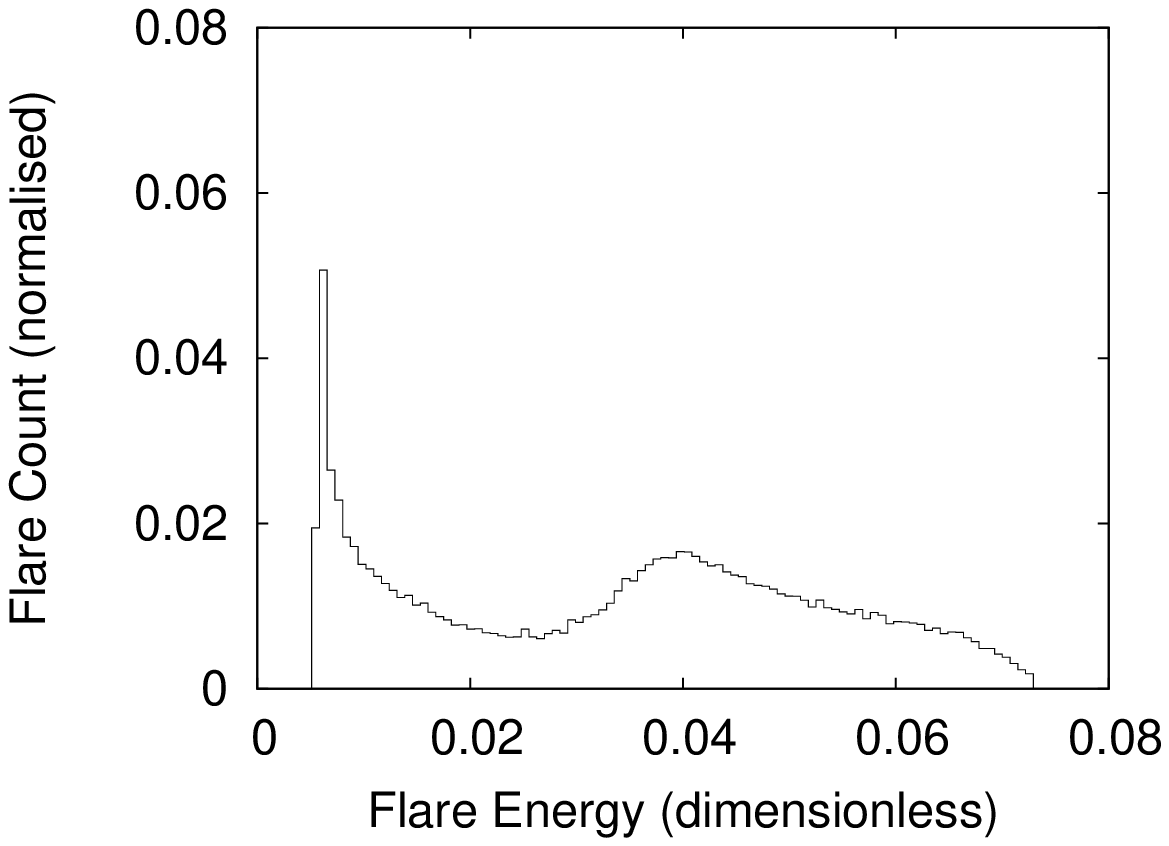}
  \caption{\small{The instability threshold and relaxation line in $\varphi$-space (left), alongside the flare energy distribution for a 10$^5$ loop ensemble performed within $\varphi$-space (right). The random walk step size, $\delta\varphi$, is approximately 0.32$\pi$, this corresponds to a step time of 200 s.}}
  \label{tw_it_wrpf}
  \vspace{-12pt}
\end{figure}

When a simulation is run, for a sequence of heating events, in $\varphi$-space, the resulting energy profile is more or less identical to that generated by an $\alpha$-space simulation, see Fig. \ref{tw_it_wrpf} (right). The translation to $\varphi$-space results in a slight reorientation of the relaxation line with respect to the threshold. Consequently, the relaxation line is closer to the part of the threshold associated with the highest energies; this explains why the distribution has a thicker tail. The energy flux is \textit{F}\,$\approx$\,4\,$\times$\,10$^6$ erg cm$^{-2}$ s$^{-1}$, which increases with step size in the same manner as shown for the $\alpha$-space simulations.

\subsection{Temporal properties}
\begin{figure}[h!]
  \vspace{-10pt}
  \center
  \includegraphics[scale=0.35]{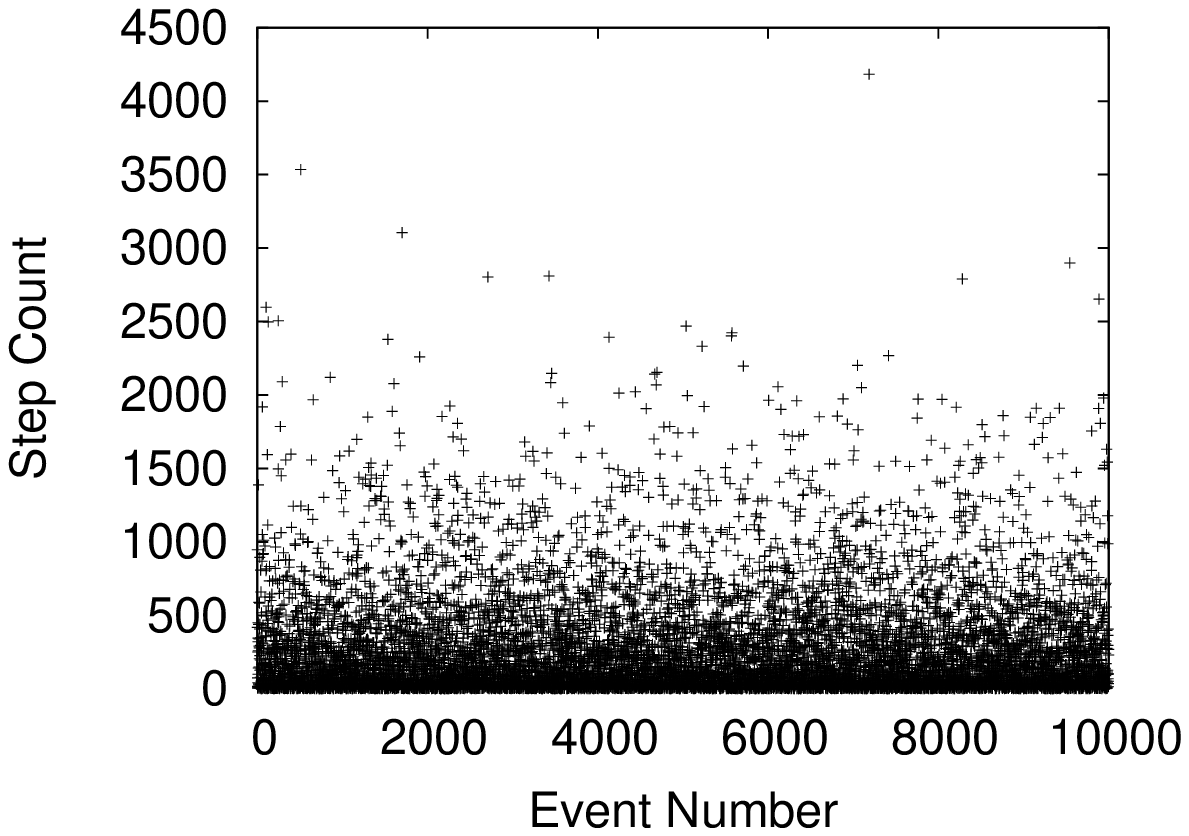}
  \includegraphics[scale=0.35]{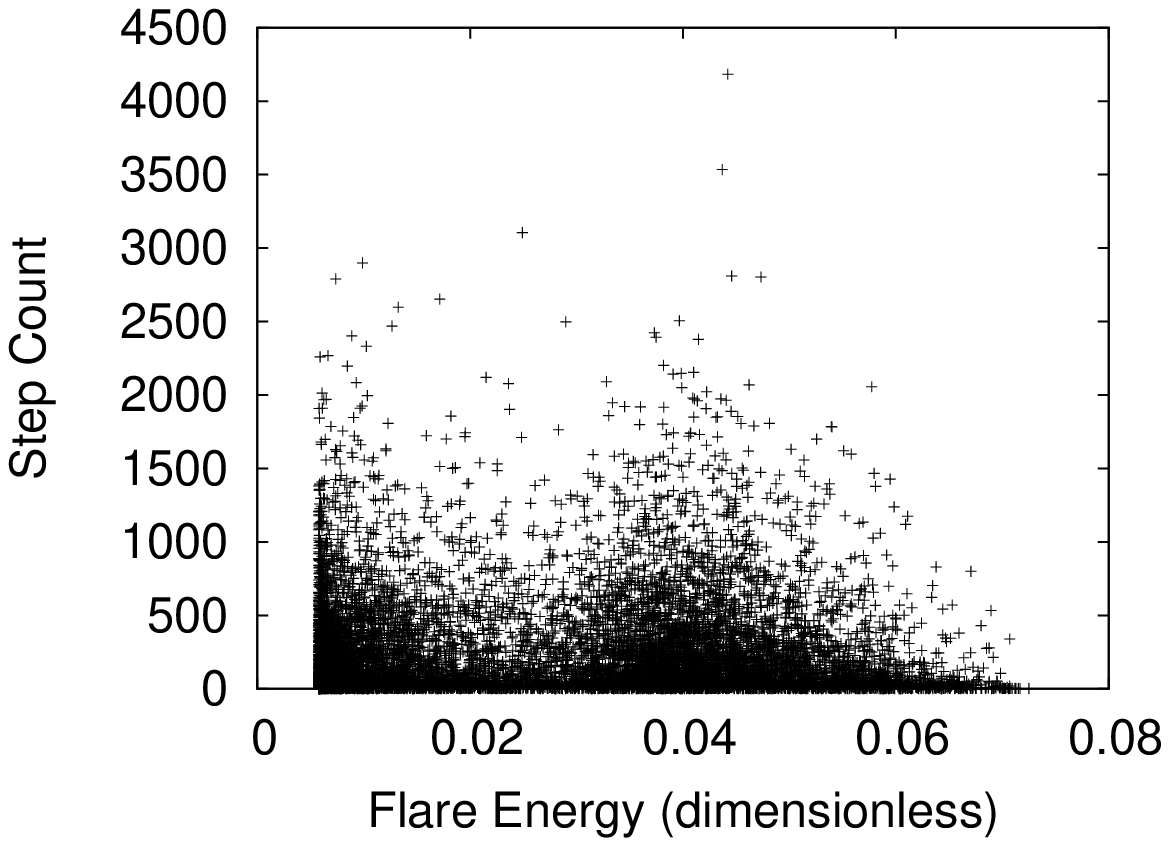}
  \includegraphics[scale=0.35]{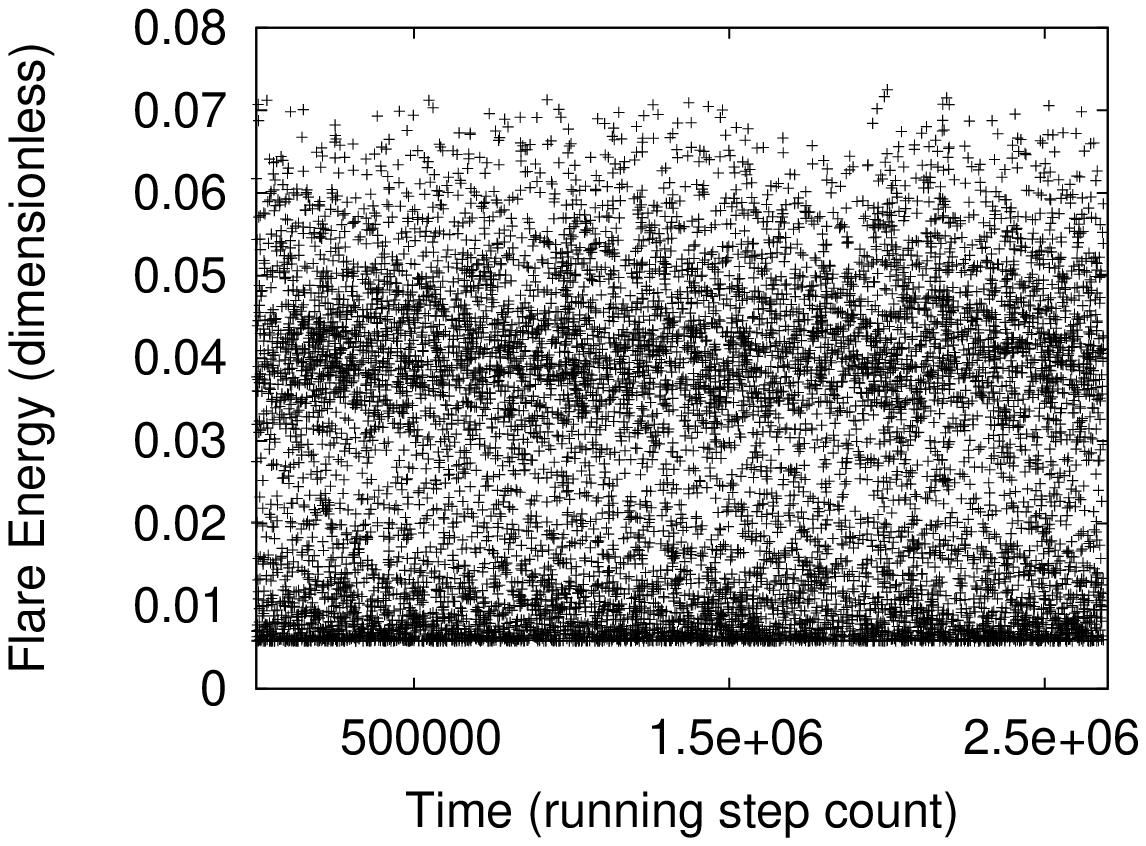}
  \includegraphics[scale=0.35]{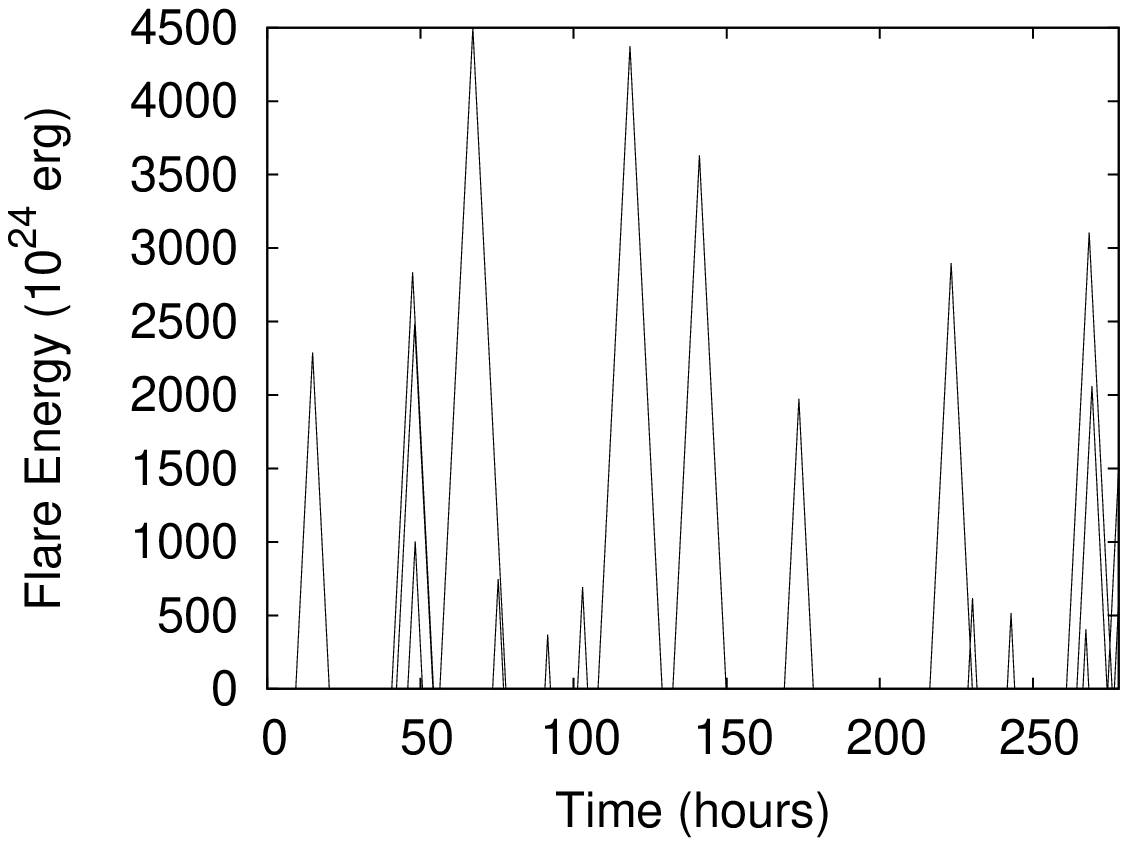}
  \caption{\small{\textit{Top Left}: the number of steps between relaxation and instability (i.e., time interval between heating events) for each event of a simulation comprising 10$^4$ events.
  \textit{Top Right}: the number of steps taken to reach the threshold against the energy released.
  \textit{Bottom Left}: the energy release as a function of time (running step count).
  \textit{Bottom Right}: the 10$^5$ event simulation produced 10$^5$ flares of varying energies. The size of the flares are shown in a way that is reminiscent of actual flare/microflare/nanoflare observations: the bigger the event, the wider the base of the triangle used to represent that flare. The figure covers a time sequence equal to 5000 steps, taken from a random position within the simulation data.}}
  \label{temp_prop}
  \vspace{-10pt}
\end{figure}

We can gain some insight into the temporal distribution of heating events by assuming that each step of a random walk in $\alpha$-space takes the same arbitrary unit of time, corresponding to photospheric driving. The time axes of Fig. \ref{temp_prop} are shown in terms of random walk step number (i.e., step count or running step count). The time unit $\tau$ is the time taken for one step, which was estimated to be 200 s, see Sect. \ref{sec:random_walk}. 

The probability that a flare event will have occurred after a particular number of steps in $\alpha$-space is invariant with time, see Fig. \ref{temp_prop} (top left). This is also true for the probability that a flare event will have a particular energy: the flare energy is not dependent on where in the simulation it occurs. The two horizontal bands shown in the bottom left panel are consistent with the peaks shown in Figs. \ref{wrpf_lts}-\ref{wrpf_step}.

There does appear to be a slight relationship between the flare interval time (i.e., the number of steps taken to reach the threshold) and the flare energy, see Fig. \ref{temp_prop} (top right); but, no positive correlation is evident. Observations strongly suggest that these two properties are uncorrelated (Wheatland 2000). The top right panel echoes the shape of the energy release distributions. The instability threshold has an energy release gradient, which means that sections of threshold that have a more or less constant energy release (low gradient) will be visited by more loops. The higher the number of visiting loops, the wider the range of step counts associated with that section of the threshold. Thus, Fig. \ref{temp_prop} (top right) is in agreement with observations.

Finally, the simulated energy releases can also be represented as a time series of flare energies, again see Fig. \ref{temp_prop} (bottom left).

\subsection{Critical magnetic shear and coronal heating requirements}
Parker (1988) explains that a magnetic flux bundle rooted in the photosphere (i.e., a coronal loop) will be stressed as photospheric motions move the footpoints in a direction transverse to the original magnetic field. The coronal heating requirements for Active Regions and for the Quiet Sun can be converted to equivalent magnetic stresses, which can in turn be converted to magnetic shears (\textit{B}$_{\perp}/$\textit{B}$_z$); 
\begin{eqnarray}
  \label{magnetic_shear_approx}
  \frac{B_{\perp}}{B_z} & \approx & \frac{4\pi F}{B_z^2\,v_{\theta}}\,,
\end{eqnarray}
where \textit{B}$_\perp$ is the transverse field, \textit{B}$_z$ is the original axial field, \textit{F} is the coronal heating requirement and \textit{v}$_{\theta}$ is the photospheric flow velocity. We set \textit{v}$_{\theta}$\,=\,1 km s$^{-1}$ for consistency and find that the necessary magnetic shear is approximately 0.12 ($\sim$7$^{\circ}$) for Active Regions and 0.38 ($\sim$21$^{\circ}$) for the Quiet Sun.

\begin{figure}[h!]
  \vspace{-5pt}
  \center
  \includegraphics[scale=0.45]{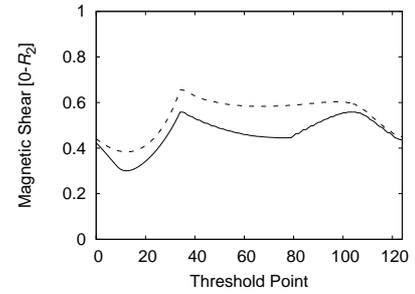}
  \caption{\small{The mean absolute (solid) and root mean square (dashed) of the magnetic shear along the instability threshold. The shears are calculated over the loop volume, 0-\textit{R}$_2$.}}
  \label{it_a2p_shear_rms_mabs}
  \vspace{-10pt} 
\end{figure}

Thus, for coronal heating to \textit{work}, there must be some mechanism which restricts the shear to around these levels. If dissipation occurs at lower levels of shear the energy flux cannot maintain coronal temperatures. Conversely, if the shear can build up to much larger levels, the energy input would be higher than required. Our model provides an explanation for these critical shear values. We calculate two types of mean magnetic shear along the threshold, the mean absolute shear and the root mean square shear:
\begin{eqnarray}
  \label{magnetic_shear}
  S_{MABS} & = & \frac{2}{R_2^2}\int_{0}^{R_2}r\,\,\Bigg|\frac{B_{\theta}}{B_z}\Bigg| \, dr\,,\\ 
  \nonumber \\ 
  S_{RMS} & = & \sqrt{\frac{2}{R_2^2}\int_{0}^{R_2}r\,\,\Bigg(\frac{B_{\theta}}{B_z}\Bigg)^2 \, dr}\,\,.
\end{eqnarray}
Both quantities are area-averaged. The plot (Fig. \ref{it_a2p_shear_rms_mabs}) shows that the values of \textit{S}$_{MABS}$ and \textit{S}$_{RMS}$ are comparable with those derived above.
In general, we find that the average shear at the threshold is slightly higher than the limit suggested by Parker. Perhaps more shear is required, since not all of the magnetic energy released during relaxation will be converted to heat. Fig. \ref{it_a2p_shear_rms_mabs} suggests that the apparent limiting value for magnetic shear is determined by the linear instability.

We should mention the work of Dahlburg et al. (2009), who provide an alternative explanation of this shear dependency involving an explosive "secondary instability". They have used a 3D viscoresistive MHD code to simulate the shearing of a line-tied flux tube. The results of this code have revealed that the secondary instability requires a critical shear for onset.

\section{Summary and conclusions}
We have used a model based on relaxation theory to determine the energy release of a sequence of heating events generated by random photospheric footpoint motions. This is a "stress-relax" scenario, in which free energy is continually built-up in the corona due to photospheric driving - repeated relaxations occur whenever a critical condition is reached. We postulate that heating events are triggered by the onset of an ideal kink instability. Energy release then occurs by fast magnetic reconnection in current sheets generated during the nonlinear phase of the instability, and this is calculated by assuming a helicity-conserving relaxation to a minimum energy state. The energy release of any individual heating event depends on the current profile at the point where the instability threshold is crossed - thus heating events of a range of sizes are found.

The initial stressed fields are modelled using piecewise-constant profiles of $\alpha$, yielding a two-parameter family of current profiles. Photospheric driving thus yields a random walk of the current profile. In order to pursue the calculation, the ideal instability threshold of a line-tied loop has been determined. The modelled current profiles incorporate regions of positive and negative twist as well as monotonic twist profiles, as have been more commonly studied. We have shown that it is not possible to determine the onset of instability by any simple criterion such as achieving a critical twist or critical helicity. Nevertheless, Fig. \ref{rdtwpf} suggests that the location of the instability is coincident with the peak twist (the largest absolute twist). 

A distribution of heating events has been calculated, both for the case of a single loop which undergoes repeated stressing and relaxation, and for an ensemble of identical loops which are randomly stressed (and for intermediate cases). For a sufficiently large number of events, a statistically-stable distribution of event sizes is obtained. As expected, the smallest events are the most common, and there is (as noted by Browning \& Van der Linden 2003) a minimum event size for given loop parameters, perhaps corresponding to an "elemental nanoflare". More surprisingly, there is a second peak of event frequency at intermediate magnitudes, although this is somewhat reduced in size if an ensemble of loops is considered. This can be explained as follows. The first peak (at minimum energy) occurs simply because the range of energies near the minimum naturally encompasses the largest part of the instability threshold curve (because of the flatness of the minimum). The second peak is found because the instability threshold is most likely to be crossed in the part near to the region of constant-$\alpha$.

An ensemble of identical loops - in which each individual loop starts from a randomly chosen initial state and is stressed until it undergoes a single relaxation event - is investigated in more detail. The distribution of heating events is qualitatively similar, although, the secondary peak of event frequency is much lower, and the decay of frequency with increasing energy is much flatter. A significant result is that, although the distribution is not a simple power law, the high-energy part of the distribution \textit{is} well approximated by a power law with an index of around -8.3, which is considerably steeper than the minimum required for nanoflare heating to be effective (-2).

The model requires that the field is sometimes unstable - although most of the time, the field profile will be well within the stable region. Typically, the dimensional $\alpha$ value is of magnitude 1\,-\,2, leading to dimensional values of $\alpha$\,$\approx$\,1\,-\,2\,Mm$^{-1}$ (for a loop radius of 1 Mm). Note that this is the maximum value which could be found, and usually we would expect lower values. This is consistent with observations; for example R\'egnier \& Priest (2007) find $\alpha$ magnitudes around 1\,Mm$^{-1}$. Furthermore, a consequence of the fact that the fields are predicted to fluctuate between stable and unstable states is that we predict a value for the average horizontal field component (on average, this will be somewhat less than the value at marginal stability). At threshold, we find \textit{B}$_\perp$/\textit{B}$_z$ is around 0.5, which means that its average value should be around 0.25. This agrees very well with the limit on this quantity required by Parker (1988), in order for the Poynting flux from the photosphere to match coronal heating requirements. Hence, we have an explanation for the critical shear value predicted by Parker. This also implies that the average heating flux from our model will be sufficient for coronal heating. Indeed, the fluxes derived from our results agree (especially if we raise the correlation time for photospheric motions) with the required values.

Relaxation theory makes no prediction about the spatial distribution of energy dissipation. However, this can be determined from numerical simulations. Recently, Hood et al. (2009) have shown that heating is well-distributed across the loop volume, as the current sheet, associated with the nonlinear kink instability, stretches and fragments, thereby filling the loop cross section. This has implications for the observed emission.

A further important consideration is that we study a single loop of fixed dimensions, with repeated stressing and relaxation. In practice, coronal heating events will occur in regions of field of varying sizes. A large flare will inevitably involve a large magnetic field volume, whereas nanoflares may involve small sets of fieldlines. This could be accounted for in our model by allowing the loop radius to be randomly distributed also; the effect would be to convolve a set of our energy distributions. A second limitation of the calculation is that we have assumed a uniform random walk in $\alpha$ space. This is not realistic, since photospheric twisting motions are likely to be correlated across the loop cross section and therefore changes in which the twist in the outer later is similar to that in the core are much more likely: in other words, there should be a positive correlation between changes in $\alpha_1$ (the core current) and changes in $\alpha_2$ (the current in the outer layer), rather than these being independent random variables as assumed here. This will be considered further in future work.

As well as the improvements mentioned above, the model will be enhanced such that there is, more realistically, zero net-current carried by the loop. At present, in most cases the potential layer outside the loop contains azimuthal field. A current neutralisation layer will be inserted between the loop and the envelope - external magnetic fields will have an axial direction only (Hood et al. 2009), representing a response to localised photospheric twist motions.

The model has included a number of simplifications, and at this stage is more of a "proof of principle" showing that a distribution of heating events can be produced from an (almost) \textit{ab initio} coronal heating model. These initial results demonstrate the viability of the model for further research.\\\\

\small{\textit{Acknowledgements}. We are grateful to Jim Klimchuk for helpful comments and M. R. B. thanks STFC for studentship support.}

\appendix
\section{Expressions for loop properties}
Expressions for some key quantities ($\langle\tilde{\varphi}\rangle$, \textit{K} and \textit{W}) are given here. For compactness, these are given only for $\alpha_{1}$\,$\neq$\,0 and $\alpha_{2}$\,$\neq$\,0, while special cases (e.g., $\alpha_1$\,=\,0) must be dealt with separately. Expressions for constant-$\alpha$ fields can be recovered by setting $\alpha_1$\,=\,$\alpha_2$, which gives more familiar formulae. The superscripts and subscripts that accompany each quantity term denote the upper and lower radial bounds over which the quantity is calculated.

\subsection{Average magnetic twist}
\begin{eqnarray}
  \langle\tilde{\varphi}\rangle_0^{R_1} & = & \frac{\sigma_1 L\Big[1-J_0(|\alpha_1|R_1)\Big]}{R_1 J_1(|\alpha_1|R_1)}\\
  \nonumber   &   &\\
  \nonumber   &   &\\
  \langle\tilde{\varphi}\rangle_{R_1}^{R_2} & = & \frac{\sigma_2 L\Big[F_0(|\alpha_2|R_1)-F_0(|\alpha_2|R_2)\Big]}{\Big[R_2 F_1(|\alpha_2|R_2)-R_1 F_1(|\alpha_2|R_1)\Big]}\\
  \nonumber   &   &\\
  \nonumber   &   &\\
  \langle\tilde{\varphi}\rangle_{R_2}^{R_3} & = & \frac{2\sigma_2 L R_2\Big[B_2 J_1(|\alpha_2|R_2) + C_2 Y_1(|\alpha_2|R_2)\Big]\log\Big(\frac{R_3}{R_2}\Big)}{B_2 F_0(|\alpha_2|R_2)\Big[R_3^{\,2} - R_2^{\,2}\Big]} 
\end{eqnarray}

\subsection{Magnetic helicity}
\begin{eqnarray}  
  \nonumber K_0^{R_1} & = & \sigma_1\frac{2\pi L B_1^{\,2}}{|\alpha_1|}\bigg[R_1^{\,2} J_0^{\,2}(|\alpha_1|R_1)+R_1^{\,2} J_1^{\,2}(|\alpha_1|R_1)\\
  &   & \,\,\,\,\,\,\,\,\,\,\,\,\,\,\,\,\,\,\,\,\,\,\,\,\,\,\,-\,2\frac{R_1}{|\alpha_1|}J_0(|\alpha_1|R_1)J_1(|\alpha_1|R_{1})\bigg]\\
  \nonumber   &   &\\
  \nonumber   &   &\\
  \nonumber K_{R_1}^{R_2} & = & \frac{2\sigma_2\pi LB_2^{\,2}}{|\alpha_2|}\Bigg[R_2^{\,2}F_0^{\,2}(|\alpha_2|R_2)+R_2^{\,2} F_1^{\,2}(|\alpha_2|R_2)\\
  \nonumber   &   &\\
  \nonumber   &   & \,\,\,\,\,\,\,\,\,\,\,\,\,\,\,\,\,\,\,\,\,\,\,\,\,\,-\,2\frac{R_2}{|\alpha_2|}F_0(|\alpha_2|R_2)F_1(|\alpha_2|R_2)\\
  \nonumber   &   &\\
  \nonumber   &   & \,\,\,\,\,\,\,\,\,\,\,\,\,\,\,\,\,\,\,\,\,\,\,\,\,\,-\,R_1^{\,2} F_0^{\,2}(|\alpha_2|R_1)\,-\,R_1^{\,2} F_1^{\,2}(|\alpha_2|R_1)\\
  \nonumber   &   &\\
  \nonumber   &   & \,\,\,\,\,\,\,\,\,\,\,\,\,\,\,\,\,\,\,\,\,\,\,\,\,\,+\,2\frac{R_1}{|\alpha_2|}F_0(|\alpha_2|R_1)F_1(|\alpha_2|R_1)\\
  \nonumber   &   &\\
  \nonumber   &   & \,\,\,\,\,\,\,\,\,\,\,\,\,\,\,\,\,\,\,\,\,\,\,\,\,\,+\,\frac{2 B_1 R_1 J_1(|\alpha_1|R_1)}{B_2}\Big[F_0(|\alpha_2|R_1)-F_0(|\alpha_2|R_2)\Big]\\
  &   & \,\,\,\,\,\,\,\,\,\,\,\,\,\,\,\,\,\,\,\,\,\,\,\,\,\,\,\,\,\,\,\,\times\,\Bigg(\frac{1}{|\alpha_1|}-\frac{\sigma_{1,2}}{|\alpha_2|}\Bigg)\Bigg]
\end{eqnarray}
\begin{eqnarray}  
  K_{R_2}^{R_3} & = & 2\sigma_2 L C_3 R_2\Bigg[\Big(\psi_{R_2} - \pi B_3 R_2^{\,2}\Big)\log\Bigg(\frac{R_3}{R_2}\Bigg) + \frac{\pi B_3}{2}\Big(R_3^{\,2} - R_2^{\,2}\Big)\Bigg]
\end{eqnarray}

\subsection{Magnetic energy}
\begin{eqnarray}
  \nonumber W_0^{\,R_1}     & = &\frac{L\pi B_1^{\,2}}{\mu_0}\Bigg[R_1^{\,2} J_0^{\,2}(|\alpha_1|R_1)+R_1^{\,2} J_1^{\,2}(|\alpha_1|R_1)\\
  \nonumber               &   &\\
                          &   &\,\,\,\,\,\,\,\,\,\,\,\,\,\,\,\,\,-\frac{R_1}{|\alpha_1|}J_0(|\alpha_1|R_1)J_1(|\alpha_1|R_1)\Bigg]\\
  \nonumber               &   &\\
  \nonumber               &   &\\
  \nonumber W_{R_1}^{\,R_2} & = &\frac{L\pi B_2^{\,2}}{\mu_0}\Bigg[R_2^{\,2} F_0^{\,2}(|\alpha_2|R_2)+R_2^{\,2} F_1^{\,2}(|\alpha_2|R_2)\\
  \nonumber               &   &\\
  \nonumber               &   &\,\,\,\,\,\,\,\,\,\,\,\,\,\,\,\,\,-\,\frac{R_2}{|\alpha_2|}F_0(|\alpha_2|R_2)F_1(|\alpha_2|R_2)\\
  \nonumber               &   &\\
  \nonumber               &   &\,\,\,\,\,\,\,\,\,\,\,\,\,\,\,\,\,-\,R_1^{\,2} F_0^{\,2}(|\alpha_2|R_1)-R_1^{\,2} F_1^{\,2}(|\alpha_2|R_1)\\
  \nonumber               &   &\\
                          &   &\,\,\,\,\,\,\,\,\,\,\,\,\,\,\,\,\,+\,\frac{R_1}{|\alpha_2|}F_0(|\alpha_2|R_1)F_1(|\alpha_2|R_1)\Bigg]\\
  \nonumber   &   &\\
  \nonumber   &   &\\
  W_{R_2}^{\,R_3} & = &\,\frac{L\pi}{\mu_0}\Bigg[\frac{B_3^{\,2}}{2}\bigg(R_3^{\,2} - R_2^{\,2}\bigg) + C_3^{\,2} R_2^{\,2}\log\Bigg(\frac{R_3}{R_2}\Bigg)\Bigg]
\end{eqnarray}

\section{Magnetic field profiles for a selection of $\alpha$-space points}
\begin{figure}[h!]
  \vspace{-10pt}
  \center
  \includegraphics[scale=0.35]{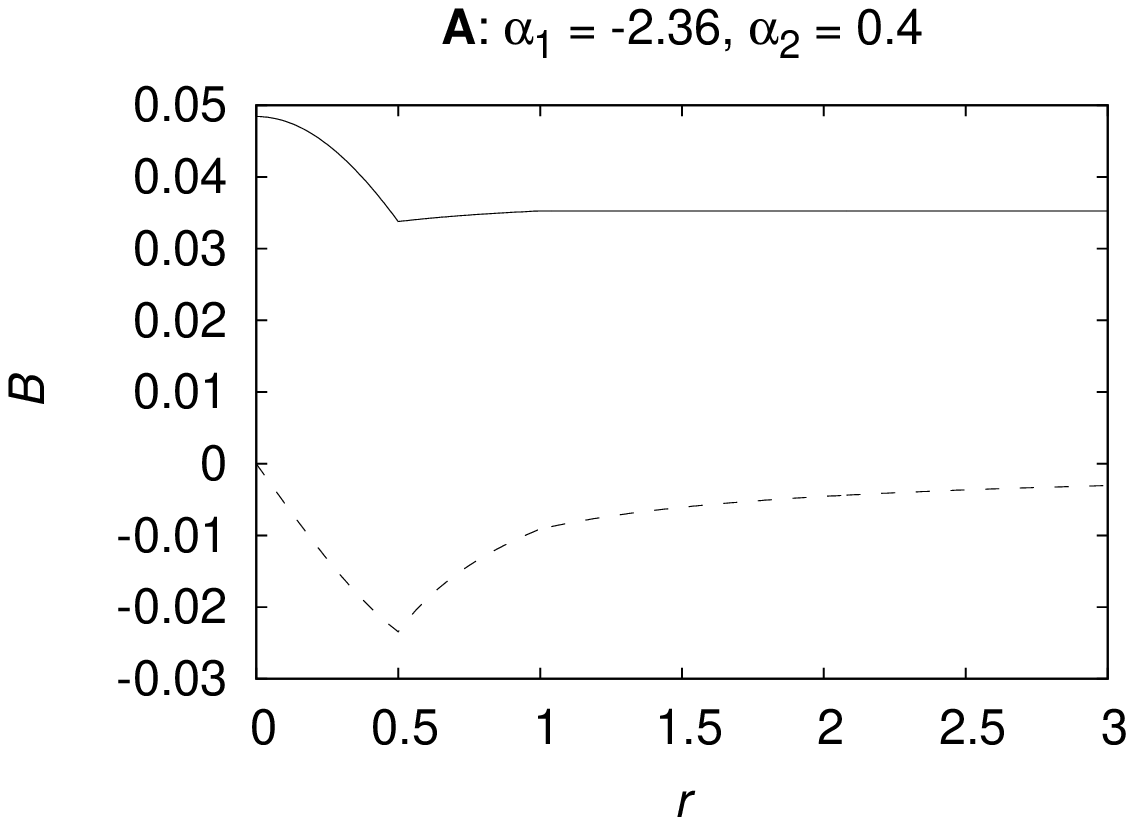}
  \includegraphics[scale=0.35]{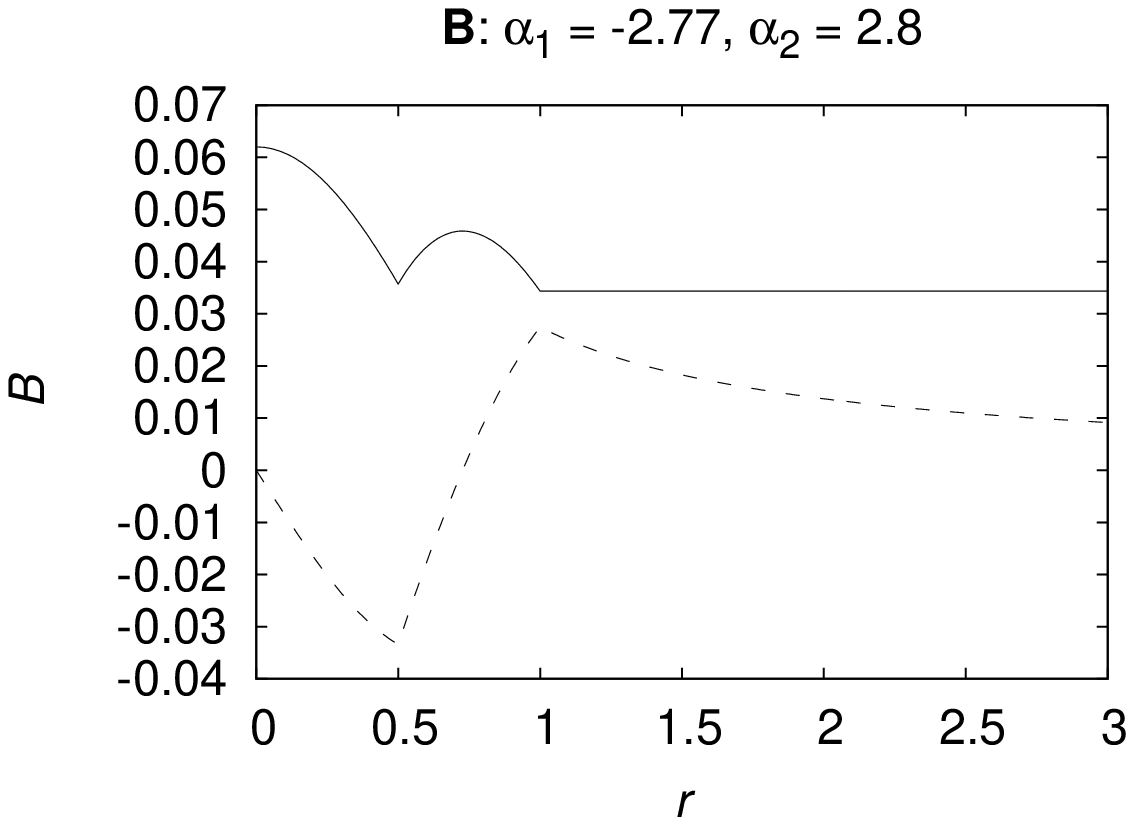}
  \center
  \includegraphics[scale=0.35]{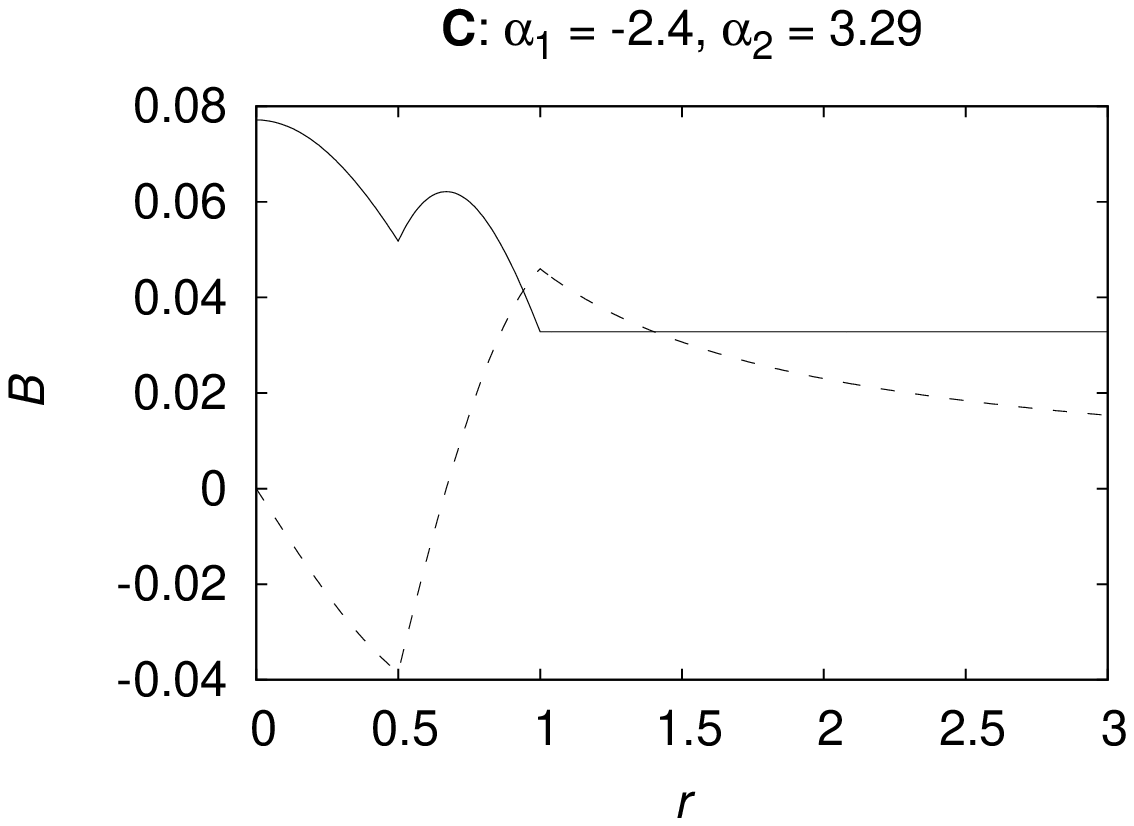}
  \includegraphics[scale=0.35]{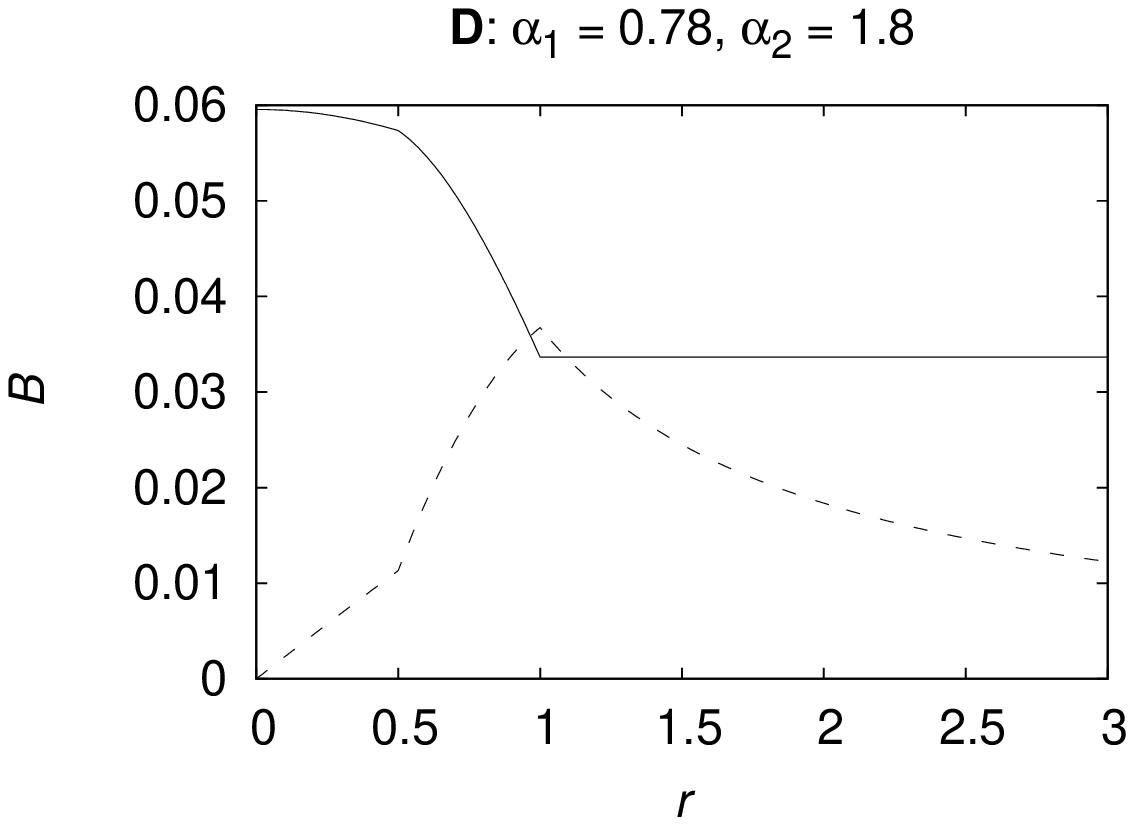}
  \center
  \includegraphics[scale=0.35]{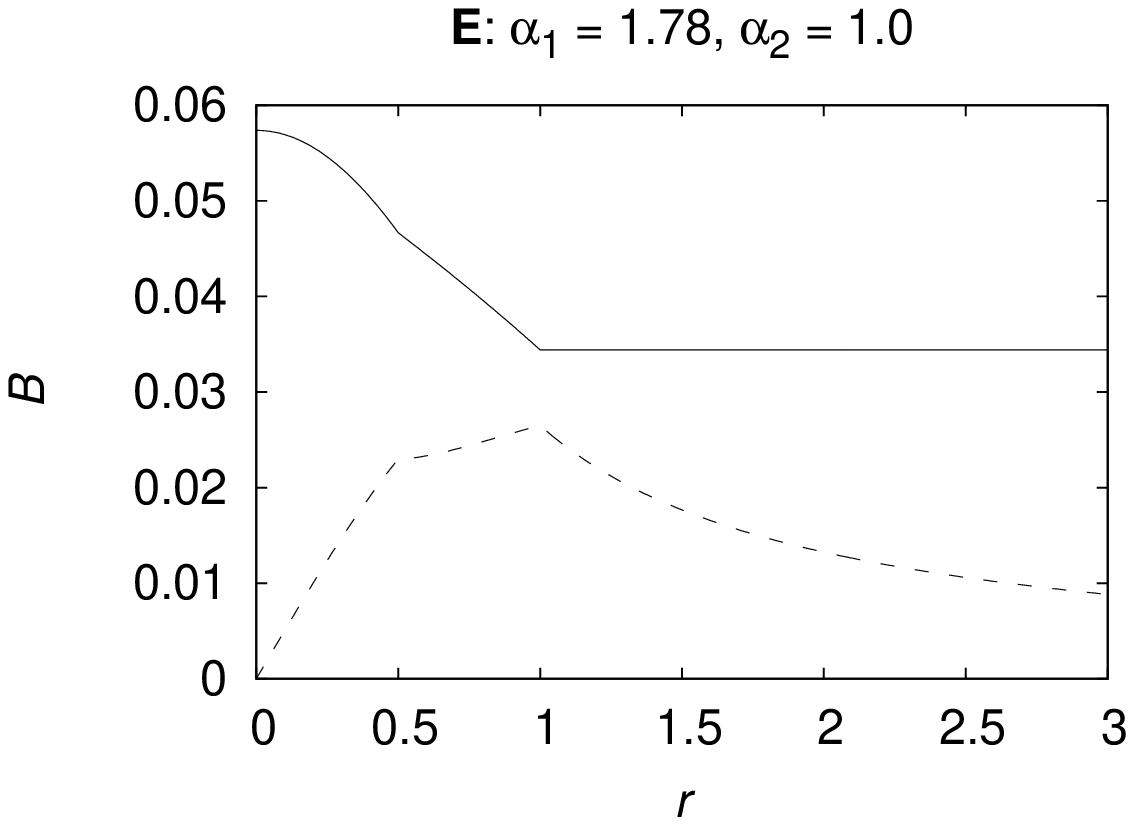}
  \includegraphics[scale=0.35]{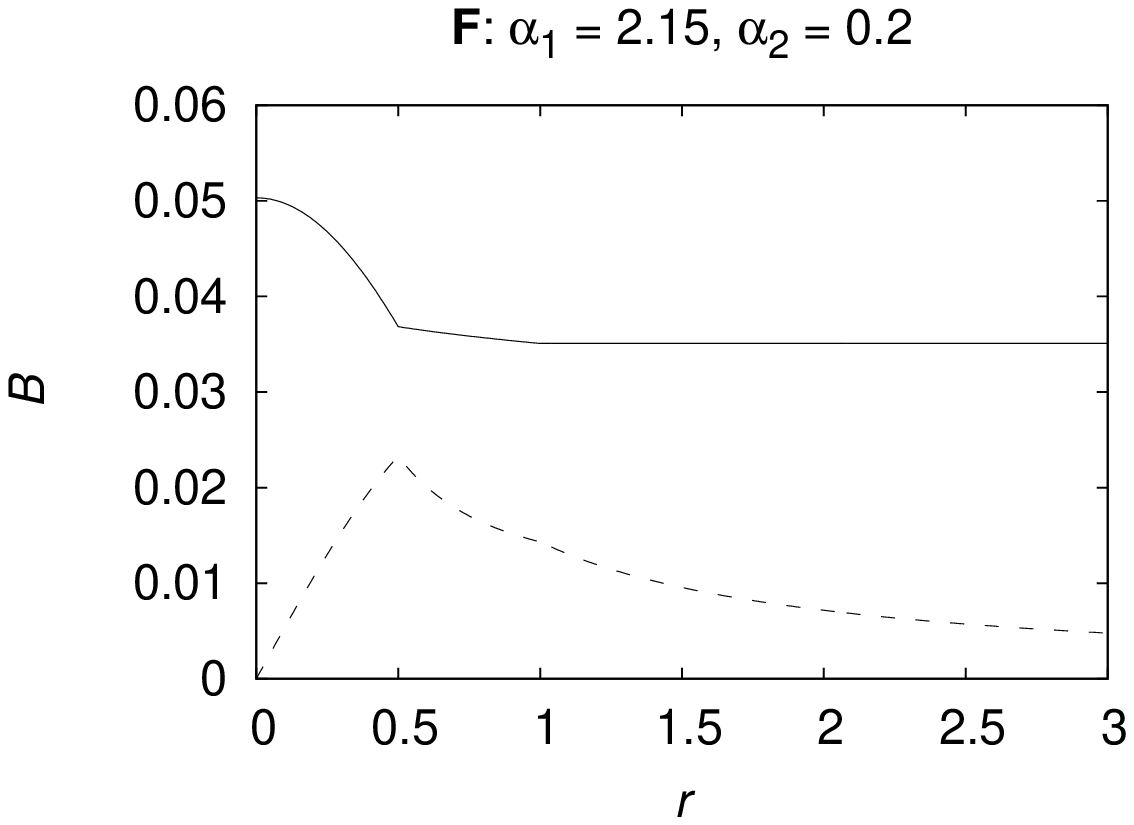}
  \caption{\small{The magnetic field profiles, \textit{B}$_z$ (solid) and \textit{B}$_\theta$ (dashed), for the six $\alpha$-space points identified in the top left panel of Fig. \ref{rdtwpf}.}}
  \label{rdbpf}
  \vspace{-10pt} 
\end{figure}


\newpage
\begin{thebibliography}{99}
\bibitem{Abramenko2006}Abramenko, V. I., Pevstov, A. A., \& Romano, P. 2006, ApJ Lett, L81, 646
\bibitem{Arber1999}Arber, T. D., Longbottom, A. W., \& Van der Linden, R. A. M. 1999, ApJ, 517, 990
\bibitem{Aschwanden2002}Aschwanden, M. J., \& Parnell, C. E. 2002, ApJ, 572, 1048
\bibitem{Baty1996}Baty, H., \& Heyvaerts, J. 1996, A\&A, 308, 935
\bibitem{Baty2000}Baty, H. 2000, A\&A, 360, 345
\bibitem{Baty2001}Baty, H. 2001, A\&A, 367, 321
\bibitem{Berger1999}Berger, M. A. 1999, Plasma Phys. Control. Fusion, 41, B167
\bibitem{Berger1991}Berger, M. A. 1991, A\&A, 252, 369
\bibitem{Berger1984}Berger, M. A., \& Field, G. B. 1984, J. Fluid. Mech., 147, 133
\bibitem{Browning1986}Browning, P. K., Sakurai, T. \& Priest, E. R. 1986, A\&A, 158, 217 
\bibitem{Browning2003}Browning, P. K., \& Van der Linden, R. A. M. 2003, A\&A, 400, 355
\bibitem{Browning2008}Browning, P. K., Gerrard, C., Hood, A. W., Kevis, R., \& Van der Linden, R. A. M. 2008, A\&A, 485, 837
\bibitem{Dahlburg2009}Dahlburg, R. B., Liu, J., Klimchuk, J. A., \& Nigro, G. 2009, ApJ, 704, 1059
\bibitem{Finn1985}Finn, J. M., \& Antonsen, T. M. 1985, Communications in Plasma Physics and Controlled Fusion, 9, 111
\bibitem{Fletcher2009}Fletcher, L. 2009, A\&A, 493, 241
\bibitem{Galsgaard1997}Galsgaard, K. and Nordlund, A. 1997, J.Geophys. Res., 102 A1, 219
\bibitem{Gerrard2001}Gerrard, C. L., Arber, T. D., Hood, A. W., \& Ven der Linden, R. A. M. 2001, A\&A, 373, 1089
\bibitem{Heyvaerts1984}Heyvaerts, J., \& Priest, E. R. 1984, A\&A, 137, 63
\bibitem{Hollweg1984}Hollweg, J. V. 1984, ApJ, 277, 392
\bibitem{Hood1979}Hood, A. W., \& Priest, E. R. 1979, Solar Phys., 64, 303
\bibitem{Hood1981}Hood, A. W., \& Priest, E. R. 1981, Geophys. Astrophys. Fluid Dynamics, 17, 297
\bibitem{Hood1992}Hood, A. W. 1992, Plasma Phys. Control. Fusion, 34, 411
\bibitem{Hood2009}Hood, A. W., Browning, P. K., \& Van der Linden, R. A. M. 2009, A\&A, 506, 913
\bibitem{Hudson1991}Hudson, H. S. 1991, Sol. Phys. 133, 357
\bibitem{Janse2007}Janse, \r{A}. M., \& Low, B. C. 2007, A\&A, 472, 957
\bibitem{Klimchuk2006}Klimchuk, J. A. 2006, Sol. Phys. 234, 41
\bibitem{Krucker1998}Krucker, S., \& Benz, A. O. 1998, ApJ, 501, L213
\bibitem{Kwon2008}Kwon, R. Y., \& Chae, J. 2008, ApJ, 677, L141
\bibitem{Lionello1998}Lionello, R., Velli, M., Einaudi, G. \& Miki\'c, Z. 1998, ApJ, 494, 840
\bibitem{Malanushenko2009}Malanushenko, A., Longcope, D. W., Fan, Y., \& Gibson, S. E. 2009, ApJ, 702, 580
\bibitem{Melrose1994}Melrose, D. B., Nicholls, J., \& Broderick, N. G. 1994, J. Plasma Phys., 51, 163
\bibitem{Mikic1990}Miki\'c, Z., Schnack, D. D., \& Van Hoven, G. 1990, ApJ, 361, 690
\bibitem{Narain1996}Narain, U., \& Ulmschieder, P. 1996, Space Sci. Rev., 75, 453
\bibitem{Parker1988}Parker, E. N. 1988, ApJ, 330, 474
\bibitem{Parnell2000}Parnell, C. E., \& Jupp, P. E. 2000, ApJ, 529, 554
\bibitem{Parnell2004}Parnell, C. E., Proceedings of the SOHO-15 Workshop (ESA SP-575), ESA Pub. Div., Noordwijk, Holland, 227.
\bibitem{Porter1994}Porter, L. J., Klimchuk, J. A., \& Sturrock, P. A. 1994, ApJ, 435, 482
\bibitem{Portier-Fozzani2001}Portier-Fozzani, F., Aschwanden, M., D\'emoulin, P., Neupert, W., \& Delaboudini\`ere. 2001, Solar Phys., 203, 289
\bibitem{Priest1987}Priest, E. R. 1987, Solar Magnetohydrodynamics, D. Reidel Publishing Company (Dordrecht, The Netherlands)
\bibitem{Priest2005}Priest, E. R., Longcope, D. W., \& Heyvaerts, J. 2005, ApJ, 624, 1057
\bibitem{Qiu2009}Qiu, J. 2009, ApJ, 692 1110
\bibitem{Regnier2007}R\'egnier, S., \& Priest, E.R. 2007, A\&A 468, 761
\bibitem{Sturrock1981}Sturrock, P. A., \& Uchida, Y. 1981, ApJ, 246, 331
\bibitem{Taylor1974}Taylor, J. B. 1974, Phys. Rev. Lett., 33, 1139
\bibitem{Taylor1986}Taylor, J. B. 1986, Rev. Mod. Phys., 58, 741
\bibitem{VanderLinden1999}Van der Linden, R. A. M., \& Hood, A. W. 1999, A\&A, 346, 303
\bibitem{Vekstein1993}Vekstein, G. E., Priest, E. R., \& Steele, C. D. C. 1993, ApJ, 417, 781
\bibitem{Velli1990}Velli, M., Einaudi, G., \& Hood, A. W. 1990, ApJ, 350, 428
\bibitem{Velli1997}Velli, M., Lionello, R., \& Einaudi, G., 1997, Sol. Phys., 172, 257
\bibitem{Wheatland2000}Wheatland, M. S. 2000, Sol. Phys., 191, 381
\bibitem{Withbroe1977}Withbroe, G. L., \& Noyes, R. W. 1977, Ann. Rev. Astr. Ap., 15, 363
\bibitem{Woltjer1958}Woltjer, L. 1958, Proc. Natl. Acad. Sci., USA, 44, 489
\bibitem{Zhang2003}Zhang, M., \& Low, B. C. 2003, ApJ, 584, 479
\bibitem{Zirker1993}Zirker, J. B., \& Cleveland, F. M. 1993, Sol. Phys., 144, 341
\end{thebibliography}
\end{document}